\documentclass[12pt, oneside, hidelinks]{article}

\usepackage[hidelinks,colorlinks=true,linkcolor=blue,citecolor=blue]{hyperref}

\usepackage[margin=0.9in]{geometry} 
\usepackage{amsmath,amsthm,amssymb, graphicx, multicol, array,hyperref, bbm, comment, subcaption,  float, booktabs, xpatch, color}
\usepackage{setspace} %change spacing
\usepackage{epigraph}%quotation
\usepackage{titlesec}
\titleformat{\section}
  {\normalfont\Large\bfseries\centering}{\thesection}{1em}{}

\usepackage{algorithm}
\usepackage{algorithmicx}
\usepackage{algpseudocode}

\titlespacing*{\section}
{0pt}{0ex}{0ex}
\titlespacing*{\subsection}
{0pt}{1ex}{0.5ex}
\def\*#1{\mathbf{#1}}
\def\+#1{\mathbb{#1}}
\def\wt#1{\widetilde{#1}}

\newcommand{\tr}{\mathrm{trace}}

\newcommand{\diag}{\mathrm{diag}}

\newcommand{\RNum}[1]{\uppercase\expandafter{\romannumeral #1\relax}}

\usepackage[T1]{fontenc}
\usepackage[utf8]{inputenc}
\usepackage{lmodern}
\usepackage[english]{babel}
\usepackage[autostyle]{csquotes}
\usepackage{amsmath}
\usepackage{natbib}
\setcitestyle{open={(},close={)}}
\usepackage{hyperref} 
\usepackage{xr}
\makeatletter
\newcommand*{\addFileDependency}[1]{
	\typeout{(#1)}
	\@addtofilelist{#1}
	\IfFileExists{#1}{}{\typeout{No file #1.}}
}
\makeatother

\usepackage{enumitem}
\setlist{itemsep=.01em}
\setlist{topsep=.5em}

\newtheorem{innercustomgeneric}{\customgenericname}
\providecommand{\customgenericname}{}

\newtheorem{theorem}{Theorem}
\newtheorem{lemma}{Lemma}
\newtheorem{proposition}{Proposition}
\newtheorem{remark}{Remark}
\newtheorem{corollary}{Corollary}

\newtheorem{assumption}{{Assumption}}

\def\beq{\begin{equation}}
\def\eeq{\end{equation}}
\def\beqr{\begin{eqnarray}}
\def\eeqr{\end{eqnarray}}
\def\beqrs{\begin{eqnarray*}}
\def\eeqrs{\end{eqnarray*}}
\def\bet{\begin{theorem}}
\def\eet{\end{theorem}}
\def\bel{\begin{lemma}}
\def\eel{\end{lemma}}
\def\bep{\begin{proposition}}
\def\eep{\end{proposition}}
\def\bg{\begin{figure}[tbph]\begin{center}}
\def\eg{\end{center}\end{figure}}

\def\bc{\begin{center}}
\def\ec{\end{center}}

\def\wt{\widetilde}
\def\wh{\widehat}

\def\var{\mbox{var}}
\def\cov{\mbox{Cov}}

\def\diag{\mbox{diag}}

\newcommand{\bA}{{\mathbf A}}

\newcommand{\bH}{{\mathbf H}}
\newcommand{\bI}{{\mathbf I}}

\newcommand{\bM}{{\mathbf M}}
\newcommand{\bN}{{\mathbf N}}

\newcommand{\bS}{{\mathbf S}}

\newcommand{\bU}{{\mathbf U}}
\newcommand{\bV}{{\mathbf V}}

\newcommand{\bX}{{\mathbf X}}
\newcommand{\bY}{{\mathbf Y}}

\newcommand{\bb}{{\mathbf b}}

\newcommand{\be}{{\mathbf e}}
\newcommand{\bff}{{\mathbf f}}

\newcommand{\bu}{{\mathbf u}}
\newcommand{\bv}{{\mathbf v}}

\newcommand{\bx}{{\mathbf x}}
\newcommand{\by}{{\mathbf y}}

\newcommand{\balpha} {\boldsymbol{\alpha}}
\newcommand{\bbeta}  {\boldsymbol{\beta}}

\newcommand{\bdelta} {\boldsymbol{\delta}}

\newcommand{\bSigma}{\boldsymbol{\Sigma}}

\newcommand{\bgamma}{\boldsymbol{\gamma}}

\newcommand{\bve}{\mbox{\boldmath$\varepsilon$}}

\newcommand{\btheta} {\boldsymbol{\theta}}

\newcommand{\bmu} {\boldsymbol{\mu}}

\newcommand{\bLambda} {\boldsymbol{\Lambda}}

\newcommand{\bD}{{\mathbf D}}

\newcommand{\ve}{{\varepsilon}}
\renewcommand{\epsilon}{{\ve}}
\renewcommand{\hat}{\widehat}

\def\wt{\widetilde}

\newcolumntype{L}[1]{>{\raggedright\arraybackslash}p{#1}}

\usepackage[margin=10pt,labelfont=bf]{caption}

\begin{document}
%	\title{Stock Market Comovements:  A Data-Driven Approach to Identifying Sparse Common Risk Factors \thanks{\scriptsize We thank.}}
\setstretch{1.3}%1.4
\title{\Large Optimal Bias-Correction and Valid Inference in High-Dimensional Ridge Regression: A Closed-Form Solution}
\date{}
\author{
Zhaoxing Gao$^1$ and Ruey S. Tsay$^2$\footnote{Corresponding author: \href{mailto:ruey.tsay@chicagobooth.edu}{ruey.tsay@chicagobooth.edu} (R.S. Tsay).  Booth School of Business, University of Chicago, 5807 S. Woodlawn Avenue, Chicago, IL 60637, USA.} \\
$^1$University of Electronic Science and Technology of China\\ $^2$University of Chicago
}
	%\author{
		%Zhaoxing Gao\thanks{\scriptsize  Center for Data Science, Zhejiang University. Email: \url{zhaoxing_gao@zju.edu.cn}. I gratefully acknowledge the research support from the NSFC. I thank Ruey S. Tsay for thoughtful and friendly guidance throughout the project, and for encouraging me to develop these results into a paper.}%\\Center for Data Science, Zhejiang University
		%\and
		%Author2\thanks{ \scriptsize  University, Department, Email: \url{abc@abc.edu}.}
		%\and
		%Author3\thanks{ \scriptsize  University, Department , Email: \url{abc@abc.edu}.}
	%}	

	%\begin{onehalfspacing}
		\begin{titlepage}
		\maketitle
			\vspace{-1cm}
			%\begin{center} {\large PRELIMINARY AND INCOMPLETE} \end{center}
			%\vspace{0.5cm}
			
			\begin{abstract}
    Ridge regression is an indispensable tool in big data analysis. 
Yet its inherent bias poses a significant and longstanding challenge, compromising both statistical efficiency and scalability across various applications. To tackle this critical issue, we introduce an iterative strategy to correct bias effectively when the dimension $p$ is less than the sample size $n$. For $p>n$, our method optimally mitigates the bias such that any remaining bias in the proposed de-biased estimator is unattainable through linear transformations of the response data. To address the remaining bias when $p>n$, we employ a Ridge-Screening (RS) method, producing a reduced model suitable for bias correction. Crucially, under certain conditions, the true model is nested within our selected one, highlighting RS as a novel variable selection approach. Through rigorous analysis, we establish the asymptotic properties and valid inferences of our de-biased ridge estimators for both $p<n$ and $p>n$, where, both $p$ and $n$ may increase towards infinity, along with the number of iterations. We further validate these results using simulated and real-world data examples. Our method offers a transformative solution to the bias challenge in ridge regression inferences across various disciplines.

				\vspace{0.5cm}
				

				%\noindent\textbf{JEL classification:} C13, C18, C55
			\end{abstract}

		\end{titlepage}

\setcounter{page}{2}
  
  \abovedisplayskip=0.1pt
\belowdisplayskip=0.1pt
Regularization theory was one of the first signs of the existence of intelligent inference.{---Vladimir N. Vapnik (p.9, \cite{vapnik2013nature})}
		\section{Introduction}
Ridge regression, or more formally $\ell_2$-regularisation estimation, is a fundamental tool in econometrics, statistics, and machine learning with applications in many fields of science, technology, engineering, mathematics, medicine, social sciences, and humanities. The idea of $\ell_2$-regularisation appeared in the early  1940s  for the stability of inverse problems; see \cite{tikhonov1943stability}.
It was first introduced to data analysis by \cite{hoerl1959optimum}  and later formulated in \cite{hoerl1970aridge,hoerl1970bridge} for providing a robust solution to some of the persistent challenges encountered in traditional linear regression techniques; see \cite{hoerl1985ridge} for a nice review. %according to the review article of \cite{hoerl1985ridge}, offering a robust solution to some of the persistent challenges encountered in traditional linear regression techniques. 
Emerging as a fundamental technique in predictive modelling, ridge regression addresses issues such as multicollinearity and overfitting, which commonly afflict predictive models dealing with high-dimensional data. Since its inception, ridge regression's practical adoption persists due to its superior performance over the least-squares estimator in various scenarios, evident in applications across neuroscience, chemistry, biology, and economics; see \cite{leonard2023large}, \cite{zahrt2019prediction}, \cite{otwinowski2014inferring}, \cite{giannone2021economic}, and \cite{abadie2019choosing}, among others, underscoring its empirical effectiveness. % despite theoretical challenges. %The evolution of ridge regression underscores the integration of diverse mathematical concepts and its subsequent impact on econometric/statistical modeling and data analysis, shaping the way researchers approach problems involving high-dimensional datasets and correlated predictors. 
From a shrinkage perspective, the ridge estimator also dominates the least-squares solutions in the sense that its mean-squared errors (MSEs) can be smaller, which provides a reasonable explanation  on the empirical effectiveness of ridge estimators.  See \cite{theobald1974generalizations}, \cite{athey2019machine}, \cite{hastie2020ridge}, \cite{hansen2022econometrics}, and a comprehensive introduction of ridge regression in  \cite{van2023lecture}.

The ridge estimator offers a closed-form expression that simplifies both theoretical and empirical analyses. It aligns with the dense modelling techniques of \cite{giannone2021economic}, which acknowledge the potential significance of all explanatory variables for prediction. Empirical studies, such as those in \cite{giannone2021economic}, indicate that dense models generally tend to outperform the sparse ones in out-of-sample economic prediction. %performance. 
Similarly, \cite{abadie2019choosing} find that the ridge estimators dominate the Lasso and the pre-testing estimators in terms of risks when the effects of different predictors on the dependent variable are ``smoothly distributed”. 
These results suggest that ridge estimators indeed constitute a crucial tool in statistical modelling and economic forecasting, especially in the big data era.

However, as highlighted in Section 2.8 of \cite{athey2019machine}, constructing valid confidence intervals remains a challenge for many regularised methods, including ridge regression, even in asymptotic settings. This long-standing challenge in ridge-type regression involves at least two critical aspects within a linear regression framework:
(a) performing hypothesis testing on specific linear combinations of the regression coefficients using the  ridge estimators; and
(b) deriving confidence or prediction intervals based on the ridge estimators in empirical applications.
The primary reason for these challenges is that the inherent bias of ridge estimators poses significant challenges, compromising both statistical efficiency and scalability across various applications. To date, the feasibility of conducting statistical inferences and hypothesis testing on ridge-type estimators without imposing additional structural constraints remains largely unexplored in the literature. This complexity arises from the ridge estimator's intrinsic bias, which complicates direct statistical inferences despite its elegant closed-form expression.
As a result, research addressing the inference challenges of ridge regression in high-dimensional settings is limited, leading to its widespread application across disciplines without comprehensive theoretical investigations.

To the best of our knowledge, there are only a few works in the literature concerning the bias and inference of ridge estimators under different scenarios. Assuming a sparse structure, \cite{shao2012estimation} proposed a threshold ridge regression method and proved its consistency. The method therein actually estimates the projected coefficient vector rather than the true one in the linear model. \cite{dobriban2018high} derived the limit of the predictive risk of ridge regression and regularised discriminant analysis in a dense random effects model. \cite{buhlmann2013statistical} proposed to use the Lasso to correct the bias of ridge estimators. However, the estimation depends on the existence of an initial estimator which must be sufficiently accurate. \cite{zhang2022ridge} adopted a similar approach as that in \cite{shao2012estimation} and proposed a threshold ridge regression and a bootstrap method to make inferences. However, the method therein still estimates the projected coefficient vector rather than the true one, and the remaining bias may not be asymptotically negligible in general.

In this paper, we introduce a systematic approach tailored specifically for mitigating the bias in ridge-type estimators for high-dimensional linear regression models. Leveraging the closed-form expression of the ridge estimators, the bias term can also be established in an analytic form. Although the bias term involves the true parameters which are unknown in practice, we found that replacing the true parameters with the ridge estimators  turns out to be an effective way to mitigate the bias.
Therefore, the proposed method employs an iterative bias-correction strategy, and the bias can be reduced substantially when the number of iterations is sufficiently large.
 Notably, it achieves complete bias correction if the covariate dimension $p$ is smaller than the sample size $n$, and can reduce considerable bias if $p$ surpasses $n$. We show that our bias-correction method is an optimal one in the sense that the bias can be completely corrected (asymptotically) when $p< n$ with a sufficient number of the proposed bias-correction iterations, and the remaining bias of the de-biased ridge estimator  when $p>n$ is unattainable through any linear transformations of the response vector.

To further combat the remaining bias in the de-biased ridge estimators when $p>n$, we introduce a novel ridge-screening (RS) method for selecting covariates prior to applying our bias correction procedure. The RS approach constructs a restricted model that inherently encompasses the true model as a subset. This is based on the assumption that only a subset of the covariates holds significance in the linear model. Specifically, we postulate that the number of significant covariates should be less than the sample size, which can appropriately diverge relative to the dimension $p$ and sample size $n$. 
Crucially, we demonstrate that under certain mild conditions, the true model is inherently nested within the selected one, establishing RS as a novel variable selection approach that offers additional value beyond bias correction. Leveraging this restricted model, our bias-correction procedure can be further applied to the restricted model and effectively rectifies the bias in the resulting ridge estimators

%we propose a novel ridge-screening (RS) method for covariate selection first before applying our bias correction procedure. This RS approach constructs a restricted model that inherently includes the true model as a subset under the assumption that only a subset of the covariates are significant in the linear mode, specifically, we require the number of significant covariates is less than the sample size which can diverge at an appropriate rate in relation to the dimension $p$ and the sample size $n$. Importantly, the selected model is shown to inherently include the true model as a subset, making RS a novel variable selection approach, which also provides independent value beyond bias rectification. 
%the number of significant covariates is , offering theoretical guarantees.
%Leveraging this restricted model, our bias-correction procedure can be further applied to the restricted model and effectively rectifies the bias in the resulting ridge estimators. %Moreover, the RS method introduces a unique approach to variable selection, extending beyond bias rectification. By integrating this method, we ensure not only bias-free estimators but also a refined selection process for predictor variables.

The derivation of the proposed methodology and theory is mainly based on a fixed design, which is similar to the setting in \cite{hansen2022modern}, but the results throughout this paper remain valid for random regressors by conditioning on the design matrix. 
Thus, this paper rigorously establishes asymptotic properties and provides valid inferences of the proposed bias-corrected estimators for both $p< n$ and $p>n$ under some relaxed and intuitive conditions. Furthermore, we delve into the bias-variance trade-off of our de-biased ridge estimators, examining its relationship with the number of iterations in the bias-correction procedure. To validate our methodology, we provide empirical evidence using both simulated and real-world data.  The prediction intervals constructed by the proposed method are indeed satisfactory for forecasting the U.S. macroeconomic series using factor-augmented regression. Moreover, the RS method can further enhance the coverage rates of these prediction intervals. These empirical validations highlight the practical efficacy and adaptability of 
the proposed approach across a wide range of high-dimensional regression settings.

The contributions of this paper are multi-fold. First, the proposed bias-correction method is simple and easy to implement. In fact, the proposed approach is a systematic procedure and the resulting de-biased estimator has a closed-form expression. Second, the optimality of the proposed bias-correction procedure consists of two aspects: (a) it achieves complete bias correction when the covariate dimension $p$ is smaller than the sample size $n$; and (b) the remaining bias of the de-biased ridge estimator in the scenario when $p>n$ is unattainable through any linear transformations of the response vector. These results distinguish our work from the existing ones that only part of the bias can be corrected in most of the aforementioned literature. Third, we also propose a novel variable selection procedure, namely the ridge-screening (RS) method, which screens out some insignificant variables based on the de-biased ridge estimator due to its optimality.  Fourth, we establish certain asymptotic properties of the de-biased ridge estimators in both scenarios when $p< n$ and $p>n$, and it is shown that the de-biased ridge estimators are asymptotically normal, which provides valid inference methods for the ridge estimators. Fifth, we develop a procedure to construct confidence and prediction intervals in ridge regressions using the proposed de-biased estimators and associated inference methods. Finally, we establish the bias-variance trade-off of the proposed approach both theoretically and through validation with simulated data. It's important to note that, unlike the scenario described in Section 2.8 of \cite{athey2019machine} where many inference approaches for regularised machine learning methods compromise predictive performance, our proposed procedure focuses  on correcting the bias and rendering the estimators suitable for statistical inferences, without adversely affecting their predictive performance.
 
We highlight that the asymptotic framework adopted in this paper is slightly different from traditional approaches in the literature. Typically, in conventional frameworks, the asymptotic properties are established when the dimension and/or the sample size are approaching infinity. However, in our study, most of the asymptotic results are derived under a different scenario that the number of iterations in the bias-correction procedure tends to infinity for any given configuration of the dimension $p$ and the sample size $n$. This configuration can be sufficiently large to encompass the framework of big data analysis. One of the primary motivations behind this choice is to demonstrate the validity of the proposed procedure by explicitly providing the exact bias and covariance terms of the de-biased estimators in this paper. This approach is also reasonable because it reflects common scenarios encountered in empirical data analysis, where datasets often have fixed dimensions and sample sizes.  Without this setting, the dimensions of the bias and covariance terms of the de-biased estimators would expand to infinity if we considered $n$ and $p$ approaching infinity, making them challenging to formulate and describe theoretically. Moreover, our asymptotic results remain valid as $n$ and $p$ approach infinity. This holds true so long as we initially allow the number of iterations to increase towards infinity at a moderate rate. In this asymptotic manner, we can symbolically retain the forms of the bias and covariance terms for a growing configuration of $(p, n)$. Therefore, it's important to emphasise that our chosen asymptotic framework does not undermine the validity of the proposed bias-correction procedures.

%We mention that the asymptotic framework considered in this paper is slightly different from the traditional ones in the literature, where the dimension and/or the sample size are usually allowed to go to infinity. Most of the asymptotic results in this paper are established under the condition that the number of iterations in the bias-correction procedure goes to infinity for any given configuration of the dimension $p$ and the sample size $n$, which can be sufficiently large to embrace the framework of big data analysis. Part of the reason is that we want to show the validity of our proposed procedure by providing the exact bias and covariance terms of the de-biased estimators explicitly in this paper. Otherwise, the dimensions of the bias and the covariance terms of the de-biased estimators will expand to infinity if we consider $n,p\rightarrow\infty$, which are difficult to describe in theory. Furthermore, our asymptotic results are still valid as $n$ and $p$ go to infinity so long as the number of iterations approaches infinity at a moderate rate if we symbolically keep the bias and covariance terms for a growing configuration of $(p,n)$.

The rest of the paper is organised as follows: Section \ref{sec2} introduces the ridge estimation and its bias-correction procedure in scenarios when $p< n$ and $p>n$, where a ridge-screening method is also introduced for variable selection.  Section \ref{sec2} presents the inference methodologies for the proposed de-biased estimators. Section \ref{sec3} examines the finite-sample performance pf the proposed method via Monte-Carlo simulations. Section~\ref{sec40} provides an empirical application of the proposed method, and Section \ref{sec4}  offers some concluding insights. All proofs and derivations for the asymptotic results are available in an online Supplementary Material.

{\bf Notation:}  We use the following notation. For a $p\times 1$ vector
$\bu=(u_1,..., u_p)'$, $\|\bu\|_2=\sqrt{\sum_{i=1}^p|u_i|^2}$ is the $\ell_1$-norm and $\|\bu\|_\infty=\max_{1\leq i\leq p}|u_i|$ is the $\ell_\infty$-norm. $\bI_p$ denotes the $p\times p$ identity matrix. For a matrix $\bH$, %its Frobenius norm is $\|\bH\|=[\tr(\bH'\bH)]^{1/2}$ and
its operator norm is $\|\bH
\|_2=\sqrt{\lambda_{\max} (\bH' \bH ) }$, where
$\lambda_{\max} (\cdot) $ denotes the largest eigenvalue of a matrix, and $\|\bH\|_{\min}$ is the square root of the minimum non-zero eigenvalue of $\bH\bH'$. $|\bH|$ denotes the absolute value of $\bH$ elementwisely. The superscript ${'}$ denotes the 
transpose of a vector or matrix. We also use the notation $a\asymp b$ to denote $a=O(b)$ and $b=O(a)$ or $a$ and $b$ are of the same order.

		\section{Models and Methodology} \label{sec2}
		
		\subsection{High-dimensional Linear Regression}\label{model_overview}
	Let $\{(\bx_1,y_1),...,(\bx_n,y_n)\}$ be a given sample of centred observable data. We consider the problem of estimating a $p$-dimensional vector $\bbeta$ from the following linear model:
 \begin{equation}\label{hlm}
     y_i=\bx_i'\bbeta+\ve_i,\,\,i=1,...,n,
 \end{equation}
where $y_i$ is a scalar response variable, $\bx_i$ is a $p$-dimensional covariate vector, and $\ve_i$ is a random error term with mean zero and finite variance.  Similar to the setting in \cite{hansen2022modern} and Ch. 29 of \cite{hansen2022econometrics}, we treat the $n\times p$ design matrix $\bX=(\bx_1,...,\bx_n)'$ consisting of $p$ covariates as a fixed one. But the estimation results throughout the paper remain valid for random regressors by conditioning on the design matrix data $\bX$. Note that Model (\ref{hlm}) can be expressed in vector form
\begin{equation}\label{v:hlm}
    \by=\bX\bbeta+\bve,
\end{equation}
where $\by=(y_1,...,y_n)'$ is an $n$-dimensional response vector,  and $\bve=(\ve_1,...,\ve_n)'$ is an $n$-dimensional vector of noises with $E(\bve)={\bf 0}$ and $\cov(\bve)=\bSigma_\ve$, where $\bSigma_\ve$ is a diagonal matrix with positive and bounded diagonal elements.  
If $p< n$ and $\bX'\bX$ is an invertible matrix, the least-squares estimator for $\bbeta$ is 
\begin{equation}\label{lse}
    \wh\bbeta_{lse}=(\bX'\bX)^{-1}\bX'\bY.
\end{equation}
This least-squares estimator $\wh\bbeta_{lse}$ is only well-defined if $(\bX'\bX)^{-1}$ exists. In a high-dimensional setting, if the columns of the design matrix $\bX$ are exactly linearly dependent, for example, this is obviously true when $p>n$, this collinearity among the columns implies that $\bX'\bX$ is singular, rendering $\wh\bbeta_{lse}$ an invalid estimator. To make the least-squares estimator in (\ref{lse}) a well-defined quantity, we modify the definition in (\ref{lse}) as
\begin{equation}\label{lse:mp}
    \wh\bbeta_{lse}=(\bX'\bX)^{+}\bX'\bY,
\end{equation}
where $(\bX'\bX)^{+}$ is the Moore-Penrose generalised inverse. It is not hard to see that the estimator in (\ref{lse:mp}) reduces to the one in (\ref{lse})
if $p<n$ and $\bX'\bX$ is invertible. Therefore, we will denote the estimator in (\ref{lse:mp}) as the least-squares solution throughout this article.

\subsection{Ridge Regression}
The ridge regression estimator was first proposed by \cite{hoerl1959optimum}; see the review article of \cite{hoerl1985ridge}. It essentially comprises of an ad-hoc fix to resolve the singularity issue of $\bX'\bX$ in the presence of many covariates. Suppose $(\bX'\bX+\lambda\bI_p)$ is invertible for a given $\lambda>0$, the ridge estimator is defined as
\begin{equation}\label{rg:e}
    \wh\bbeta(\lambda)=(\bX'\bX+\lambda\bI_p)^{-1}\bX'\by,
\end{equation}
which simply replaces $\bX'\bX$ by $\bX'\bX+\lambda\bI_p$ with a tuning parameter $\lambda>0$ in the least-squares estimator of (\ref{lse}).  From a regression point of view, the ridge estimator can be obtained in the following way. Let $\by^*=(\by',{\bf 0})'$ and $\bX^*=(\bX',\sqrt{\lambda}\bI_p)'$ be the augmented data, the ridge estimator is a solution to the following optimisation problem:
\begin{equation}
    \wh\bbeta(\lambda)=\arg\min_{\bbeta\in R^p}\left\{\|\by^*-\bX^*\bbeta\|_2^2\right\}=\arg\min_{\bbeta\in R^p}\left\{\|\by-\bX\bbeta\|_2^2+\lambda\|\bbeta\|_2^2\right\}.
\end{equation}
From the expression of the ridge estimator for a given $\lambda>0$,
it is not hard to see that
\[\wh\bbeta(\lambda)\rightarrow\wh\bbeta_{lse},\,\, \text{as}\,\, \lambda\rightarrow 0,\]
 and 
\[\lambda\wh\bbeta(\lambda)\rightarrow\bX'\by,\,\,\text{as}\,\, \lambda\rightarrow \infty,\]
which is the componentwise regression estimator if each covariate is standardised. Therefore, a large $\lambda$ would reduce the variance of the estimator, but may increase the bias. In fact, the bias increases as $\lambda$ grows. 

In this paper, we only focus on a given $\lambda>0$ and investigate the bias-correction and inference issues for $\wh\bbeta(\lambda)$. For the purpose of comparisons with the proposed approach, we first specify the initial bias of the ridge estimator of (\ref{rg:e}) in the following theorem.

\begin{theorem}\label{thm1}
If $\by$ admits a linear structure as that in (\ref{v:hlm}) with $E\bve={\bf 0}$, then the bias of the ridge estimator $\wh\bbeta(\lambda)$ in (\ref{rg:e}) is
\begin{equation}\label{bt}
    \bb_{\lambda,0}=\bbeta-E\wh\bbeta(\lambda)=\lambda (\bX'\bX+\lambda\bI_p)^{-1}\bbeta,
\end{equation}
for any given $\lambda>0$ such that $(\bX'\bX+\lambda\bI_p)$ is invertible.
\end{theorem}
\begin{remark}
    The condition for the result of Theorem~\ref{thm1} to hold is the same as that for the linear regression model. If the design matrix $\bX$ is random with either independent and identically distributed  ($i.i.d.$) or weakly dependent columns, we require $E(\bve|\bX)={\bf 0}$, and then, the bias of the ridge estimator conditioning on $\bX$ and a given $\lambda>0$ is
    \[\text{bias}[\wh\bbeta(\lambda)|\bX]=\bbeta-E[\wh\bbeta(\lambda)|\bX]=\lambda (\bX'\bX+\lambda\bI_p)^{-1}\bbeta,\]
    which is the same as that in (\ref{bt}). See also Ch. 29.6 in \cite{hansen2022econometrics} for details.
\end{remark}
%To see the bias of the ridge estimator, it follows from (\ref{v:hlm}) that
%\begin{equation}\label{decom}
%    \wh\bbeta(\lambda)=\bbeta-\lambda(\bX'\bX+\lambda\bI_p)^{-1}\bbeta+(\bX'\bX+\lambda\bI_p)^{-1}\bX'\bve,
%\end{equation}
%where $\approxeq$ is an equality if $\bX$ is a fixed design matrix, and it is an approximation if $\bX$ is random since $\bX'\bX/T$ converges to a non-random matrix  when $T$ is large enough.

From Theorem 1, the bias of the ridge estimator depends on the unknown true parameter $\bbeta$. Therefore, it is fundamentally challenging to make any statistical inference on the ridge estimator or to construct any confidence or prediction intervals involving the ridge estimator. Consequently, it is important to seek an effective way to correct or reduce the bias of a ridge estimator.

 In the next section, we will tackle this issue by proposing an iterative procedure to reduce the bias of the ridge estimator.

\subsection{Bias-Correction}\label{sec23}
We divide the discussion into two key parts depending on whether $\bX'\bX$ is invertible or singular. For simplicity and in line with the setting  in \cite{wang2016high}, we assume that $\bX'\bX$ is invertible when $p< n$ and singular when $p>n$ throughout this article. An alternative framework is to consider the scenarios that $p/n\in (0,1)$ and $p/n\in(1,\infty)$, a common setting in random matrix theory, which is also helpful in establishing the asymptotic results if we further allow $n,p\rightarrow\infty$ later. %\footnote{An alternative framework is to consider the scenarios that $p/n\in (0,1)$ and $p/n\in(1,\infty)$, a common setting in random matrix theory, which is also helpful in establishing the asymptotic results if we further allow $n,p\rightarrow\infty$ later.}. 
 This assumption is well-justified as we consider a fixed design matrix $\bX$ in this paper and  $\bX'\bX$ naturally maintains its invertibility when $p< n$. Furthermore, in extreme cases where highly correlated variables are present within $\bX$ (in a random sense), we may implement specific transformations based on prior knowledge or statistical methods such as the hierarchical clustering or the $k$-means algorithm to mitigate these correlations prior to conducting ridge regression; see the discussion in Section 4.1.2 of \cite{fan2008sure}. Consequently, we only focus on the bias-correction issue in this paper and rule out the case when some covariates in $\bX$ are highly correlated.

%Moreover, in the extreme case that there exist highly correlated variables in $\bX$ (from a random sense), we may apply the subject related transformation according to some prior knowledge or statistical transformation tools such as the hierarchical clustering or $k$-mean algorithm to create weakly correlated first before the ridge regression; see the discussion in Section 4.1.2 of \cite{fan2008sure} for detail. Therefore, we only focus on the bias-correction in this paper and rule out the case when the covariates are highly correlated in $\bX$.

%hierarchical clustering or $k$-mean algorithm ased on the correlation matrices to group variables into highly correlated groups.  
%because the singular case when $p\leq n$ is analogous to  the situation when $p>n$. %A similar setting is also adopted in \cite{wang2016high}.

%we only consider the scenarios when the dimensionality of the covariates $p$ is less than or equal to the sample size $n$ and the case when $p>n$, where we assume $\bX'\bX$ is invertible if $p\leq n$ since the singular case in this scenario is the same as the situation when $p>n$.

The ridge estimator in (\ref{rg:e}) can be written as
\begin{equation}\label{decom:1}
    \wh\bbeta(\lambda)=\bbeta-\lambda(\bX'\bX+\lambda\bI_p)^{-1}\bbeta+(\bX'\bX+\lambda\bI_p)^{-1}\bX'\bve.
\end{equation}
The bias-correction procedure is based on  this expression and (\ref{bt}) in Theorem \ref{thm1}. The rationale for the procedure is as follows. Since $\bbeta$ in the bias term of (\ref{bt}) is unknown, we first replace it by $\wh\bbeta(\lambda)$ defined in (\ref{rg:e}) and construct a first-step de-biased ridge estimator as
 \begin{equation}\label{bc:0}
\wh\bbeta_{c,1}(\lambda)=\wh\bbeta(\lambda)+\lambda(\bX'\bX+\lambda\bI_p)^{-1}\wh\bbeta(\lambda).
 \end{equation}
Plugging $\wh\bbeta(\lambda)$ from (\ref{decom:1}) into (\ref{bc:0}), we obtain:
\begin{equation}\label{bc1:rp}
    \wh\bbeta_{c,1}(\lambda)=\bbeta-\lambda^2(\bX'\bX+\lambda\bI_p)^{-2}\bbeta+(\bX'\bX+\lambda\bI_p)^{-1}\bX'\bve+\lambda(\bX'\bX+\lambda\bI_p)^{-2}\bX'\bve.
\end{equation}
Consequently, the bias term of $\wh\bbeta_{c,1}(\lambda)$ is
\begin{equation}\label{bc:1}
     \bb_{\lambda,1}=\bbeta-E(\wh\bbeta_{c,1}(\lambda))=\lambda^2(\bX'\bX+\lambda\bI_p)^{-2}\bbeta.
 \end{equation}
It is then apparent that the $\ell_2$-norm of the bias term $\bb_{\lambda,1}$ produced by $\wh\bbeta_{c,1}(\lambda)$ is smaller than that of the initial bias $\bb_{\lambda,0}$ in Theorem \ref{thm1} under some mild conditions, implying that the bias  $\bb_{\lambda,0}$ has been partially corrected by $\wh\bbeta_{c,1}(\lambda)$. To see this, we conduct a singular-value decomposition on $\bX$ or a spectral decomposition on $\bX'\bX$, and the effectiveness of the bias-correction approach depends on two observations: (a) the eigenvalues of $\lambda(\bX'\bX+\lambda\bI_p)^{-1}$ are positive and strictly less than one; and (b) the eigenvalues of $\lambda^2(\bX'\bX+\lambda\bI_p)^{-2}$ becomes smaller in the de-biased estimator of (\ref{bc1:rp}) compared to those of  $\lambda(\bX'\bX+\lambda\bI_p)^{-1}$ in $\bb_{\lambda,0}$.

Following the first correction, we replace the unknown vector $\bbeta$ in (\ref{bc:1}) by the ridge estimator $\wh\bbeta(\lambda)$ again to obtain a second-step de-biased estimator:
\begin{equation}\label{bc2:rp}
    \wh\bbeta_{c,2}(\lambda)=\wh\bbeta_{c,1}(\lambda)+\lambda^2(\bX'\bX+\lambda\bI_p)^{-2}\wh\bbeta(\lambda)=\wh\bbeta(\lambda)+\sum_{j=1}^2\lambda^j(\bX'\bX+\lambda\bI_p)^{-j}\wh\bbeta(\lambda).
\end{equation}
By a similar argument, the bias term of $\wh\bbeta_{c,2}(\lambda)$ is
\begin{equation}\label{bc:2}
    \bb_{\lambda,2}=\bbeta-E(\wh\bbeta_{c,2}(\lambda))=\lambda^3(\bX'\bX+\lambda\bI_p)^{-3}\bbeta,
\end{equation}
where the eigenvalues of $\lambda^3(\bX'\bX+\lambda\bI_p)^{-3}$ are even smaller compared to those of $\lambda^2(\bX'\bX+\lambda\bI_p)^{-2}$ in $\bb_{\lambda,1}$. Consequently, we can  show that the $\ell_2$-norm of the bias $\bb_{\lambda,2}$ is smaller than that of $\bb_{\lambda,1}$ under the same framework as before. We repeat  this procedure and denote
\begin{equation}\label{bck:rp}
    \wh\bbeta_{c,k}(\lambda)=\wh\bbeta(\lambda)+\sum_{j=1}^k\lambda^j(\bX'\bX+\lambda\bI_p)^{-j}\wh\bbeta(\lambda)
\end{equation}
as a de-biased estimator at the $k$-th step, where we define $\wh\bbeta_{c,0}(\lambda)=\wh\bbeta(\lambda)$ for $k=0$. To characterize the effect of the bias correction, we make an assumption on the singular-value decomposition (SVD) of $\bX$ first.
%%%%
\begin{assumption}\label{asm1}
    For $p< n$, $\bX'\bX$ is invertible and the SVD of $\bX$ is $\bX=\bV_1\bD_1\bU_1'$, where $\bU_1'\bU_1=\bV_1'\bV_1=\bI_p$ and $\bD_1=\diag(d_1,...,d_p)$ with $d_i>0$,  for $1\leq i\leq p$.
    \end{assumption}
Assumption~\ref{asm1} is intuitive for a fixed design $\bX$ with a large and fixed configuration of $(p,n)$. For example, the eigenvalues of $\bX'\bX$ are of order $n$ if the entries of $\bX$  are independent copies of a random variable with zero mean, unit variance, and
finite fourth moment if $p/n\in (0,1)$; see the Bai-Yin's law in \cite{bai1993limit}.  Consequently, the eigenvalues of $\lambda(\bX'\bX+\lambda\bI_p)^{-1}$ are strictly less than one, which ensures that our iterative procedure can substantially reduce the bias. 
In fact, we have the following theorem  for the de-biased ridge estimator in (\ref{bck:rp}).
\begin{theorem}\label{thm2}
     If $p<n$ and $\bX'\bX$ is invertible, then under Assumption~\ref{asm1}, the bias of the de-biased ridge estimator $\wh\bbeta_{c,k}(\lambda)$ defined in (\ref{bck:rp}) is
     \begin{equation}\label{bc:k}
         \bb_{\lambda,k}=\bbeta-E(\wh\bbeta_{c,k}(\lambda))=\lambda^{k+1}(\bX'\bX+\lambda\bI_p)^{-(k+1)}\bbeta.
     \end{equation}
     Furthermore, for any configuration of $(p,n)$ with $\|\bbeta\|_2<C_{n,p}<\infty$, if the number of iterations $k$ satisfies $\max_{1\leq j\leq p}C_{n,p}(\frac{\lambda}{d_j^2+\lambda})^{k+1}\rightarrow 0$, we have
     \[\bb_{\lambda,k}\rightarrow {\bf 0},\,\, \text{as}\,\, k\rightarrow\infty.\]
\end{theorem}
\begin{remark}\label{rm2}
    %(i) As previously mentioned, we only consider the scenario where $\bX'\bX$ is invertible if $p\leq n$, and it is inherently singular if $p>n$. While it's conceivable that some columns of $\bX$ might be strongly correlated, causing $\bX'\bX$ to become singular when $p\leq n$ in practice, the approach to handle this situation aligns with the ridge-screening method introduced later for the case when  $p>n$. Consequently, throughout this article, we treat $\bX'\bX$ as an invertible matrix if $p\leq n$. \\
    (i) We observe that the assumptions required for Theorem~\ref{thm2} are quite minimal. We only need the fundamental assumptions inherent to linear regression models, ensuring that the bias term can be asymptotically eliminated with a sufficient number of iterations.\\
    (ii) The requirement for the $\ell_2$-norm of $\bbeta$ to be finite stems from the expectation of the response having a limited number of significant covariates in linear regression models. Without this restriction, as the number of covariates grows, each coefficient's contribution might become sufficiently small, allowing the response's variance to remain finite. It's important to highlight that the condition for the number of iterations $k$ can still be met even if $\|\bbeta\|_2\rightarrow\infty$ at a polynomial rate. This is because $(\frac{\lambda}{d_j^2+\lambda})^{k+1}$  decays exponentially, given that $|\frac{\lambda}{d_j^2+\lambda}|<1$, for $1\leq j\leq p$. Thus, the number of iterations $k$ can be selected to scale logarithmically with the dimension $p$.\\
    (iii) As discussed in the Introduction section, the asymptotic results in Theorem~\ref{thm2} are established for any given 
$(p,n)$ as 
$k\rightarrow\infty$.  This approach is also reasonable because it reflects common scenarios encountered in practical data analysis, where datasets often have fixed dimensions and sample sizes.
%This flexibility is particularly beneficial since we frequently work with specific datasets characterized by fixed dimensions and sample sizes. Our methodology is specifically designed to address the bias introduced by ridge estimators.
As a matter of fact, if Assumption~\ref{asm1} holds for  increasing $p$ and $n$ with $p<n$, the results remain applicable when $k\rightarrow\infty$ first, followed by
$n,p\rightarrow\infty$. In addition, under the framework in \cite{bai1993limit} that $d_i^{2}\asymp n$, the asymptotic results hold simultaneously as $n,p,k\rightarrow\infty$ if $p/n\in(0,1)$ and the penalty parameter $\lambda\asymp n$  because $\max_{1\leq j\leq p}|\frac{\lambda}{d_j^2+\lambda}|<1$ still holds in this setting.\\
%This is contingent upon the number of iterations $k$ satisfying the conditions outlined in Theorem~\ref{thm2}. Such a condition can be easily met, as elaborated in Remark \ref{rm2}(iii). \\
(iv) Although the convergence in Theorem~\ref{thm1} is based on a given $\lambda>0$, it can be readily shown that 
\[\sup_{\lambda\in[\lambda_1,\lambda_2]}\|\bb_{\lambda,k}\|_2\rightarrow{0},\,\,\text{as}\,\, k\rightarrow\infty,\,\,\text{for}\,\,0<\lambda_1\leq \lambda_2<\infty,\]
implying that the convergence to zero is uniformly  for a range of $\lambda$. Similar argument also applies to the asymptotic results in Theorem~\ref{thm3}-\ref{thm5} below.
   % As explained in the {\it Introduction} section, the asymptotic results in Theorem~\ref{thm2} are established for any given $(p,n)$ as $k\rightarrow\infty$. This is particularly useful since we often deal with a specific data set with fixed dimension and sample size, and our procedure aims to correct the bias incurred by the ridge estimators.
    %On the other hand, if Assumption~\ref{asm1} holds for diverging dimension $p$, the results still hold if we allow $n,p,k\rightarrow\infty$ so long as the number of iterations $k$ satisfies the condition in Theorem~\ref{thm2}, which can be easily achieved as described in Remark \ref{rm2}(iii) above.
\end{remark}
Theorem~\ref{thm2} implies that we can  completely correct the bias term incurred by the ridge estimator $\wh\bbeta(\lambda)$ if $p<n$ so long as we conduct a sufficient number of iterations. This result is particularly useful for empirical data analysis when $n$ and $p$ are given.%, and we only need to conduct a sufficiently large number of bias-corrections on the ridge estimators.

To investigate the performance of the de-biased estimator $\wh\bbeta_{c,k}(\lambda)$ when $p>n$ under which $\bX'\bX$ is singular, we first perform a singular-value-decomposition on $\bX$. Suppose the true rank of $\bX$ is $\text{rank}(\bX)=p^*\leq  \min(p,n)=n$, where we can simply set $p^*=n$ since we deal with a given and fixed configuration of $(p,n)$ without including highly correlated covariates. However, the results in Theorem~\ref{thm3} still hold for $p^*<n$.
By abuse of notation, there exist semi-orthogonal matrices $\bV_1\in R^{n\times p^*}$ and $\bU_1\in R^{p\times p^*}$, and a diagonal matrix $\bD_1=\diag(d_1,...,d_{p^*})$ with $d_1\geq ...\geq d_{p^*}>0$ such that
\begin{equation}\label{svd:x}
\bX=\bV_1\bD_1\bU_1'\,\,\text{and}\,\,\bX'\bX=\bU_1\bD_1^2\bU_1'.
\end{equation}
%\[\bX=\bV_1\bD_1\bU_1'\,\,\text{and}\,\,\bX'\bX=\bU_1\bD_1^2\bU_1'.\]
Since $\bX'\bX+\lambda\bI_p$ is symmetric and invertible, there also exists an orthogonal complement matrix $\bU_2\in R^{p\times (p-p^*)}$ of $\bU_1$ such that
\begin{equation}\label{svd:xxi}
 \bX'\bX+\lambda\bI_p=\bU\bD\bU', \,\,\text{where}\,\, \bU=[\bU_1,\bU_2]\,\,\text{and}\,\,\bD=\diag(\bD_1^2+\lambda\bI_{p^*},\lambda\bI_{p-p^*}).
\end{equation}
We formulate the above description in Assumption~\ref{asm2} below.
%%%%%%%%
\begin{assumption}\label{asm2}
    Rank($\bX$)$=p^*\leq \min(p,n)=n$ if $p>n$, and the design matrix $\bX$ has a SVD $\bX=\bV_1\bD_1\bU_1'$ such that  $\bX'\bX+\lambda\bI_p=\bU\bD\bU'$, where $\bD$, $\bD_1$, $\bU$, $\bU_1$,  $\bV_1$, and $\bU_2$ are defined in (\ref{svd:x}) and (\ref{svd:xxi}).
\end{assumption}
Then, we have the following theorem.
\begin{theorem}\label{thm3}
    If $p>n$ and $\bX'\bX$ is singular, but $\bX'\bX+\lambda\bI_p$ is invertible for a given $\lambda>0$. Under Assumption~\ref{asm2}, the bias term $\bb_{\lambda,k}$ of $\wh\bbeta_{c,k}(\lambda)$ is the same as (\ref{bc:k}) in Theorem~\ref{thm2}. If the number of iterations $k$ satisfies $\max_{1\leq j\leq p^*}C_{n,p}(\frac{\lambda}{d_j^2+\lambda})^{k+1}\rightarrow 0$, where the constant $C_{n,p}$ is the same as that in Theorem~\ref{thm2}, we have
    \[\bb_{\lambda,k}\rightarrow\bU_2\bU_2'\bbeta,\,\,\text{as}\,\, k\rightarrow \infty,\]
    where $\bU_2$ is defined in (\ref{svd:xxi}).
\end{theorem}
%%%%%
\begin{remark}\label{rm3}
   (i) The requirement for the number of iterations $k$ is the same as that in Theorem~\ref{thm2}, and therefore, we omit the illustrations to save space.\\
   (ii) The bias term in Theorem~\ref{thm3} corroborates the assertion in \cite{shao2012estimation} that ridge regression primarily estimates 
 $\bU_1\bU_1'\bbeta$ rather than $\bbeta$.\\
   (iii) Similar to Theorem~\ref{thm2}, the asymptotic results in Theorem~\ref{thm3} are established for any given configuration of $(p,n)$ as $k\rightarrow\infty$, because we can explicitly specify the bias term $\bU_2\bU_2'\bbeta$ for fixed $p$ and $n$. If we additionally allow $n,p\rightarrow\infty$, the result in Theorem~\ref{thm3} can be reformulated as
   \[ \bb_{\lambda,k}-\bU_2\bU_2'\bbeta\rightarrow{\bf 0},\,\,\text{as}\,\, n,p,k\rightarrow \infty,\]
   so long as $\max_{1\leq j\leq p^*}C_{n,p}(\frac{\lambda}{d_j^2+\lambda})^{k+1}\rightarrow 0$ and $d_j>0$, for $1\leq j\leq p^*$, where $p^*$ is the rank of $\bX$.\\
   (iv) It is possible that the rank of $\mathbf{X}$ is $p^*<p$ in the case of $p < n$. In such situations, we can obtain similar results as those outlined in Theorem~\ref{thm3}. The approach to handling this scenario is similar to that for $p>n$, and therefore, we omit further discussion on this issue and focus on the previously mentioned setting.
\end{remark}
%%%%%%
A key insight from Theorem~\ref{thm3} is that there is a remaining bias term $\bU_2\bU_2'\bbeta$ that cannot be corrected by the proposed method. This is understandable since the projection of $\bbeta$ on the singular directions $\bU_2$ is not captured in Model (\ref{v:hlm}) according to the singular-value-decomposition of $\bX$ in 
(\ref{svd:x}). If the vector $\bbeta$ belongs to the space spanned by the columns of $\bU_1$, there will be no need to correct the bias since $\bU_2'\bbeta={\bf 0}$. Otherwise, it is challenging to make such a correction for the bias in Theorem~\ref{thm3}.  As a matter of fact, we have the following theorem regarding the bias term in Theorem~\ref{thm3}.
\begin{theorem}\label{thm4}
    Under the conditions in Theorem~\ref{thm3}, there is no linear transformation matrix $\bS\in R^{p\times n}$ such that $E[\bS\by]=\bU_2\bU_2'\bbeta$.
\end{theorem}
We focus on linear transformations of the data  in Theorem~\ref{thm4} because the least-squares estimator and the ridge estimator are all linear combinations of the data $\by$. Theorem~\ref{thm4} indicates that the remaining bias term $\bU_2\bU_2'\bbeta$ is unattainable through any linear transformation of the data $\by$, showing that the proposed approach does its best to 
correct the bias. The results in Theorems~\ref{thm2}--\ref{thm4} also indicate that our bias-correction method is an optimal one for any given $(p,n)$.

Simulation results (e.g., Figure~\ref{fig-4}) in Section~\ref{sec3} suggest that the uncorrectable bias $\bU_2\bU_2'\bbeta$ can be significant even when the number of nonzero elements in $\bbeta$ is small. This underscores that any method that only  corrects part of the bias may lead to poor inference performance.

\subsection{Ridge Screening}\label{sec24}
To address further the uncorrectable bias identified in Theorem~\ref{thm3} and to ensure valid statistical inferences when $p>n$, we impose some additional structures on the parameters or covariates. Without these additional structures, as demonstrated in Theorem~\ref{thm4}, the bias-correction becomes challenging. In the subsequent analysis, we use $C$ or $c$ to denote a generic constant, the specific value of which may vary across different contexts.

In this section, we propose a ridge-screening approach to select the significant variables in Model (\ref{v:hlm}). It is feasible for both scenarios when $p<n$ and $p>n$. Therefore, it can also be treated as a new variable selection approach especially useful in a high-dimensional setting, which is of independent interest to statisticians, econometricians, and data scientists. We only focus on the scenario when $p>n$ and $\bX\bX'$ is singular. Note that the reason for the remaining bias term in Theorem~\ref{thm3} that cannot be corrected is that some projection directions of the coefficients $\bbeta$ are not captured in Model (\ref{v:hlm}). For example, this is the case when some parameters in $\bbeta$ are redundant if certain covariates in $\bX$ are strongly correlated.  Therefore, it is reasonable to assume that some covariates in $\bX$ are not useful in the linear regression (\ref{v:hlm}), and hence, they can be dropped before establishing valid ridge estimators. To embrace the sparsity assumption in high-dimensional data analysis, we assume the true parameter vector $\bbeta$ belongs to the following submodel class %{\color{red} (I changed the definition a bit below.)}
\begin{equation}\label{submodel}
   \mathcal{M}_0=\{1\leq i\leq p, |\beta_i| > \kappa\},  
\end{equation}    
with cardinality $|\mathcal{M}_0|=s^*< \min(p,n)$, where $\kappa$ is a small positive 
number. It is obvious that we require $s^*< p^*$, which is the rank of $\bX$ defined in Assumption~\ref{asm2}. This 
cardinality assumption is natural and it reflects the idea that many coefficient parameters are relatively small among the $p$-dimensional vector $\bbeta$. We treat them as zero elements only for ease of exploitation.  According to the results in Theorem~\ref{thm2} and Theorem~\ref{thm3}, we have that
\begin{equation}\label{rd:decom}
    E\wh\bbeta_{c,k}(\lambda)=\bbeta-\bb_{\lambda,k}=\bbeta-\bU_2\bU_2'\bbeta+o(1),
\end{equation}
as $k\rightarrow\infty$ for a given $\lambda>0$.

It is intuitive to expect that the components of $\wh\bbeta_{c,k}(\lambda)$ corresponding to positions in the submodel $\mathcal{M}_0$ will be greater than those at positions in $\mathcal{M}_0^c$, the complement set of  $\mathcal{M}_0$. This is due to the following reason: Assuming the number of nonzero elements in $\bbeta$ is finite or relatively small compared to 
$p$, then the $\ell_2$-norm of the  $p$-dimensional dense vector $\bU_2\bU_2'\bbeta$ is also of finite or relatively small order.  Consequently, the magnitude of each projected coordinate in $\bU_2\bU_2'\bbeta$ is of a smaller order compared to the nonzero elements in $\bbeta$ on average.
Therefore, we propose a Ridge-Screening (RS) method that selects the submodel class
\begin{equation}\label{rs:sub}
    \mathcal{M}_k(\lambda^*)=\{1\leq i\leq p:|\wh\beta_{c,k,i}(\lambda^*)|\,\,\text{are among the largest $n^*$ of all $|\wh\beta_{c,k,i}(\lambda^*)|$'s}\}, %\footnote{We clarify the notation used here. For $p>n$, $s^*$ is the number of nonzero elements in $\bbeta$, $p^*$ denotes the rank of $\bX$, and $n^*$ is the number of covariates selected by the RS method. Obviously, we require that $s^*<n^*\leq p^*\leq\min(p,n)=n$ for $p>n$. We can also simply set $p^*=n$ since we deal with a fixed design.},
\end{equation}
where $\wh\bbeta_{c,k}(\lambda^*)=(\wh\beta_{c,k,1}(\lambda^*),...,\wh\beta_{c,k,p}(\lambda^*))'$ and we use a different penalty $\lambda^*>0$ to distinguish it from the one utilised in the subsequent step.  We clarify the notation used here. For $p>n$, $s^*$ is the number of nonzero elements in $\bbeta$, $p^*$ denotes the rank of $\bX$, and $n^*$ is the number of covariates selected by the RS method. Obviously, we require that $s^*<n^*\leq p^*\leq\min(p,n)=n$ for $p>n$. We can also simply set $p^*=n$ since we deal with a fixed design.

The ranking method to derive the submodel in (\ref{rs:sub}) is similar to the approach presented in Section 2.2 of \cite{fan2008sure}. However, the method in \cite{fan2008sure} is based on marginal correlations between the response and the features, while (\ref{rs:sub}) is a utilisation of a de-biased estimator, which is more proximate to the true parameter than the conventional ridge estimator in Eq. (5) of \cite{fan2008sure}. Furthermore, our methodology is not constrained by a specific choice of  $\lambda^*$, and the optimal one can be chosen by an information criterion or a cross-validation method as discussed at the end of Section~\ref{sec24}.

In practice, we may select $n^*< \min(p,n)=n$ in (\ref{rs:sub}) through cross-validation along with $\lambda^*$ %, or even $n^*=n$,
because the actual design matrix with $n^*$ columns of covariates is no longer singular in such cases. Additionally, we can show that the model with $n^*$ covariates asymptotically encompasses the submodel $\mathcal{M}_0$ in (\ref{submodel}) under some mild conditions, i.e., the true submodel $\mathcal{M}_0$ in (\ref{submodel}) is nested within the one with $n^*$ covariates. 

Next, we restrict the design matrix to the submodel class $\mathcal{M}_k(\lambda^*)$ and denote the restricted design as $\bX_{\mathcal{M}_k}\in R^{n \times n^*}$. The new ridge estimator becomes 
\begin{equation}\label{r:rd}
    \wh\bbeta_{\mathcal{M}_k}(\lambda)=(\bX_{\mathcal{M}_k}'\bX_{\mathcal{M}_k}+\lambda\bI_{n^*})^{-1}\bX_{\mathcal{M}_k}'\by,
\end{equation}
where $n^*< n$, and $\lambda$ can be different from the $\lambda^*$ used in the ridge screening approach in (\ref{rs:sub}). When $\lambda^*$ and $n^*$
  are optimally selected using the method outlined at the end of Section~\ref{sec24} below, we can set 
$\lambda=\lambda^*$
  because it is optimal for the restricted ridge estimator with a given set of $n^*$ significant variables. We observe that the eigenvalues of $\bX_{\mathcal{M}_k}'\bX_{\mathcal{M}_k}$ are strictly greater than zero for a given configuration of $(p,n)$. This essentially reduces the scenario to the case when $p<n$ in Section \ref{sec23}. Consequently, we can employ the bias-correction procedure detailed in Section~\ref{sec23} over $l$ iterations to obtain a de-biased estimator:
\begin{equation}\label{db:l}
    \wh\bbeta_{\mathcal{M}_k,l}(\lambda)=\wh\bbeta_{\mathcal{M}_k}(\lambda)+\sum_{j=1}^l\lambda^j(\bX_{\mathcal{M}_k}'\bX_{\mathcal{M}_k}+\lambda\bI_{n^*})^{-j}\wh\bbeta_{\mathcal{M}_k}(\lambda).
\end{equation}
To establish the asymptotic properties of the RS method and the de-biased estimator for the restricted one in (\ref{db:l}), we make a few intuitive assumptions below.
%%%%%%%
\begin{assumption}\label{asm3}
    The nonzero singular values of $\bX$ in (\ref{svd:x}) are of order $\sqrt{n}$ and the penalty parameters $\lambda^*\asymp\lambda\asymp n$.
\end{assumption}
%%%
\begin{assumption}\label{asm30}
    For any submatrix $\bX_{\mathcal{M}_k}$ of $\bX$ with dimension $n^*< n$, all the eigenvalues of $\bX_{\mathcal{M}_k}'\bX_{\mathcal{M}_k}$, denoted as $d_{\mathcal{M}_k,j}^2$, for $1\leq j\leq n^*$, are of order $n$.
\end{assumption}
%%%%%
\begin{assumption}\label{asm4}
    For $i\in \mathcal{M}_0$ defined in (\ref{submodel}), $\min_{i\in\mathcal{M}_0}|\beta_i|\geq \kappa\geq C_1n^{-\tau}$ for some $0< \tau< 1/2$, and the magnitude of the $i$-th projected coordinate $|(\bU_2\bU_2'\bbeta)_i|\leq C_2|\beta_i|$ for $C_2<1$,  and $\|\bbeta\|_2^2\leq C_3s^*$, where $s^*$ is the number of nonzero elements in $\bbeta$.
\end{assumption}
%%%%%%%%%
\begin{assumption}\label{asm5}
    Assume $\bve$ is a sub-Gaussian random variable in the sense that
    \[P(|\bv'\bve|\geq x)\leq C\exp(-x^2),\]
    for any $\|\bv\|_2^2=c_*>0$, which is a finite positive constant.
\end{assumption}
Assumptions~\ref{asm3}  and \ref{asm30} are natural conditions about the orders of the singular values of $\bX\in R^{n\times p}$ and $\bX_{\mathcal{M}_k}$, and the penalty parameter $\lambda^*$ (or $\lambda$) is comparable to the magnitude of $d_j^2$ (or $d_{\mathcal{M}_k,j}^2$). Similar to the illustrations for Assumption~\ref{asm1}, the magnitude in Assumptions~\ref{asm3} and \ref{asm30} can be easily verified if 
the entries of $\bX$  are independent copies of a random variable with zero mean, unit variance, and
finite fourth moment under the setting  $p/n\in(1,\infty)$; see \cite{bai1993limit}. In fact, the orders specified in Assumptions~\ref{asm3} and \ref{asm30} are only employed to establish the validity of the ridge-screening method in Theorem~\ref{thm5} below, and they are not the only ones capable of achieving this, so long as the rates can be  properly controlled in the proof of Theorem~\ref{thm5}. 
%Theorem~\ref{thm5} can also be established under different magnitudes on the these order
%These orders are not unique conditions required to establish Theorem~\ref{thm5} within an asymptotic setting. 
For any fixed $(p,n)$, which can be large,  the efficacy of the bias-correction method and the inference methods in Section~\ref{sec25} below remain valid so long as the nonzero singular values and the penalty parameters are strictly greater than zero.
%Assumption~\ref{asm30} ensures that each iteration of our bias-correction procedure applied to the restricted model can reduce the bias, which is similar to that in the scenario of $p\leq n$. 
Assumption~\ref{asm4} indicates that the minimum nonzero element in $\bbeta$ cannot be too small, and the norm of the $i$-th projected coordinate $(\bU_2\bU_2'\bbeta)_i$ is bounded by the  magnitude of its original coordinate $|\beta_i|$ (up to a small constant).  This is actually a reasonable and intuitive assumption.  For instance, in Assumption~\ref{asm4}, if $\|\bbeta\|_2\leq C\sqrt{s^*}$ (at most) because there are only $s^*$ nonzero elements in $\bbeta$, then $\|\bU_1\bU_1'\bbeta\|_2=O_p(\sqrt{s^*})$, but it is a $p$-dimensional vector, meaning that the magnitude of each $(\bU_1\bU_1'\bbeta)_i$ is of order $\sqrt{s^*/p}$ on average. We may postulate that $\sqrt{s^*/p}=o(n^{-\tau})$ in such a case and
%Suppose $p^*$, the number of nonzero elements in $\bbeta$, is finite or relatively smaller than $p$. For instance, we may postulate that $p^*/p=o(n^{-\tau})$ and  assume each entry of the orthogonal matrix $\bU_2$ is of order $p^{-1/2}$, because the length of each column in $\bU_2$ is 1.  Then the magnitude of each projected coordinate is of order $p^*/p$, which is smaller than that of $\beta_i$. 
Assumption~\ref{asm4} is even slightly weaker than this situation as we can allow that the projected coordinate is of the same order as that of the original one.
%In fact, we can relax it to either $|(\bU_2\bU_2'\bbeta)_i|\leq C|\beta_i|$ or $|\beta_i|\leq C|(\bU_2\bU_2'\bbeta)_i|$ for $C<1$, the proof for Theorem~\ref{thm5} still holds. 
Assumption~\ref{asm5} is a general condition that includes Gaussian distributions as a special case.

We have the following theorem regarding the de-biased estimator in (\ref{db:l}) after the ridge-screening.
\begin{theorem}\label{thm5} Let Assumptions \ref{asm1}-\ref{asm5} hold.\\
    (i) Assume the true parameter $\bbeta$ belongs to the submodel $\mathcal{M}_0$ in (\ref{submodel}). If $\frac{\log(p)}{n^{1-2\tau}}\rightarrow 0$, the number of iterations $k$ in the first stage satisfies $\max_{1\leq j\leq p^*}s^*(\frac{\lambda^*}{d_j^2+\lambda^*})^{k+1}\rightarrow 0$, and $n^*$, the number of selected elements in (\ref{rs:sub}),  satisfies that $n^*\geq s^*$ \text{when}  $p<n$, and
    \[\frac{n^*}{Cn^{-2\tau}s^*+Cn^{2\tau-2}p\log(p)}\rightarrow\infty\,\,(\text{as}\,\, n,p\rightarrow\infty)
    \quad\text{when}\quad p>n,\]
    then we have
    \[P(\mathcal{M}_0\subset \mathcal{M}_{k}(\lambda^*))\rightarrow 1,\,\text{as $k\rightarrow\infty$},\]
    where we use $\subset$ in the sense that $\mathcal{M}_{k}(\lambda^*)$ may contain more parameters than $\mathcal{M}_0$.\\
    (ii) Conditioning on the event of $\{\mathcal{M}_0\subset \mathcal{M}_{k}(\lambda^*)\}$, for a properly chosen $\lambda>0$, the bias of the de-biased and restricted ridge estimator $\wh\bbeta_{\mathcal{M}_k,l}(\lambda)$ is 
    \begin{equation}\label{bc:kl}
         \bb_{\lambda,k,l}=\bbeta_{\mathcal{M}_k}-E(\wh\bbeta_{\mathcal{M}_k,l}(\lambda))=\lambda^{l+1}(\bX_{\mathcal{M}_k}'\bX_{\mathcal{M}_k}+\lambda\bI_{n^*})^{-(l+1)}\bbeta_{\mathcal{M}_k},
    \end{equation}
     where $\bbeta_{\mathcal{M}_k}\in R^{n^*}$ is the true value of $\bbeta$ restricted on the submodel $\mathcal{M}_k$. That is,  $\bbeta_{\mathcal{M}_k}$ consists of the nonzero elements in $\bbeta$ and some zero ones associated with their original indexes in $\mathcal{M}_k\setminus \mathcal{M}_0$.
    If the number of iterations $l$ satisfies $\max_{1\leq j\leq n^*}s^*(\frac{\lambda}{d_{\mathcal{M}_k,j}^2+\lambda})^{l+1}\rightarrow 0$, we have
     \[\bb_{\lambda,k,l}\rightarrow {\bf 0},\,\, \text{as}\,\, k,l\rightarrow\infty.\]
     %where $\bbeta_{\mathcal{M}_k}\in R^{n^*}$ is the true value of $\bbeta$ restricted on the submodel $\mathcal{M}_k$, where the elements of $\bbeta_{\mathcal{M}_k}$ with original indexes in $\mathcal{M}_k\setminus \mathcal{M}_0$ are set to zero.
\end{theorem}
\begin{remark}
  (i)   Theorem~\ref{thm5} implies that the ridge-screening method is applicable in both  $p<n$ and $p>n$ scenarios. Remarkably, there is no need to specify $s^*$, which is the number of nonzero elements in the true $\bbeta$. For $p<n$, we can simply set $p^*=p$ as Assumption~\ref{asm2} and choose $n^*=p$ variables, which consists of all the covariates. For $p>n$, we  may choose $n^*\asymp n$ under the assumption that $s^*/n\rightarrow 0$ and $p\log(p)/n^{3-2\tau}\rightarrow 0$ in an asymptotic sense, which is reasonable because $s^*$ is often small and we can adopt $p/n=c\in (1,\infty)$ in the setting of random matrix theory. \\
  %(ii) In practice, the true parameter $p^*$ is unknown, but it can be restricted to a specific interval based on some prior knowledge from the data. According to the discussion above, it is a reasonable assumption that $p^*/n^{1+2\tau}\rightarrow 0$, therefore, Theorem~\ref{thm5} implies that the ridge screening method can correctly recover the true structure of the linear model of (\ref{v:hlm}) no matter $p\leq n$ or $p>n$ if some covariates in $\bX$ are not significant.\\
  (ii) Theorem~\ref{thm5} establishes that our proposed method from Section~\ref{sec23} can completely correct the bias of the restricted ridge estimators, which is particularly helpful when $p>n$. This, in conjunction with the results of Theorem~\ref{thm2}, confirms the versatility of our bias-correction approaches in addressing both scenarios of $p<n$ and $p>n$.\\
  (iii) %In practice, for correcting bias and facilitating statistical inferences with ridge estimators when 
  For $p > n$, our discoveries from Theorems \ref{thm4}-\ref{thm5} underscore the significance of initially reducing all covariates to $n^*$ significant ones, where $n^*< \min(p,n)$. This reduction, combined with the proposed bias-correction procedure for the restricted ridge estimators, enables subsequent valid statistical inferences using the de-biased ridge estimators, as detailed in Section \ref{sec25} below.
\end{remark}
%The requirement in Theorem~\ref{thm5} can be verified if we choose $n^*\asymp n$ in (\ref{rs:sub}) and let $p^*/n^{1+2\tau}\rightarrow 0$ and $p\log(p)/n^{3-2\tau}\rightarrow 0$, which can happen no matter $p<n$ or $p>n$ with a small $0<\tau<1$ in an asymptotic framework.
%Theorem~\ref{thm5} implies that the ridge screening method can correctly recover the true structure of the linear model of (\ref{v:hlm}) no matter $p\leq n$ or $p>n$ if some covariates in $\bX$ are not significant. 
%In addition, Theorem~\ref{thm5} indicates that the bias of the restricted ridge estimators can be fully corrected by our proposed method in Section~\ref{sec23}, which is particularly helpful when $p>n$.

To conclude this subsection, we provide a summary of the pseudo-code for the proposed bias-correction procedures in Algorithm \ref{pless:a1} and Algorithm \ref{plarge:a2}, corresponding to the scenarios of $p < n$ and $p > n$, respectively. For the convergence criterion in Algorithm~\ref{pless:a1}, we can choose a small threshold $\eta>0$, say $\eta=10^{-2}$, and the convergence of the algorithm is determined by checking whether the following inequality hold:
\begin{equation}\label{con:cri}
    \|\wh\bbeta_{c,k}(\lambda)-\wh\bbeta_{c,k-1}(\lambda)\|_2\leq \eta.
\end{equation}
It is important to note that the convergence is guaranteed because the non-random sequence $\bA_k:=\sum_{j=1}^k\lambda^j(\bX'\bX+\lambda\bI_p)^{-j}$ converges, as demonstrated in the proof of  Theorem~\ref{thm2} in the Appendix. Additionally, we emphasise that the number of iterations is typically not large, as only a logarithmic order of the dimension is required to ensure convergence, as discussed in Remark~\ref{rm2}. Furthermore, the computation is efficient since we only need to perform the SVD on $\bX'\bX+\lambda\bI_p$ once, as we do for the classical ridge estimator. After this single SVD, the iterative process can be applied to the diagonal singular-value matrix in the expansion.

Finally, we briefly discuss the methods for selecting the unknown parameters in Algorithms~\ref{pless:a1} and \ref{plarge:a2} of the proposed approaches. Firstly, the sole unknown parameter in Algorithm~\ref{pless:a1} is $\lambda$. Throughout this paper, we assume $\lambda$ to be given, as our focus is mainly on the bias-correction issue for ridge estimators. Empirically, one can employ the widely-accepted information criteria (e.g., AIC or BIC) or utilise cross-validation methods to determine an appropriate $\lambda$. For a comprehensive understanding, readers are referred to Section 1.8 of \cite{van2023lecture}. Secondly, in the ridge-screening method of Algorithm~\ref{plarge:a2} when $p>n$, the parameters $(\lambda^*,n^*)$ are unknown. Here, one can employ the aforementioned information criteria or cross-validation methods, as discussed in Sections 1.8.1-1.8.3 of \cite{van2023lecture}, to simultaneously determine $(\lambda^*,n^*)$. This can be achieved by selecting a candidate for 
$\lambda^*$ and then determining the optimal $n^*$ such that the pair 
$(\lambda^*,n^*)$ minimises the information criterion or yields the best prediction performance on test sets using the restricted ridge estimator $\wh\bbeta_{\mathcal{M}_k}(\lambda^*)$. The final choice of 
$(\lambda^*,n^*)$ can be determined by evaluating the performance across each grid point of the penalty parameters. After identifying the significant variables through the RS method, the parameter $\lambda$ used in bias-correction is equal to 
$\lambda^*$, as 
$\lambda^*$ represents the optimal choice for the associated 
$n^*$. Given that these methods are well-established in the literature, we omit further details here to save space.

%Finally, we briefly discuss the ways to choose the unknown parameters in Algorithms~\ref{pless:a1}-\ref{plarge:a2} of the proposed methods. First, the only unknown parameter in Algorithm~\ref{pless:a1} is $\lambda$, which is assumed given throughout this paper since we only focus on the bias-correction issue for ridge estimators. In a given application, we can adopt the widely used information criterion (e.g. AIC or BIC) or the cross-validation method to choose an appropriate $\lambda$. We refer the readers to Section 1.8 of \cite{van2023lecture} for details.  Second, note that $(\lambda^*,n^*)$ are unknown in the ridge-screening method when $p>n$ in Algorithm~\ref{plarge:a2}, and we can follow the information criterion or the cross-validation method in Section 1.8-1.9 of \cite{van2023lecture} and choose $(\lambda^*,n^*)$ simulateneously. Once the significant variables have been selected by the ridge-screening method, we can follow the selection of $\lambda$ as that when $p\leq n$. Since these methods are commonly used in the literature, and we omit the details to save space.

%the unknown parameters are $(\lambda^*,n^*)$ and $\lambda$ in Algorithm~\ref{plarge:a2}.

%Similarly, we can also make use of the information criterion (e.g. AIC or BIC) or the cross-validation method to simultaneously choose  

%%%%%%%%%%%%%
\begin{algorithm}[ht]
\caption{Iterative bias-correction of ridge estimators when $p<n$}\label{pless:a1}
{\bf Input:} Design matrix $\bX\in R^{n\times p}$, response vector $\by\in R^{n}$, and penalty $\lambda>0$;
%{\bf Output:} A de-biased ridge estimator $\wh\bbeta_{c,k}(\lambda)$
\begin{algorithmic}[1]
\State{Construct a ridge estimator $\wh\bbeta(\lambda)$, set $k=1$;}
 \While{Not Convergent}
      \State  {form $\wh\bbeta_{c,k}(\lambda)=\wh\bbeta(\lambda)+\sum_{j=1}^k\lambda^j(\bX'\bX+\lambda\bI_p)^{-j}\wh\bbeta(\lambda)$;}
      %\State     compute $\wh\bA_{i,1}$ consisting of the leading $r_1^0$ eigenvectors of $\wh\bM_{i,1}^*$
      % \State{form the matrix $\wh\bM_{i,2}^*$ based on $\bY_t'\wh\bA_{i,1}$}
      % \State{compute $\wh\bP_{i+1,1}$ consisting of the leading $r_2^0$ eigenvectors of $\wh\bM_{i,2}^*$}
       \State{$k\gets k+1$;}
\EndWhile
%\State{End{\bf while}}
\State{$k\gets k-1$;}
%\State{Output }
\State{END}
\end{algorithmic}
{\bf Output:} A de-biased ridge estimator $\wh\bbeta_{c,k}(\lambda)$.
\end{algorithm}
%%%%%%%

%%%%%%%%%%%%%
\begin{algorithm}[ht]
\caption{Ridge-screening and bias-correction of ridge estimators when $p> n$}\label{plarge:a2}
{\bf Input:} Design matrix $\bX\in R^{n\times p}$, response vector $\by\in R^{n}$, and penalty parameters $\lambda^*,\lambda>0$;
%{\bf Output:} A de-biased ridge estimator $\wh\bbeta_{c,k}(\lambda)$
\begin{algorithmic}[1]
\State{Apply Algorithm~\ref{pless:a1} and obtain an initial de-biased ridge estimator $\wh\bbeta_{c,k}(\lambda^*)$, set $k_1\gets k$; }
 \State{Identify the indexes of the largest $n^*< \min(n,p)$ elements among $|\wh\bbeta_{c,k_1}(\lambda^*)|$;}
%\State{End{\bf while}}
\State{Define $\bX_{\mathcal{M}_{k_1}}$ as a restricted design matrix of $\bX$ on the indexes found in Step 2;}
\State{Initialize the new design $\bX_{\mathcal{M}_{k_1}}\in R^{n\times n^*}$, response vector $\by\in R^n$, and penalty $\lambda>0$;}
\State{Apply Algorithm~\ref{pless:a1} again, set $l\gets k$, where $k$ is the one in Step 6 of Algorithm~\ref{pless:a1};}
%\State{Output }
\State{END}
\end{algorithmic}
{\bf Output:} A de-biased ridge estimator $\wh\bbeta_{c,k_1,l}(\lambda)$.
\end{algorithm}
%%%%%%%

  \subsection{Inference}\label{sec25}
In this section, we briefly introduce the inference method for the de-biased ridge estimators in Sections~\ref{sec23} and \ref{sec24}. The following theorem can be derived immediately based on the proofs of Theorems~\ref{thm2} and \ref{thm5}.

\begin{theorem}\label{thm6}Assume $\bve\sim N({\bf 0},\bSigma_\ve)$, where $\bSigma_\ve$ is a diagonal covariance matrix.\\
(i) If $p<n$ and $\bX'\bX$ is invertible, then under Assumption~\ref{asm1}, we have
\[\wh\bbeta_{c,k}(\lambda)-\bbeta\sim_d N(\bmu_{1,k}(\lambda),\bSigma_{1,k}(\lambda)),\]
where $\sim_d$ denotes the exact distribution, 
$\bmu_{1,k}(\lambda)=-\lambda^{k+1}(\bX'\bX+\lambda\bI_p)^{-(k+1)}\bbeta$,
and
\[\bSigma_{1,k}(\lambda)=\sum_{j=0}^k \lambda^j(\bX'\bX+\lambda\bI_p)^{-(j+1)}\bX'\bSigma_\ve\bX\sum_{j=0}^k \lambda^j(\bX'\bX+\lambda\bI_p)^{-(j+1)}.\]
Furthermore, for any given  configuration of $(p,n)$ with $p<n$, we have
\[\sqrt{n}(\wh\bbeta_{c,k}(\lambda)-\bbeta)\longrightarrow_d N\left({\bf 0},\bSigma_{1}(\lambda)\right),\,\,\text{as}\quad k\rightarrow\infty,\]
where $\bSigma_1(\lambda)$ is the asymptotic limit of $n\bSigma_{1,k}(\lambda)$.\\
%if the number of iteration $k$ satisfies $\max_{1\leq j\leq p}C_{n,p}p(\frac{\lambda}{d_j^2+\lambda})^{k+1}\rightarrow 0$, as $n,p,k\rightarrow\infty$, where $C_{n,p}$ is defined in Theorem~\ref{thm2}.\\
(ii) If $ p > n$, assume that Assumptions~\ref{asm2}-\ref{asm5} hold and a true model in (\ref{submodel}). Suppose $k$ is sufficiently large in the ridge-screening such that the event $\{\mathcal{M}_0\subset \mathcal{M}_{k}(\lambda^*)\}$ holds. Then the estimator in (\ref{db:l}) has the following limiting distribution
\[\wh\bbeta_{\mathcal{M}_k,l}(\lambda)-\bbeta_{\mathcal{M}_k}\sim_d N(\bmu_{2,k,l},\bSigma_{2,k,l}(\lambda)),\]
where $\bmu_{2,k,l}(\lambda)=-\lambda^{l+1}(\bX_{\mathcal{M}_k}'\bX_{\mathcal{M}_k}+\lambda\bI_{n^*})^{-(l+1)}\bbeta_{\mathcal{M}_k}$,
and
\[\bSigma_{2,k,l}(\lambda)=\sum_{j=0}^l \lambda^j(\bX_{\mathcal{M}_k}'\bX_{\mathcal{M}_k}+\lambda\bI_{n^*})^{-(j+1)}\bX_{\mathcal{M}_k}'\bSigma_\ve\bX_{\mathcal{M}_k}\sum_{j=0}^l \lambda^j(\bX_{\mathcal{M}_k}'\bX_{\mathcal{M}_k}+\lambda\bI_{n^*})^{-(j+1)}.\]
Furthermore, for any given  configuration of $(p,n)$ with $p> n$, we have
\[\sqrt{n}(\wh\bbeta_{\mathcal{M}_k,l}(\lambda)-\bbeta_{\mathcal{M}_k})\longrightarrow_d N\left({\bf 0},\bSigma_{2}(\lambda)\right),\,\,\text{as}\quad k,l\rightarrow\infty,\]
where $\bSigma_2(\lambda)$ is the asymptotic limit of $n\bSigma_{2,k,l}(\lambda)$.\\
%\begin{align*}
% \wh\bbeta_{\mathcal{M}_k,l}(\lambda)-&\bbeta_{\mathcal{M}_k}\\
% &\longrightarrow_d N\left({\bf 0},\sum_{j=0}^l \lambda^j(\bX_{\mathcal{M}_k}'\bX_{\mathcal{M}_k}+\lambda\bI_p)^{-(j+1)}\bX_{\mathcal{M}_k}'\bSigma_\ve\bX_{\mathcal{M}_k}\sum_{j=0}^l \lambda^j(\bX_{\mathcal{M}_k}'\bX_{\mathcal{M}_k}+\lambda\bI_p)^{-(j+1)}\right),
%\end{align*}
%for sufficiently large $k,l>0$.
\end{theorem}
\begin{remark}\label{rm5}
(i) The Assumption of $\bve\sim N({\bf 0},\bSigma_\ve)$ can be relaxed to $\{\ve_i,i=1,...,T\}$ is a martingale difference sequence with finite variances, and the results in Theorem~\ref{thm6} still hold if we make use of martingale central limit theorems in \cite{hall2014martingale} as $n\rightarrow\infty$ under Assumptions~\ref{asm3}-\ref{asm30}.\\
   (ii) Equation (\ref{bck:rp}) and Theorem~\ref{thm6} indicate that
\begin{align}\label{dis:1}
  \sqrt{n} [\wh\bbeta(\lambda)-\bbeta&+\sum_{j=1}^k\lambda^j(\bX'\bX+\lambda\bI_p)^{-j}\wh\bbeta(\lambda)]\notag\\
   &\overset{\sim}{\longrightarrow}_d N\left({\bf 0},n\sum_{j=0}^k \lambda^j(\bX'\bX+\lambda\bI_p)^{-(j+1)}\bX'\bSigma_\ve\bX\sum_{j=0}^k \lambda^j(\bX'\bX+\lambda\bI_p)^{-(j+1)}\right),
\end{align}    
and 
\begin{align}\label{dis:2}
   &\sqrt{n}[\wh\bbeta_{\mathcal{M}_k}(\lambda)-\bbeta_{\mathcal{M}_k}+\sum_{j=1}^l\lambda^j(\bX_{\mathcal{M}_k}'\bX_{\mathcal{M}_k}+\lambda\bI_{n^*})^{-j}\wh\bbeta_{\mathcal{M}_k}(\lambda)]\notag\\
   &\overset{\sim}{\longrightarrow}_d N\left({\bf 0},n\sum_{j=0}^k \lambda^j(\bX_{\mathcal{M}_k}'\bX_{\mathcal{M}_k}+\lambda\bI_{n^*})^{-(j+1)}\bX_{\mathcal{M}_k}'\bSigma_\ve\bX_{\mathcal{M}_k}\sum_{j=0}^k \lambda^j(\bX_{\mathcal{M}_k}'\bX_{\mathcal{M}_k}+\lambda\bI_{n^*})^{-(j+1)}\right),
\end{align}   
for $p<n$ and $p>n$, respectively, where $\overset{\sim}{\longrightarrow}_d$ denotes approximate equivalence in distribution for sufficiently large $k$  and a given configuration of $(p,n)$. Therefore, we can make use of the approximations in (\ref{dis:1}) and (\ref{dis:2}) to make statistical inference.\\
(iii) Under the assumption that $\lambda\asymp n$ and the nonzero singular values of $\bX$ are of order $\sqrt{n}$,  we can easily show that the variance terms in (\ref{dis:1}) and (\ref{dis:2}) are of order $O(1)$. Consequently,
\[\wh\bbeta_{c,k}(\lambda)-\bbeta=O_p(n^{-1/2})\quad \text{and}\quad \wh\bbeta_{\mathcal{M}_k,l}(\lambda)-\bbeta_{\mathcal{M}_k}=O_p(n^{-1/2}),\]
implying that our de-biased estimators are convergent with the standard 
rate $\sqrt{n}$.\\
(iv) In practice,  it is often assumed that  the error term $\bve$ is homoskedastic with $\bSigma_\ve=\sigma^2\bI_n$, which simplifies the inference. 
Consequently, for sufficiently large $k$ and $l$, a consistent estimator for $\sigma$ can be obtained as:
\[\wh\sigma=\sqrt{\frac{1}{n}\|\by-\bX\wh\bbeta_{c,k}(\lambda)\|_2^2}\quad \text{or}\quad \wh\sigma=\sqrt{\frac{1}{n}\|\by-\bX_{\mathcal{M}_k}\wh\bbeta_{\mathcal{M}_k,l}(\lambda)\|_2^2},\]
depending on whether $p<n$ or $p>n$. Then, the variance terms in (\ref{dis:1}) and (\ref{dis:2}) can be estimated from the data.
\end{remark}
Hence, by correcting the bias, we can proceed to make statistical inferences and construct confidence intervals using distributions in (\ref{dis:1}) and (\ref{dis:2}) without the need to identify the sparse structure in Model (\ref{hlm}). This is possible because all biases can be approximated by the data, and the covariance terms in the limiting distributions are of full rank, and they can be estimated from the available data.
  %%%%%%%%%%%%%
\subsection{Time Series Ridge Regression}

In this section, we briefly illustrate the application of the proposed bias-correction method in time series ridge regression. In particular, we consider the following Autoregressive (AR) model:
\begin{equation}\label{yt:ts}
    y_t=\beta_1 y_{t-1}+...+\beta_p y_{t-p}+\ve_t=\bx_t'\bbeta+\ve_t,t=p+1,...,n,
\end{equation}
where $\bx_t=(y_{t-1},...,y_{t-p})'$ is the covariate vector consisting of the $p$ lagged variables of $y_t$, and $\bbeta=(\beta_1,...,\beta_p)'$ is the associated parameter vector. Note that $E(y_t) = 0$ for simplicity. We restrict our consideration to cases where $p<n$. It's uncommon to encounter situations where the number of lagged regressors in an AR model exceeds the sample size. As a matter of fact, it is rare to see that the order of an AR model 
exceeds 20 for non-seasonal time series in empirical applications. See, for instance, 
\cite{Woodward2012}.

Let $\wt\by=(y_{p+1},...,y_n)'$, $\wt\bX=(\bx_{p+1},...,\bx_{n})'$, and $\wt\bve=(\ve_{p+1},...,\ve_{p})'$, then, Model (\ref{yt:ts}) can be expressed as
\begin{equation}
    \wt\by_t=\wt\bX\bbeta+\wt\bve,
\end{equation}
where we assume $\cov(\wt\bve)=\sigma^2\bI_{n-p}$ for simplicity. For a proper $\lambda>0$, the de-biased estimator in (\ref{bck:rp}) can be written as
\[\wh\bbeta_{c,k}(\lambda)
   =\bbeta-\lambda^{k+1}(\wt\bX'\wt\bX+\lambda\bI_p)^{-(k+1)}\bbeta+\sum_{j=0}^k\lambda^j(\wt\bX'\wt\bX+\lambda\bI_p)^{-(j+1)}\wt\bX'\wt\bve.\]
For weakly stationary time series sequence $\{y_t\}$, the ergodic theorem guarantees that
\[\frac{1}{n}\wt\bX'\wt\bX=\frac{1}{n}\sum_{t=p+1}^n\bx_t\bx_{t}'\rightarrow_p E(\bx_t\bx_t'),\,\,\text{as}\,\, n\rightarrow\infty.\]
Suppose the covariance matrix $E(\bx_t\bx_t')$ admits a similar spectral decomposition as $\bX'\bX/n$ in Assumption~\ref{asm1} above, and  the the magnitude of the penalty $\lambda$ satisfies $\lambda\asymp n$ as that in Assumption~\ref{asm3}, we can show that
\[\sum_{j=0}^k\lambda^{j+1}(\wt\bX'\wt\bX+\lambda\bI_p)^{-(j+1)}=O_p(1),\,\,\text{and}\quad\frac{1}{\lambda}\wt\bX'\wt\bve=O_p(\frac{1}{\sqrt{n}}).\]
For large $n$, it follows that
\begin{equation}\label{bc:ts}
    \wh\bbeta_{c,k}(\lambda)-\bbeta=o_p(1),\,\,\text{as}\quad k\rightarrow\infty,
\end{equation}
and 
\begin{equation}\label{rate:ts}
    \sqrt{n}(\wh\bbeta_{c,k}(\lambda)-\bbeta)=\sqrt{n}\sum_{j=0}^k\lambda^j(\wt\bX'\wt\bX+\lambda\bI_p)^{-(j+1)}\wt\bX'\wt\bve=O_p(1)\,\,\text{as}\quad k\rightarrow\infty.
\end{equation}
Equation (\ref{bc:ts}) suggests that the bias-correction procedure remains applicable to time series regression with weakly dependent regressors. Additionally, Equation (\ref{rate:ts}) confirms that the asymptotic distribution of Theorem~\ref{thm6}(i) continues to hold for time series data as well.

%%%%%%%%%
\subsection{Construction of Confidence and Prediction Intervals}\label{sec27}
In this section, we explore methods for constructing confidence and prediction intervals based on ridge regression. For simplicity, %we concentrate on the scenario where $p\leq n$, as the approach can be analogously applied when $p>n$. Additionally, 
we assume  $\bSigma_\ve=\sigma^2\bI_n$ to facilitate the illustration.

We begin by outlining the construction of confidence intervals for the mean response  $E(y|\bx_0)=\bx_0'\bbeta$ with a given covariate $\bx_0$ when $p<n$. Note that $\wh y_0=\bx_0'\wh\bbeta(\lambda)$ on a ridge regression, then, by (\ref{dis:1}) in Remark~\ref{rm5}(ii) with a sufficiently large $k$, we have
\begin{equation}\label{y0:hat:ci}
    \bx_0'\wh\bbeta(\lambda)-\bx_0'\bbeta+\bx_0'\sum_{j=1}^k\lambda^j(\bX'\bX+\lambda\bI_p)^{-j}\wh\bbeta(\lambda)\overset{\sim}{\longrightarrow}_d N\left({\bf 0},\bx_0'\bSigma_{1,k}(\lambda)\bx_0\right),
\end{equation}
where 
\[\bSigma_{1,k}(\lambda)=\sigma^2\sum_{j=0}^k \lambda^j(\bX'\bX+\lambda\bI_p)^{-(j+1)}\bX'\bX\sum_{j=0}^k \lambda^j(\bX'\bX+\lambda\bI_p)^{-(j+1)}.\]
In practice, we may replace $\sigma$ in $\bSigma_{1,k}(\lambda)$ by $\wh\sigma$, which is estimated from the data as that in Remark \ref{rm5}(iv), and we denote the estimated variance in (\ref{y0:hat:ci}) by $\wh\bSigma_{1,k}(\lambda)$. Denote $z_{\alpha}$ the critical value of a standard normal distribution such that its tail probability is $\alpha$, then, the $(1-\alpha)$-confidence interval for $\bx_0'\bbeta$  is $[L_1,U_1]$, where
\begin{equation}\label{l1}
   L_1=\bx_0'\wh\bbeta(\lambda)+\bx_0'\sum_{j=1}^k\lambda^j(\bX'\bX+\lambda\bI_p)^{-j}\wh\bbeta(\lambda)-z_{\alpha/2}\sqrt{\bx_0'\wh\bSigma_{1,k}(\lambda)\bx_0},
\end{equation}
and
\begin{equation}\label{u1}
   U_1=\bx_0'\wh\bbeta(\lambda)+\bx_0'\sum_{j=1}^k\lambda^j(\bX'\bX+\lambda\bI_p)^{-j}\wh\bbeta(\lambda)+z_{\alpha/2}\sqrt{\bx_0'\wh\bSigma_{1,k}(\lambda)\bx_0}.
\end{equation}
Next, we discuss the way to construct prediction intervals for a future value $y_{n+1}$ with a new covariate $\bx_{n+1}$. According to Ch. 2 of \cite{montgomery2012introduction}, by a similar argument as that for the confidence intervals, we can construct the $(1-\alpha)$-prediction intervals of the future value $y_{n+1}$ as $[L_2,U_2]$, where
\begin{equation}\label{l2}
   L_2=\bx_0'\wh\bbeta(\lambda)+\bx_0'\sum_{j=1}^k\lambda^j(\bX'\bX+\lambda\bI_p)^{-j}\wh\bbeta(\lambda)-z_{\alpha/2}\sqrt{\bx_0'\wh\bSigma_{1,k}(\lambda)\bx_0+\wh\sigma^2},
\end{equation}
and
\begin{equation}\label{u2}
   U_2=\bx_0'\wh\bbeta(\lambda)+\bx_0'\sum_{j=1}^k\lambda^j(\bX'\bX+\lambda\bI_p)^{-j}\wh\bbeta(\lambda)+z_{\alpha/2}\sqrt{\bx_0'\wh\bSigma_{1,k}(\lambda)\bx_0+\wh\sigma^2}.
\end{equation}
The values presented in (\ref{l1})-(\ref{u2}) are directly estimated from the data. Consequently, we are equipped to construct valid confidence and prediction intervals in ridge regression via employing the proposed bias-correction techniques. Analogously, confidence and prediction intervals can also be established for models utilising de-biased ridge estimators post ridge-screening when $p>n$. For instance, the 
 $(1-\alpha)$-prediction interval of the future value $y_{n+1}$  produced by the restricted ridge estimator is $[L_3,U_3]$ with
\begin{equation}\label{l3}
   L_3=\bx_{0,k}'\wh\bbeta_{\mathcal{M}_k}(\lambda)+\bx_{0,k}'\sum_{j=1}^l\lambda^j(\bX_{\mathcal{M}_k}'\bX_{\mathcal{M}_k}+\lambda\bI_{n^*})^{-j}\wh\bbeta_{\mathcal{M}_k}(\lambda)-z_{\alpha/2}\sqrt{\bx_{0,k}'\wh\bSigma_{2,k,l}(\lambda)\bx_{0,k}+\wh\sigma^2},
\end{equation}
and
\begin{equation}\label{u3}
U_3=\bx_{0,k}'\wh\bbeta_{\mathcal{M}_k}(\lambda)+\bx_{0,k}'\sum_{j=1}^l\lambda^j(\bX_{\mathcal{M}_k}'\bX_{\mathcal{M}_k}+\lambda\bI_{n^*})^{-j}\wh\bbeta_{\mathcal{M}_k}(\lambda)+z_{\alpha/2}\sqrt{\bx_{0,k}'\wh\bSigma_{2,k,l}(\lambda)\bx_{0,k}+\wh\sigma^2},
\end{equation}
 where $\bx_{0,k}$ is the $n^*$-dimensional sub-vector of the new covariate $\bx_0$ selected by the ridge-screening method.

%%%%%%%%%%%%%%%%
\subsection{Bias-Variance Trade-off}
The bias-variance trade-off is a fundamental concept in machine learning and statistics. It refers to the delicate balance between two sources of error in a predictive model. It is a way of analysing a learning algorithm's expected errors. Therefore, it would be interesting if we can derive the bias-variance trade-off of our de-biased estimators as we increase the number of iterations.

In this section, we study the MSE of the de-biased ridge estimator $\wh\bbeta_{c,k}(\lambda)$ in the $k$-th iteration  for $p<n$ with a given $\lambda>0$. The analysis of the scenario with $p>n$ is similar, and hence, we only focus on the case of $p<n$. The MSE of the $k$-th de-biased ridge estimator is defined as 
\begin{align}\label{mse:bt}
    \text{MSE}(\wh\bbeta_{c,k}(\lambda))=&E[\wh\bbeta_{c,k}(\lambda)-\bbeta]'[\wh\bbeta_{c,k}(\lambda)-\bbeta]\notag\\
    =&E[\wh\bbeta_{c,k}(\lambda)-E\wh\bbeta_{c,k}(\lambda)]'[\wh\bbeta_{c,k}(\lambda)-E\wh\bbeta_{c,k}(\lambda)]+[E\wh\bbeta_{c,k}(\lambda)-\bbeta]'[E\wh\bbeta_{c,k}(\lambda)-\bbeta]\notag\\
    =&\var(\wh\bbeta_{c,k}(\lambda))+\text{bias}(\wh\bbeta_{c,k}(\lambda))^2.
\end{align}
For simplicity, we assume $\bSigma_\ve=\sigma^2\bI_n$. By Assumption~\ref{asm1} and Theorem~\ref{thm6}(i),
\begin{align}\label{var:btc}
    \var({\wh\bbeta_{c,k}(\lambda)})=&E[\bve'\bX\sum_{j=0}^k\lambda^j(\bX'\bX+\lambda\bI_p)^{-(k+1)}\sum_{j=0}^k\lambda^j(\bX'\bX+\lambda\bI_p)^{-(k+1)}\bX'\bve]\notag\\
    =&E[\bve'\bV_1\bD_{k}(\lambda)\bV_1'\bve]=\sigma^2\tr[\bD_{k}(\lambda)],
\end{align}
where $\bD_{k}(\lambda)=\diag(d_{k,1}(\lambda),...,d_{k,p}(\lambda))$ with
\[d_{k,i}(\lambda)=\frac{1}{d_i^2}\left[1-(\frac{\lambda}{\lambda+d_i^2})^{k+1}\right]^2.\]
By a similar argument, we can show that
\begin{align}\label{bias:sq}
 \text{bias}(\wh\bbeta_{c,k}(\lambda))^2=&  \bbeta'\lambda^{k+1}(\bX'\bX+\lambda\bI_p)^{-(k+1)}\lambda^{k+1}(\bX'\bX+\lambda\bI_p)^{-(k+1)}\bbeta\notag\\
 =&\bbeta'\bU_1\bLambda_{k}(\lambda)\bU_1'\bbeta,
\end{align}
where $\bLambda_{k}(\lambda)=\diag(\gamma_{k,1}(\lambda),...,\gamma_{k,p}(\lambda))$ with
\[\gamma_{k,i}(\lambda)=[\frac{\lambda}{\lambda+d_i^2}]^{2(k+1)}.\]
As the number of iterations $k$ increases, we can readily see that the bias term in (\ref{bias:sq}) decreases while the variance term in (\ref{var:btc}) increases. Therefore, it is interesting to see whether there is a bias-variance trade-off in the proposed bias-correction procedure. 

In the following theorem, we provide some sufficient conditions under which we have a bias-variance trade-off for the proposed method.
%%%%%%%%%%%%%
\begin{theorem}\label{thm7}
Let Assumption~\ref{asm1} hold and $\bU_1'\bbeta=(\delta_1,...,\delta_p)'$, 
where $p<n$.\\
    (i) If $\frac{\delta_i^2d_i^2}{\sigma^2}<1$ and $\frac{\delta_i^2d_i^2}{\sigma^2}+(\frac{\lambda}{\lambda+d_i^2})^{k^*+1}\geq 1$, for $1\leq i\leq p$, and some $k^*\geq 1$, then, as the number of iterations $k$ increases, the MSE of $\wh\bbeta_{c,k}(\lambda)$ will initially decrease to its minimum value and subsequently rise to a stable level. In particular, the minimum of the MSE can be achieved at the $k$-th iteration for some $k\in [\lfloor k_1 \rfloor,\lfloor k_2\rfloor]$, where
    \[k_1=\min\left\{\frac{\log(1-\frac{\delta_i^2d_i^2}{\sigma^2})}{\log(\frac{\lambda}{\lambda+d_i^2})}-1:1\leq i\leq p\right\},k_2=\max\left\{\frac{\log(1-\frac{\delta_i^2d_i^2}{\sigma^2})}{\log(\frac{\lambda}{\lambda+d_i^2})}-1:1\leq i\leq p\right\},\]
and $\lfloor x\rfloor$ is the largest integer that does not exceed $x$.\\
    (ii) If $\frac{\delta_i^2d_i^2}{\sigma^2}>1$, for $1\leq i\leq p$, then, as the number of iterations $k$ increases, the MSE of $\wh\bbeta_{c,k}(\lambda)$ will decrease to a stable level.
\end{theorem}
\begin{remark}
    (i) Theorem~\ref{thm7} indicates there exists a bias-variance trade-off in the proposed bias-correction procedure under certain conditions. That is, if the design matrix and the penalty satisfy the criteria outlined in Theorem~\ref{thm7}(i), the MSE of the de-biased estimators can attain a minimum value by balancing the bias and variance terms as the number of iterations $k$, or equivalently, the "model complexity", increases.\\
    (ii) Theorem~\ref{thm7}(ii) suggests the possibility of a monotonically decreasing MSE as the number of iterations $k$ increases. This phenomenon arises because the increase in the variance term is not comparable to the decrease in the bias term. Consequently, this finding is intriguing as it indicates the potential for a de-biased estimator to achieve an even smaller MSE than classical ridge estimators.\\
    (iii) We highlight that the conditions specified in Theorem \ref{thm7}(ii) can be easily satisfied under certain scenarios. For instance, assuming $d_i^2\asymp {n}$ as detailed in Assumption~\ref{asm3}, and $\delta_i^2\asymp O(p^{-1})$ or $O(1)$ depending on whether $\bbeta$ is a sparse or dense vector with bounded elements (given that each entry in $\bU_1$ is of order $p^{-1/2}$, it becomes evident that $\delta_i^2d_i^2/\sigma^2$ is at least of order $O(n/p)$, which can exceed $1$. Similarly, the condition in Theorem~\ref{thm7}(i) can also be satisfied if $d_i$ is of a smaller rate. We omit the details to save space. 
The simulation results presented in Section~\ref{sec3} further corroborate our findings.
    
\end{remark}
%%%%%%%%%%
From Theorem 2 in \cite{theobald1974generalizations}, the MSE of the classical ridge estimator can be smaller than that of the least-squares estimator for certain properly chosen $\lambda>0$. In our framework, we observe a similar paradigm concerning the number of iterations $k>0$, as described in the following corollary
%%%%%%%%
%%%%%%%%%%%
\begin{corollary}
Under the conditions outlined in Theorem~\ref{thm7}, there exists an iteration number 
$k$ such that
    \[\text{MSE}(\wh\bbeta_{c,k}(\lambda))<\text{MSE}(\wh\bbeta_{c,0}(\lambda))=\text{MSE}(\wh\bbeta(\lambda)),\]
     suggesting that the proposed de-biased estimators can further minimise the MSEs while also correct a portion of the bias compared to classical ridge estimators.
\end{corollary}

  %%%%%%%%%%%%%%%%%%
  \section{Monte Carlo Simulations }\label{sec3}
We conduct Monte-Carlo experiments to demonstrate the performance of the proposed procedure by considering two scenarios where $p<n$ and $p>n$. Our goal is to study the effect of the bias-correction procedure and the approximation of the asymptotic normalities established in Section~\ref{sec25}.\\

{\noindent\bf Example 1.} In this example, we employ the data-generating process in (\ref{v:hlm}) for different settings of $(p,n)$ with $p<n$. For each configuration of  $(p,n)$, we set the seed number in \texttt{R} software to 1234 and generate an $n\times p$ matrix $\bH$, where its elements are drawn independently from a Uniform distribution $U(-2,2)$. We then perform a singular-value decomposition on $\bH$ and obtain its left and right singular vectors $\bM\in R^{n\times p}$ and $\bN\in R^{p\times p}$, then we let $\bX=\bM\bN'$. The first $p/2$ elements of $\bbeta$ are generated from $U(-2,-1)$ independently, and the remaining $p/2$ elements are from $U(1,2)$. In each replication, the noise $\bve$ is generated from multivariate normal distribution $N({\bf 0},\bI_n)$. We consider the scenarios where $(p,n)=(50,100)$, $(50,400)$, $(100,200)$, and $(100,500)$. The choices of $\lambda$ are $\lambda=0.05n$, $0.1n$, $0.3n$, and $0.5n$. 1000 replications are used in each setting throughout the experiments.

We report the estimation errors of the ridge estimators and the de-biased ones 
in Table~\ref{Table-a1}. For each configuration of $(p,n,\lambda)$, we define the empirical mean squared  errors (MSEs) as 
\begin{equation}\label{sse0}
    MSE(\bb_\lambda)=\frac{1}{1000}\sum_{j=1}^{1000}\|\wh\bbeta^{(j)}(\lambda)-\bbeta\|_2^2
\end{equation}
and
\begin{equation}\label{ssek}
      MSE(\bb_{\lambda,k})=\frac{1}{1000}\sum_{j=1}^{1000}\|\wh\bbeta_{c,k}^{(j)}(\lambda)-\bbeta\|_2^2,  
\end{equation}
where $\wh\bbeta^{(j)}(\lambda)$ and $\wh\bbeta_{c,k}^{(j)}(\lambda)$ are the ridge estimators and the de-biased one with $k$ iterations in the $j$-th replication, respectively. The standard errors $\wh\sigma_0$ and $\wh\sigma_{k}$
are estimated via
\begin{equation}\label{sigma:e}
    \wh\sigma_0=\sqrt{\frac{1}{1000}\sum_{j=1}^{1000}\frac{1}{n}\|\by-\bX\wh\bbeta^{(j)}(\lambda)\|_2^2}\,\,\text{and}\,\, \wh\sigma_{k}=\sqrt{\frac{1}{1000}\sum_{j=1}^{1000}\frac{1}{n}\|\by-\bX\wh\bbeta_{c,k}^{(j)}(\lambda)\|_2^2}.
\end{equation}
Thus, we evaluate the estimations of the standard errors by the ridge estimators and the de-biased ones.

\begin{figure}[ht]
\begin{center}
%{\includegraphics[width=8cm,height=14cm]{Vol-loadings.pdf}}
{\includegraphics[width=14cm,height=8cm]{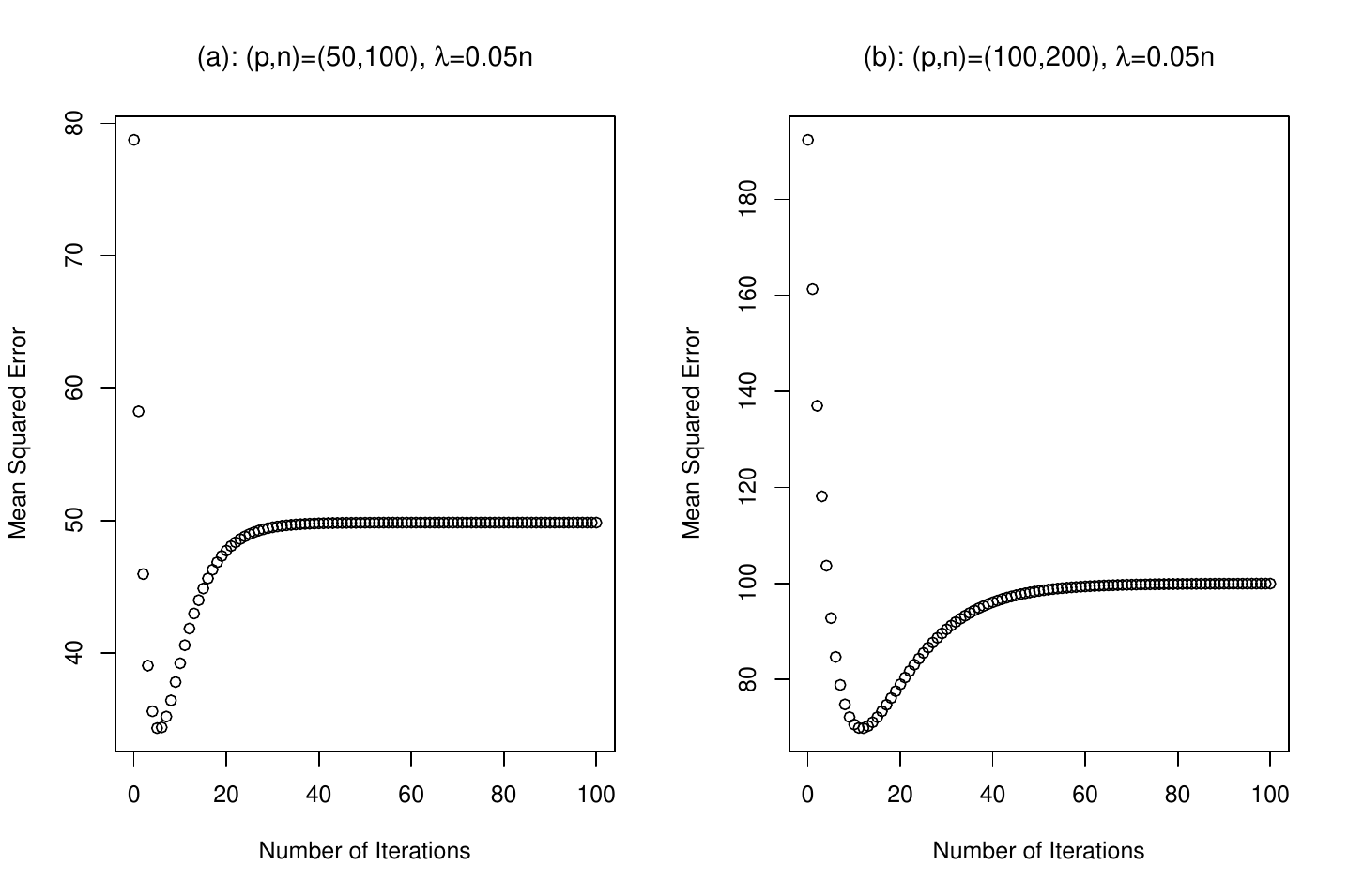}}
\caption{  The bias-variance trade-off reflected by the empirical MSEs of the de-biased ridge estimators for different number of iterations in Example 1, where we consider $\lambda=0.05n$ for $(p,n)=(50,100)$ and $(100,200)$ in (a) and (b), respectively.  1000 replications are used in the experiments.  }\label{fig-0}
\end{center}
\end{figure}

For each setting in Table~\ref{Table-a1}, we see that the MSE first decreases as the number of iterations increases, but it increases if we use too many iterations to correct the bias terms. This is understandable because it is in line with the bias-variance trade-off in the machine learning literature. See, for example, \cite{hastie2009elements}. Specifically, we plot the MSEs for $(p,n)=(50,100)$ and $(100,200)$ with $\lambda=0.05n$ in Figure~\ref{fig-0}, where we can clearly see that the MSEs can achieve a minimum point as we increase the number of iterations, or equivalently, the model complexity, and the MSE will increase if we try to completely correct the biases. Finally, the MSE becomes stable, which is consistent with our asymptotic theory. Furthermore, comparing 
$\hat{\sigma}$ and $\hat{\sigma}_{100}$, we see that the standard error estimated using the debiased ridge estimator is closer to the true one (unity), showing the effectiveness of our bias-correction procedure.

%In particular, the estimated standard errors are more accurate than those by the classic ridge estimators without bias correction, implying that our approach can produce more accurate inference results.

 %%%
\begin{table}[htp]
\caption{ Empirical mean squared errors (MSEs) when $p<n$ in Example 1, where the MSEs are defined in (\ref{sse0}) and (\ref{ssek}) for $\bb_{\lambda}$ and $\bb_{\lambda,k}$, respectively. The number of iterations $k=1,5,10,20,50,100$, and $\wh\sigma_0$ and $\wh\sigma_{100}$ are estimated by (\ref{sigma:e}). 1000 replications are used in the experiments.} 
          \label{Table-a1}
{\begin{center}
\begin{tabular}{cccccccccc}
\toprule
&\multicolumn{9}{c}{$\lambda=0.05n$}\\
\cline{2-10}
$(p,n)$&$\bb_{\lambda}$&$\bb_{\lambda,1}$&$\bb_{\lambda,5}$&$\bb_{\lambda,10}$&$\bb_{\lambda,20}$&$\bb_{\lambda,50}$&$\bb_{\lambda,100}$&$\wh\sigma_0$&$\wh\sigma_{100}$\\
\hline
(50,100)&78.8&58.3&34.3&39.2&47.7&49.8&49.9&1.29&1.19\\
(50,400)&105.3&95.9&67.9&48.4&35.6&42.9&49.3&1.11&1.06\\
(100,200)&192.4&161.3&92.8&70.6&79.0&98.5&100.0&1.47&1.23\\
(100,500)&217.4&201.4&151.2&111.4&76.7&79.0&96.3&1.15&1.05\\
\midrule
&\multicolumn{9}{c}{$\lambda=0.1n$}\\
\cline{2-10}
$(p,n)$&$\bb_{\lambda}$&$\bb_{\lambda,1}$&$\bb_{\lambda,5}$&$\bb_{\lambda,10}$&$\bb_{\lambda,20}$&$\bb_{\lambda,50}$&$\bb_{\lambda,100}$&$\wh\sigma_0$&$\wh\sigma_{100}$\\
\hline
(50,100)&92.6&77.6&44.8&34.5&39.2&49.1&49.8&1.34&1.19\\
(50,400)&110.4&105.2&87.2&70.3&49.4&35.1&42.9&1.12&1.05\\
(100,200)&210.5&191.5&135.5&96.4&70.9&85.7&98.5&1.50&1.23\\
(100,500)&210.4&217.2&186.5&155.8&113.8&71.6&79.0&1.16&1.03\\
\midrule
&\multicolumn{9}{c}{$\lambda=0.3n$}\\
\cline{2-10}
$(p,n)$&$\bb_{\lambda}$&$\bb_{\lambda,1}$&$\bb_{\lambda,5}$&$\bb_{\lambda,10}$&$\bb_{\lambda,20}$&$\bb_{\lambda,50}$&$\bb_{\lambda,100}$&$\wh\sigma_0$&$\wh\sigma_{100}$\\
\hline
(50,100)&104.6&98.1&76.8&58.6&40.2&36.7&46.4&1.39&1.18\\
(50,400)&114.1&112.2&105.1&97.1&83.2&55.8&37.9&1.12&1.04\\
(100,200)&224.3&217.1&191.0&163.9&124.3&75.4&74.1&1.52&1.20\\
(100,500)&231.9&228.8&217.1&203.5&179.5&127.6&85.3&1.16&1.03\\
\midrule
&\multicolumn{9}{c}{$\lambda=0.5n$}\\
\cline{2-10}
$(p,n)$&$\bb_{\lambda}$&$\bb_{\lambda,1}$&$\bb_{\lambda,5}$&$\bb_{\lambda,10}$&$\bb_{\lambda,20}$&$\bb_{\lambda,50}$&$\bb_{\lambda,100}$&$\wh\sigma_0$&$\wh\sigma_{100}$\\
\hline
(50,100)&107.3&103.2&88.6&74.0&54.1&34.7&39.2&1.40&1.14\\
(50,400)&114.9&113.7&109.3&104.1&94.6&72.3&50.3&1.12&1.05\\
(100,200)&227.2&222.8&206.0&187.3&156.1&99.8&71.2&1.53&1.20\\
(100,500)&233.1&231.3&224.1&215.4&199.4&159.8&115.9&1.16&1.06\\
\bottomrule
\end{tabular}
  \end{center}}
\end{table}
%%%%%%%%%%%%%%%%%%%%%%%%%%%%%%%%%%%%%%%%%%%%%%%%%%

%%%%%%%%%%%%%%%%%%%%%%%%%%%%%
%%%%%%%%%%%%%%%%%%%%%

%%%%%%%%%%%%%%%%%%%%%%%%%%%%%%%%%%%%%%%%%%%%%%%

Next, we study the consistency of the de-biased estimators using the proposed method. For simplicity, we investigate the average estimation errors (AEEs) of the ridge and 
de-biased estimators. These AEE are defined, respectively, as
\begin{equation}\label{aee0}
    AEE(\bb_\lambda)=\frac{1}{\sqrt{p}}\|\frac{1}{1000}\sum_{j=1}^{1000}\wh\bbeta^{(j)}(\lambda)-\bbeta\|_2
\end{equation}
and
\begin{equation}\label{aeek}
    AEE(\bb_{\lambda,k})=\frac{1}{\sqrt{p}}\|\frac{1}{1000}\sum_{j=1}^{1000}\wh\bbeta_{c,k}^{(j)}(\lambda)-\bbeta\|_2,
\end{equation}
based on the 1000 simulations. These measures, where we use $\ell_2$-norm 
to quantify the empirical biases. The AEEs are reported in Table~\ref{Table-a2}.
From the table, %Table~\ref{Table-a2}, 
we can see a decreasing pattern for each $(p,n)$ as the number of iterations increases, implying that our de-biased estimators are consistent to  the true parameters for sufficiently large $k$.

%%%
\begin{table}[htp]
\caption{Empirical average estimation errors (AEEs) when $p<n$ in Example 1, where 
$p$ is the number of predictors, $n$ is the sample size, 
and AEEs are defined in (\ref{aee0}) and (\ref{aeek}) for $\bb_{\lambda}$ and $\bb_{\lambda,k}$, respectively. The numbers of bias-correction iterations used are $k=1,5,10,20,50$, and $100$. The results are based on 1000 replications of the experiments.} 
          \label{Table-a2}
{\begin{center}
\begin{tabular}{cccccccc}
\toprule
&\multicolumn{7}{c}{$\lambda=0.05n$}\\
\cline{2-8}
$(p,n)$&$\bb_{\lambda}$&$\bb_{\lambda,1}$&$\bb_{\lambda,5}$&$\bb_{\lambda,10}$&$\bb_{\lambda,20}$&$\bb_{\lambda,50}$&$\bb_{\lambda,100}$\\
\hline
(50,100)&1.24&1.01&0.50&0.20&0.04&0.03&0.03\\
(50,400)&1.45&1.38&1.14&0.89&0.55&0.13&0.03\\
(100,200)&1.38&1.26&0.86&0.53&0.21&0.03&0.03\\
(100,500)&1.47&1.42&1.21&1.00&0.67&0.21&0.04\\
\midrule
&\multicolumn{7}{c}{$\lambda=0.1n$}\\
\cline{2-8}
$(p,n)$&$\bb_{\lambda}$&$\bb_{\lambda,1}$&$\bb_{\lambda,5}$&$\bb_{\lambda,10}$&$\bb_{\lambda,20}$&$\bb_{\lambda,50}$&$\bb_{\lambda,100}$\\
\hline
(50,100)&1.36&1.23&0.84&0.52&0.20&0.03&0.03\\
(50,400)&1.49&1.45&1.31&1.16&0.91&0.44&0.13\\
(100,200)&1.45&1.38&1.14&0.89&0.55&0.13&0.03\\
(100,500)&1.50&1.47&1.36&1.23&1.01&0.56&0.21\\
\midrule
&\multicolumn{7}{c}{$\lambda=0.3n$}\\
\cline{2-8}
$(p,n)$&$\bb_{\lambda}$&$\bb_{\lambda,1}$&$\bb_{\lambda,5}$&$\bb_{\lambda,10}$&$\bb_{\lambda,20}$&$\bb_{\lambda,50}$&$\bb_{\lambda,100}$\\
\hline
(50,100)&1.45&1.40&1.23&1.04&0.75&0.28&0.05\\
(50,400)&1.51&1.50&1.45&1.39&1.28&1.00&0.66\\
(100,200)&1.50&1.47&1.38&1.27&1.08&0.66&0.29\\
(100,500)&1.52&1.51&1.47&1.42&1.33&1.09&0.78\\
\midrule
&\multicolumn{7}{c}{$\lambda=0.5n$}\\
\cline{2-8}
$(p,n)$&$\bb_{\lambda}$&$\bb_{\lambda,1}$&$\bb_{\lambda,5}$&$\bb_{\lambda,10}$&$\bb_{\lambda,20}$&$\bb_{\lambda,50}$&$\bb_{\lambda,100}$\\
\hline
(50,100)&1.47&1.44&1.33&1.20&0.98&0.54&0.20\\
(50,400)&1.52&1.51&1.48&1.44&1.37&1.18&0.92\\
(100,200)&1.51&1.49&1.43&1.36&1.24&0.92&0.56\\
(100,500)&1.53&1.52&1.49&1.47&1.41&1.25&1.02\\
\bottomrule
\end{tabular}
  \end{center}}
\end{table}
%%%%%%%%%%%%%%%%%%%%%%%%%%%%%%%%%%%%%%%%%%%%%%%%%%

\begin{figure}[ht]
\begin{center}
%{\includegraphics[width=8cm,height=14cm]{Vol-loadings.pdf}}
{\includegraphics[width=14cm,height=8cm]{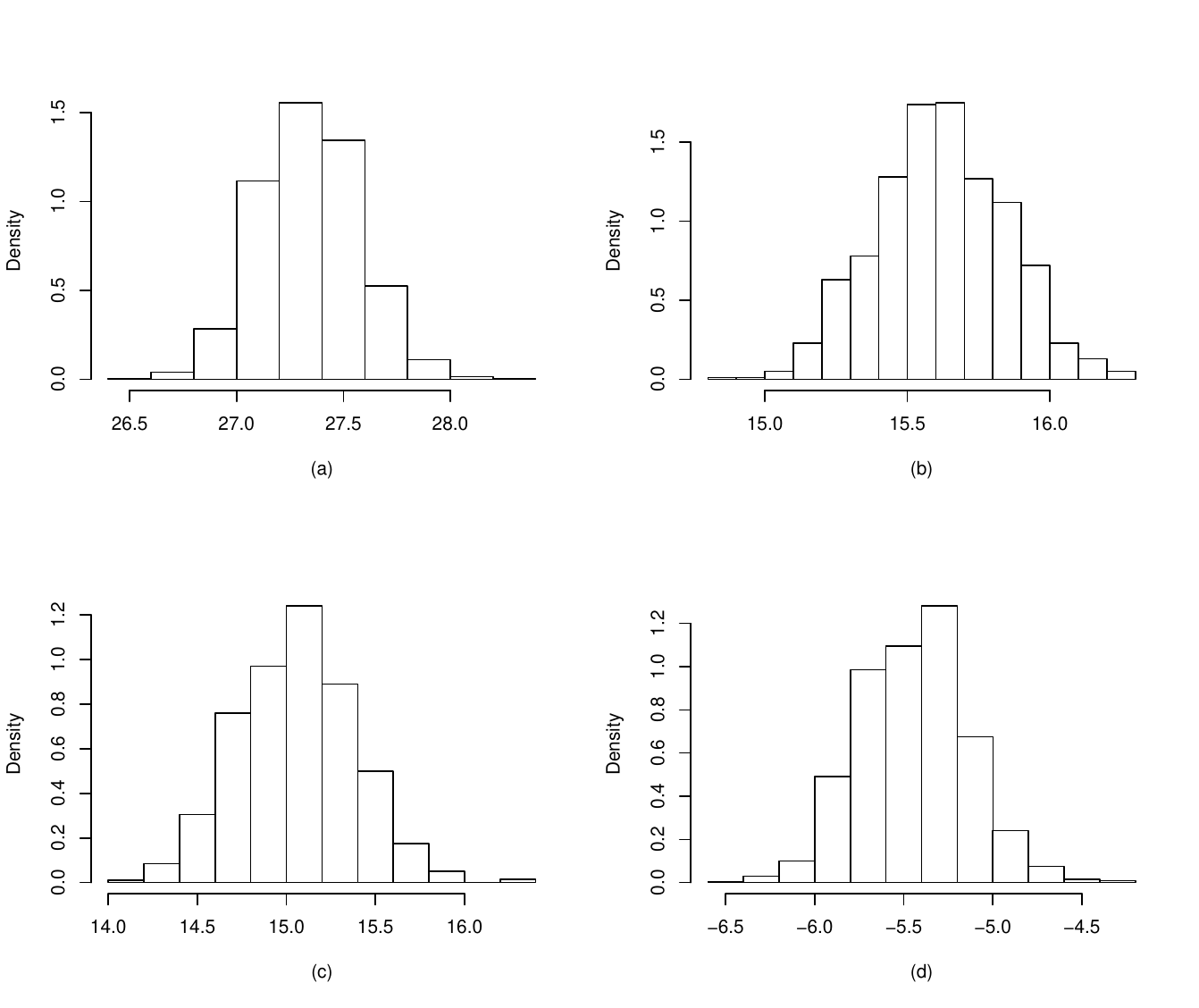}}
\caption{  Empirical histograms of (a) $\sqrt{n}\be_1'(\wh\bbeta(\lambda)-\bbeta)$; (b)  $\sqrt{n}\be_2'(\wh\bbeta(\lambda)-\bbeta)$; (c) $\sqrt{n}\btheta_1'(\wh\bbeta(\lambda)-\bbeta)$; and (d) $\sqrt{n}\btheta_2'(\wh\bbeta(\lambda)-\bbeta)$ in Example 1, where $(p,n)=(100,200)$ and $\lambda=0.3n$. 1000 replications are used in the experiments.  }\label{fig-1}
\end{center}
\end{figure}

\begin{figure}[ht]
\begin{center}
%{\includegraphics[width=8cm,height=14cm]{Vol-loadings.pdf}}
{\includegraphics[width=14cm,height=8cm]{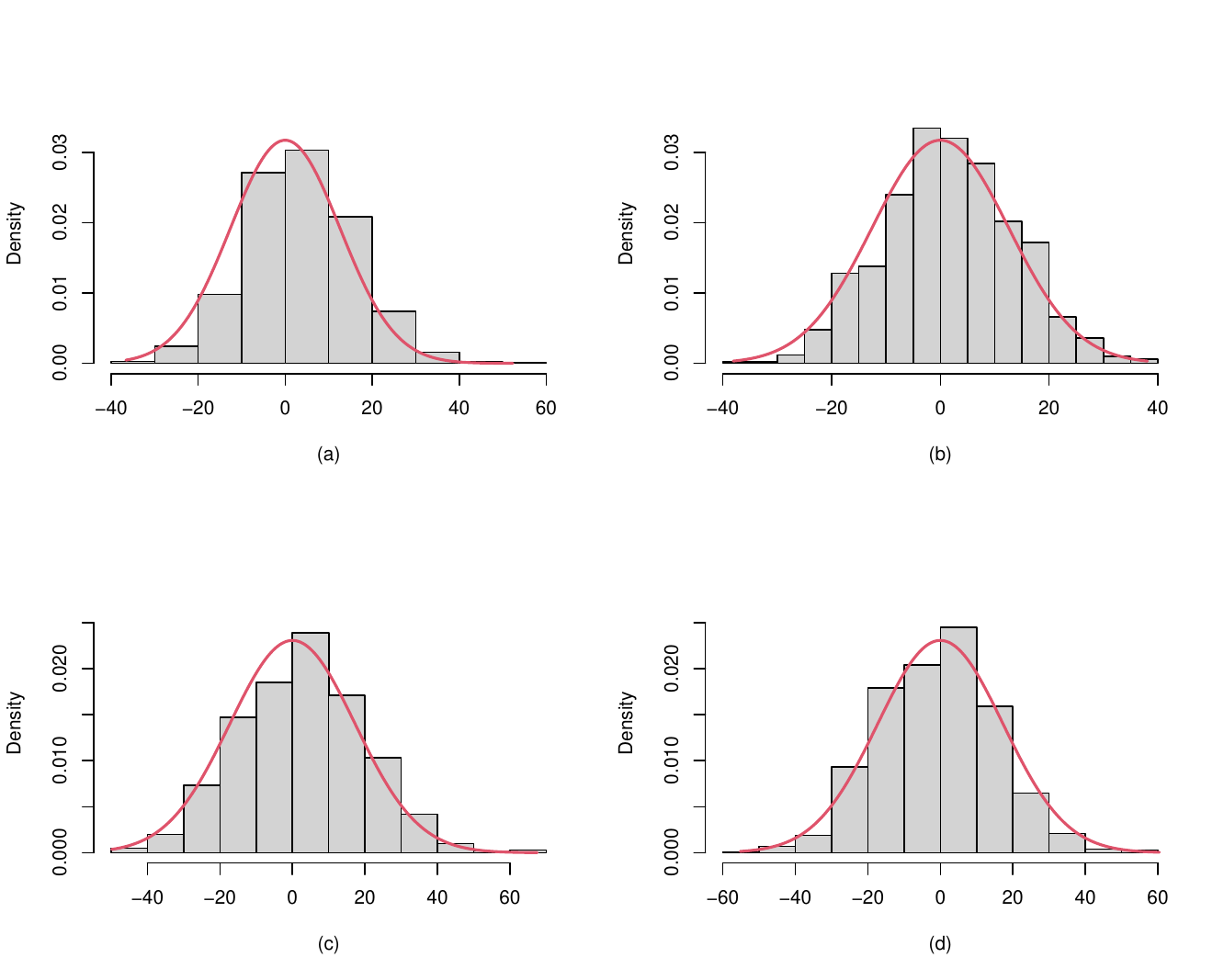}}
\caption{Empirical histograms of (a) $\sqrt{n}\be_1'(\wh\bbeta_{c,k}(\lambda)-\bbeta)$; (b)  $\sqrt{n}\be_2'(\wh\bbeta_{c,k}(\lambda)-\bbeta)$; (c) $\sqrt{n}\btheta_1'(\wh\bbeta_{c,k}(\lambda)-\bbeta)$; and (d) $\sqrt{n}\btheta_2'(\wh\bbeta_{c,k}(\lambda)-\bbeta)$ in Example 1. We choose 
$k=120$ iterations for the de-biased estimators with $(p,n)=(100,200)$ and $\lambda=0.3n$. A normal density curve is plotted based on the limiting distribution outlined in Theorem~\ref{thm6}(i).  1000 replications are used in the experiments. }\label{fig-2}
\end{center}
\end{figure}

Finally, we investigate the performance of statistical inference using the de-biased estimators given in Theorem~\ref{thm6} for $p<n$. For simplicity, we only consider the case when $(p,n)=(100,200)$ and $\lambda=0.3n$, and can produce similar results for other settings. Let $\be_i$ be the $i$-th standard unit vector of $R^p$, 
\[\btheta_1=(0.8,-1,0.5,{\bf 0}_{p-3}')',\,\,\text{and}\,\,\btheta_2=(-1,0.5,0.8,{\bf 0}_{p-3}')',\]
where ${\bf 0}_s$ is an $s$-dimensional  vector of zeros. Figure~\ref{fig-1} presents the histograms of $\sqrt{n}\be_1'(\wh\bbeta(\lambda)-\bbeta)$, $\sqrt{n}\be_2'(\wh\bbeta(\lambda)-\bbeta)$, $\sqrt{n}\btheta_1'(\wh\bbeta(\lambda)-\bbeta)$, and $\sqrt{n}\btheta_2'(\wh\bbeta(\lambda)-\bbeta)$ obtained from 1000 experiments. From Figure~\ref{fig-1}, we see that the biases of the empirical means in all the histogram plots are significantly large and diverge from zero, implying that the traditional ridge estimators are not appropriate for making statistical inference  as they overlook the bias terms. In contrast, we plot the empirical histograms of the de-biased estimators in Figure~\ref{fig-2} under the same setting.  From Figure~\ref{fig-2}, we see that the finite sample performance is quite satisfactory, and the bias effect of the de-biased estimators is small. In addition, the curve of the normal distribution in Theorem~\ref{thm6}(i) is added to the corresponding histograms, where the standard error term $\sigma$ is estimated from the data as that in Remark~\ref{rm5}(iv) and Table~\ref{Table-a1}. From these curves, we can further confirm that the proposed inference method is valid for the de-biased ridge estimators, indicating that one can make valid inferences using the  ridge estimators with our bias-correction method in practice.\\

%%%%%%%%%%%%%%%%%%%%%%%%%%%%%
%%%%%%%%%%%%%%%%%%%%%

%%%%%%%%%%%%%%%%%%%%%%%%%%%%%
%%%%%%%%%%%%%%%%%%%%%

{\noindent\bf Example 2.} In this example, we consider the scenario of $p>n$. The seed number used in this example is the same as that in Example 1. Each row of the design matrix $\bX$ is generated independently from multivariate standard normal distribution $N({\bf 0},\bI_p)$. For each configuration of $(p,n)$, the first $5$ elements in $\bbeta$ are generated from $U(-5,-2)$,  the $6$th to the $10$th elements are from $U(2,5)$, and the remaining ones are zeros. In other words, we consider a sparse vector $\bbeta$ with $p^*=10$ in (\ref{submodel}). The dimensions and the sample sizes are $(p,n)=(150,120), (150,140), (220,180)$, and $(220,200)$, and we consider $\lambda=0.1n$, $0.3n$, and $0.8n$ in each replication for any given $(p,n)$. For simplicity, we choose $n^*=40$ in the ridge screening of (\ref{rs:sub}) with $k=100$ iterations. $1000$ replications are used for each configuration of $(p,n)$ throughout the experiments.

In Table~\ref{Table-a3}, we report the empirical MSEs of the ridge estimators, the de-biased estimators before applying the ridge-screening method, and those after the ridge-screening with $k=100$ iterations in the first stage of bias-corrections.
Specifically, the upper panel in Table~\ref{Table-a3} for each $\lambda$ presents 
the MSEs of the ridge estimators and the de-biased estimators with 100 iterations before applying the ridge-screening approach. We see that the MSEs of the de-biased estimators become quite stable as we increase the number of iterations. This result is consistent with our findings that there exists some minor bias that cannot be corrected from the data. In addition, the estimated standard errors are not close to unity, which is the true value in the experiments. In the lower panel of Table~\ref{Table-a3} for each $\lambda$, we apply the ridge-screening approach with $k=100$ iterations and sort out the largest $n^*=40$ components of $\wh\bbeta_{c,100}(\lambda)$ to obtain the restricted ridge estimators. We further apply the bias-correction approach, and the de-biased estimators are obtained in another $100$ iterations. From the lower panels of Table~\ref{Table-a3} for each $\lambda$, we see that the bias of the de-biased estimators decreases sharply to a relatively stable value for each configuration of $(p,n)$, which is understandable since the selected model consists of more parameters than those in the true one. In addition, the estimation errors are significantly smaller than those without the ridge-screening approach, and the estimated standard errors after bias correction are closer to the true one than those of the methods without using the ridge-screening and the bias-correction. We also note that the estimated standard errors are still larger than the true one (unity) when $\lambda=0.8n$ after the ridge-screening, but they are much closer to the true ones than those produced by the original ridge estimators. In addition, the result can be optimised by choosing more appropriate tuning parameter $\lambda$ (e.g. $\lambda=0.1n$ or $0.3n$) such that the estimated standard errors are close to one. Furthermore, we also note that there is also a bias-variance trade-off in the MSEs of the de-biased estimators with or without the ridge-screening. Figure~\ref{fig-20} in Sec.~\ref{secB} of the Supplement plots the the empirical MSEs of the de-biased ridge estimators for different number of iterations for $\lambda=0.5n$ and $(p,n)=(150,120)$, where Part (a) plots the MSEs of the de-baised ridge estimators  before applying the ridge-screening method, and Part (b) provides the MSEs of the de-biased  ridge estimators after the ridge-screening.  From the plots, it is clear that there is an optimal number of iteration that minimises the MSE, which is in line with our asymptotic results in Theorem~\ref{thm7}.

We also study the performance of the proposed ridge-screening method in Table~\ref{Table-a30} of the Supplement, where the empirical probability (EP) is calculated by
\begin{equation}
 EP(\mathcal{M}_0\subset \mathcal{M}_{k}(\lambda^*))=\frac{1}{1000}\sum_{i=1}^{1000}\frac{|\mathcal{M}_0\cap \mathcal{M}_{k}^i(\lambda^*)|}{10},
\end{equation}
where $\mathcal{M}_{k}^i(\lambda^*)$ is the recovery in the $i$-th experiment and $|\mathcal{M}_0\cap \mathcal{M}_{k}^i(\lambda^*)|$ is the cardinality of the set $\mathcal{M}_0\cap \mathcal{M}_{k}^i(\lambda^*)$. From Table~\ref{Table-a30}, 
the proposed method provides satisfactory performance in variable selections. The comparison results with other variable selection approaches, such as the LASSO, are not provided in this experiment due to the empirical probability of correct recoveries by the proposed RS method being $100\%$ in all settings of Table~\ref{Table-a30}. %{\footnote{The comparison results with other variable selection approaches, such as the LASSO, are not provided in this experiment due to the empirical probability of correct recoveries by the proposed RS method being $100\%$ in all settings of Table~\ref{Table-a30}.}}.
%, where we consider $\lambda=0.5n$ for $(p,n)=(150,120)$ in Example 2, where (a) plots the MSEs of the de-baised ridge estimators  before applying the ridge-screening method, and (b) provides the MSEs of the de-biased  ridge estimators after the ridge-screening.  

%%%%%%%%%%%%%%%%

%%%%%%%%%%%%

 %%%
\begin{table}[htp]\scriptsize
\caption{Empirical mean squared  errors (MSEs) when $p> n$ in Example 2, where the MSEs are similarly defined as those in (\ref{sse0}) and (\ref{ssek}) for $\bb_{\lambda}$, $\bb_{\lambda,k}$, and $\bb_{\lambda,k,l}$. For each $\lambda^*=\lambda$, the upper panel reports the MSEs before ridge-screening, and the lower one presents the MSEs after ridge-screening with $k=100$ and $n^*=40$ in the variable selection. The number of iterations is set to $k=1,5,10,20,50$, and $100$ for the first-stage de-biased estimation, and  $l=1,5,10,20,50,$ and $100$ for the second-stage bias-correction following the ridge-screening.  $\wh\sigma_0$, $\wh\sigma_{100}$, $\wh\sigma_{k,0}$ and $\wh\sigma_{k,100}$ are similarly estimated by the method in (\ref{sigma:e}). 1000 replications are used in the experiments.} 
          \label{Table-a3}
{\begin{center}
\begin{tabular}{cccccccccc}
\toprule
&\multicolumn{9}{c}{$\lambda^*=0.1n$ (before ridge-screening)}\\
\cline{2-10}
$(p,n)$&$\bb_{\lambda}$&$\bb_{\lambda,1}$&$\bb_{\lambda,5}$&$\bb_{\lambda,10}$&$\bb_{\lambda,20}$&$\bb_{\lambda,50}$&$\bb_{\lambda,100}$&$\wh\sigma_0$&$\wh\sigma_{100}$\\
\hline
(150,120)&30.64&26.70&28.64&28.73&28.73&28.73&28.73&1.66&2.14\\
(150,140)&14.84&10.87&11.67&11.68&11.68&11.68&11.68&2.00&2.07\\
(220,180)&50.10&45.30&46.08&46.12&46.12&46.12&46.12&1.75&2.24\\
(220,200)&30.90&24.09&25.02&25.04&25.04&25.04&25.04&1.82&2.03\\
\midrule
&\multicolumn{9}{c}{$\lambda=0.1n$, $k=100$ (after ridge-screening)}\\
\cline{2-10}
$(p,n)$&$\bb_{\lambda,k}$&$\bb_{\lambda,k,1}$&$\bb_{\lambda,k,5}$&$\bb_{\lambda,k,10}$&$\bb_{\lambda,k,20}$&$\bb_{\lambda,k,50}$&$\bb_{\lambda,k,100}$&$\wh\sigma_{k,0}$&$\wh\sigma_{k,100}$\\
\hline
(150,120)&11.34&5.44&5.50&5.50&5.50&5.50&5.50&1.37&0.84\\
(150,140)&8.95&5.52&5.62&5.62&5.62&5.62&5.62&1.25&0.82\\
(220,180)&10.89&3.51&3.48&3.48&3.48&3.48&3.48&1.58&0.91\\
(220,200)&8.49&3.40&3.45&3.45&3.45&3.45&3.45&1.50&0.90\\
\midrule
\midrule
&\multicolumn{9}{c}{$\lambda^*=0.3n$ (before ridge-screening)}\\
\cline{2-10}
$(p,n)$&$\bb_{\lambda}$&$\bb_{\lambda,1}$&$\bb_{\lambda,5}$&$\bb_{\lambda,10}$&$\bb_{\lambda,20}$&$\bb_{\lambda,50}$&$\bb_{\lambda,100}$&$\wh\sigma_0$&$\wh\sigma_{100}$\\
\hline
(150,120)&39.96&31.59&34.59&35.23&35.25&35.25&35.25&2.48&3.46\\
(150,140)&23.83&15.88&18.09&18.39&18.39&18.39&18.39&2.33&3.10\\
(220,180)&60.94&51.27&52.98&53.43&53.45&53.45&53.45&2.63&3.68\\
(220,200)&45.13&32.61&35.01&35.45&35.46&35.46&35.46&2.86&3.62\\
\midrule
&\multicolumn{9}{c}{$\lambda=0.3n$, $k=100$ (after ridge-screening)}\\
\cline{2-10}
$(p,n)$&$\bb_{\lambda,k}$&$\bb_{\lambda,k,1}$&$\bb_{\lambda,k,5}$&$\bb_{\lambda,k,10}$&$\bb_{\lambda,k,20}$&$\bb_{\lambda,k,50}$&$\bb_{\lambda,k,100}$&$\wh\sigma_{k,0}$&$\wh\sigma_{k,100}$\\
\hline
(150,120)&24.95&12.36&12.55&12.55&12.55&12.55&12.55&2.60&1.31\\
(150,140)&17.94&9.56&10.04&10.05&10.05&10.05&10.05&2.39&1.21\\
(220,180)&31.66&13.39&12.73&12.73&12.73&12.73&12.73&3.01&1.44\\
(220,200)&24.13&9.39&9.51&9.51&9.51&9.51&9.51&3.00&1.30\\
\midrule
\midrule
&\multicolumn{9}{c}{$\lambda^*=0.8n$ (before ridge-screening)}\\
\cline{2-10}
$(p,n)$&$\bb_{\lambda}$&$\bb_{\lambda,1}$&$\bb_{\lambda,5}$&$\bb_{\lambda,10}$&$\bb_{\lambda,20}$&$\bb_{\lambda,50}$&$\bb_{\lambda,100}$&$\wh\sigma_0$&$\wh\sigma_{100}$\\
\hline
(150,120)&53.47&40.79&44.04&47.60&48.03&48.04&48.04&3.89&5.72\\
(150,140)&36.98&24.48&29.14&31.97&32.23&32.23&32.23&3.90&5.35\\
(220,180)&76.32&61.94&64.70&68.21&68.59&68.60&68.60&4.15&6.17\\
(220,200)&64.95&46.37&50.70&54.71&55.11&55.11&55.11&4.63&6.53\\
\midrule
&\multicolumn{9}{c}{$\lambda=0.8n$, $k=100$ (after ridge-screening)}\\
\cline{2-10}
$(p,n)$&$\bb_{\lambda,k}$&$\bb_{\lambda,k,1}$&$\bb_{\lambda,k,5}$&$\bb_{\lambda,k,10}$&$\bb_{\lambda,k,20}$&$\bb_{\lambda,k,50}$&$\bb_{\lambda,k,100}$&$\wh\sigma_{k,0}$&$\wh\sigma_{k,100}$\\
\hline
(150,120)&44.82&27.37&26.15&26.49&26.50&26.50&26.50&4.23&2.75\\
(150,140)&31.97&17.88&18.98&19.42&19.43&19.43&19.43&4.22&2.53\\
(220,180)&59.55&35.56&32.39&32.55&32.56&32.56&32.56&4.80&2.86\\
(220,200)&51.11&26.92&25.79&26.20&26.21&26.21&26.21&5.14&2.93\\
\bottomrule
\end{tabular}
  \end{center}}
\end{table}
%%%%%%%%%%%%%%%%%%%%%%%%%%%%%%%%%%%%%%%%%%%%%%%%%%
%%%%%%%%%%%%%%%%%%%%%%%%%%%%%
%\begin{figure}[ht]
%\begin{center}
%{\includegraphics[width=8cm,height=14cm]{Vol-loadings.pdf}}
%%\caption{  The bias-variance trade-off shown by the empirical MSEs of the de-biased ridge estimators for different number of iterations in Example 2, where we consider $\lambda=0.5n$ for $(p,n)=(150,120)$, where Part (a) plots the MSEs of the de-baised ridge estimators  before applying the ridge-screening method, and Part (b) provides the MSEs of the de-biased  ridge estimators after the ridge-screening.  1000 replications are used in the experiments.  }\label{fig-20}
%\end{center}
%\end{figure}
%%%%%%%%%%%%%%%%%%%%%

%%%%%%%%%%%%%%%%%%%%%%

Similarly to the experiments in Example 1, we present the empirical average estimation errors (AEEs) in Table~\ref{Table-a4}. The settings of $(p,n)$, $\lambda^*$, $\lambda$, $k$, and $l$
are the same as those in Table~\ref{Table-a3}. From Table~\ref{Table-a4}, we see that the average estimation errors before ridge-screening are also quite stable as we increase the number of iterations, which is similar to the findings in Table~\ref{Table-a3}. After we apply the ridge-screening approach to the de-biased estimators in the upper panel of Table~\ref{Table-a4} with $k=100$ for each $\lambda^*$, the bias terms (in the lower panel for each $\lambda$) decrease substantially from their previous values, and the biases also decrease to stable values as we increase the number of iterations in the second-stage bias correction. Notably, the bias terms are all significantly smaller than those in the upper panel without the ridge-screening procedure. This finding is in agreement with our theoretical results in Theorem~\ref{thm5} and Theorem~\ref{thm7}.

%%%
\begin{table}[htp]\scriptsize
\caption{Empirical average estimation errors (AEEs) when $p> n$ in Example 2, where the AEEs are similarly defined as those in (\ref{sse0}) and (\ref{ssek}) for $\bb_{\lambda}$, $\bb_{\lambda,k}$, and $\bb_{\lambda,k,l}$. For each $\lambda$, the upper panel reports the AEES before ridge-screening, and the lower one presents the AEEs after ridge-screening with $k=100$ and $n^*=40$ in the variable selection. The number of iterations is set to $k=1,5,10,20,50$, and $100$ for the first-stage de-biased estimation, and  $l=1,5,10,20,50,$ and $100$ for the second-stage bias-correction following the ridge-screening.   1000 replications are used in the experiments.} 
          \label{Table-a4}
{\begin{center}
\begin{tabular}{cccccccc}
\toprule
&\multicolumn{7}{c}{$\lambda^*=0.1n$ (before ridge-screening)}\\
\cline{2-8}
$(p,n)$&$\bb_{\lambda}$&$\bb_{\lambda,1}$&$\bb_{\lambda,5}$&$\bb_{\lambda,10}$&$\bb_{\lambda,20}$&$\bb_{\lambda,50}$&$\bb_{\lambda,100}$\\
\hline
(150,120)&0.44 &0.41& 0.42& 0.42& 0.42& 0.42 &0.42\\
(150,140)&0.30& 0.24& 0.25& 0.25& 0.25& 0.25& 0.25\\
(220,180)&0.47 &0.44& 0.44& 0.44& 0.44 &0.44 &0.44\\
(220,200)&0.37 &0.32 &0.32& 0.32 &0.32& 0.32& 0.32\\
\midrule
&\multicolumn{7}{c}{$\lambda=0.1n$, $k=100$ (after ridge-screening)}\\
\cline{2-8}
$(p,n)$&$\bb_{\lambda,k}$&$\bb_{\lambda,k,1}$&$\bb_{\lambda,k,5}$&$\bb_{\lambda,10}$&$\bb_{\lambda,k,20}$&$\bb_{\lambda,k,50}$&$\bb_{\lambda,k,100}$\\
\hline
(150,120)&0.26& 0.17& 0.17& 0.17 &0.17& 0.17 &0.17\\
(150,140)&0.22& 0.16 &0.16 &0.16& 0.16& 0.16& 0.16\\
(220,180)& 0.21& 0.11 &0.11& 0.11& 0.11& 0.11& 0.11\\
(220,200)&0.19& 0.11& 0.11 &0.11 &0.11& 0.11 &0.11\\
\midrule
\midrule
&\multicolumn{7}{c}{$\lambda^*=0.3n$ (before ridge-screening)}\\
\cline{2-8}
$(p,n)$&$\bb_{\lambda}$&$\bb_{\lambda,1}$&$\bb_{\lambda,5}$&$\bb_{\lambda,10}$&$\bb_{\lambda,20}$&$\bb_{\lambda,50}$&$\bb_{\lambda,100}$\\
\hline
(150,120)&0.51 &0.45& 0.47& 0.47& 0.47& 0.47 &0.47\\
(150,140)&0.39& 0.31& 0.33& 0.33& 0.33& 0.33& 0.33\\
(220,180)&0.52 &0.48& 0.48& 0.48& 0.48 &0.48 &0.48\\
(220,200)&0.45 &0.38 &0.39& 0.39 &0.39& 0.39& 0.39\\
\midrule
&\multicolumn{7}{c}{$\lambda=0.3n$, $k=100$ (after ridge-screening)}\\
\cline{2-8}
$(p,n)$&$\bb_{\lambda,k}$&$\bb_{\lambda,k,1}$&$\bb_{\lambda,k,5}$&$\bb_{\lambda,10}$&$\bb_{\lambda,k,20}$&$\bb_{\lambda,k,50}$&$\bb_{\lambda,k,100}$\\
\hline
(150,120)&0.40& 0.27& 0.27& 0.27 &0.27& 0.27 &0.27\\
(150,140)&0.34& 0.24 &0.24 &0.25& 0.25& 0.25& 0.25\\
(220,180)& 0.37& 0.24 &0.23& 0.23& 0.23& 0.23& 0.23\\
(220,200)&0.33& 0.20& 0.20 &0.20 &0.20& 0.20 &0.20\\
\midrule
\midrule
&\multicolumn{7}{c}{$\lambda^*=0.8n$ (before ridge-screening)}\\
\cline{2-8}
$(p,n)$&$\bb_{\lambda}$&$\bb_{\lambda,1}$&$\bb_{\lambda,5}$&$\bb_{\lambda,10}$&$\bb_{\lambda,20}$&$\bb_{\lambda,50}$&$\bb_{\lambda,100}$\\
\hline
(150,120)&0.60 &0.52& 0.53& 0.56& 0.56& 0.56 &0.56\\
(150,140)&0.50& 0.40& 0.43& 0.45& 0.45& 0.45& 0.45\\
(220,180)&0.59 &0.53& 0.54& 0.55& 0.55 &0.55 &0.55\\
(220,200)&0.54 &0.46 &0.47& 0.49 &0.49& 0.49& 0.49\\
\midrule
&\multicolumn{7}{c}{$\lambda=0.8n$, $k=100$ (after ridge-screening)}\\
\cline{2-8}
$(p,n)$&$\bb_{\lambda,k}$&$\bb_{\lambda,k,1}$&$\bb_{\lambda,k,5}$&$\bb_{\lambda,10}$&$\bb_{\lambda,k,20}$&$\bb_{\lambda,k,50}$&$\bb_{\lambda,k,100}$\\
\hline
(150,120)&0.54& 0.42& 0.41& 0.41 &0.41& 0.41 &0.41\\
(150,140)&0.46& 0.34 &0.35 &0.35& 0.35& 0.35& 0.35\\
(220,180)& 0.52& 0.40 &0.38& 0.38& 0.38& 0.38& 0.38\\
(220,200)&0.48& 0.34& 0.33&0.33 &0.33& 0.33 &0.33\\
\bottomrule
\end{tabular}
  \end{center}}
\end{table}
%%%%%%%%%%%%%%%%%%%%%%%%%%%%%%%%%%%%%%%%%%%%%%%%%%

\begin{figure}[ht]
\begin{center}
%{\includegraphics[width=8cm,height=14cm]{Vol-loadings.pdf}}
{\includegraphics[width=14cm,height=8cm]{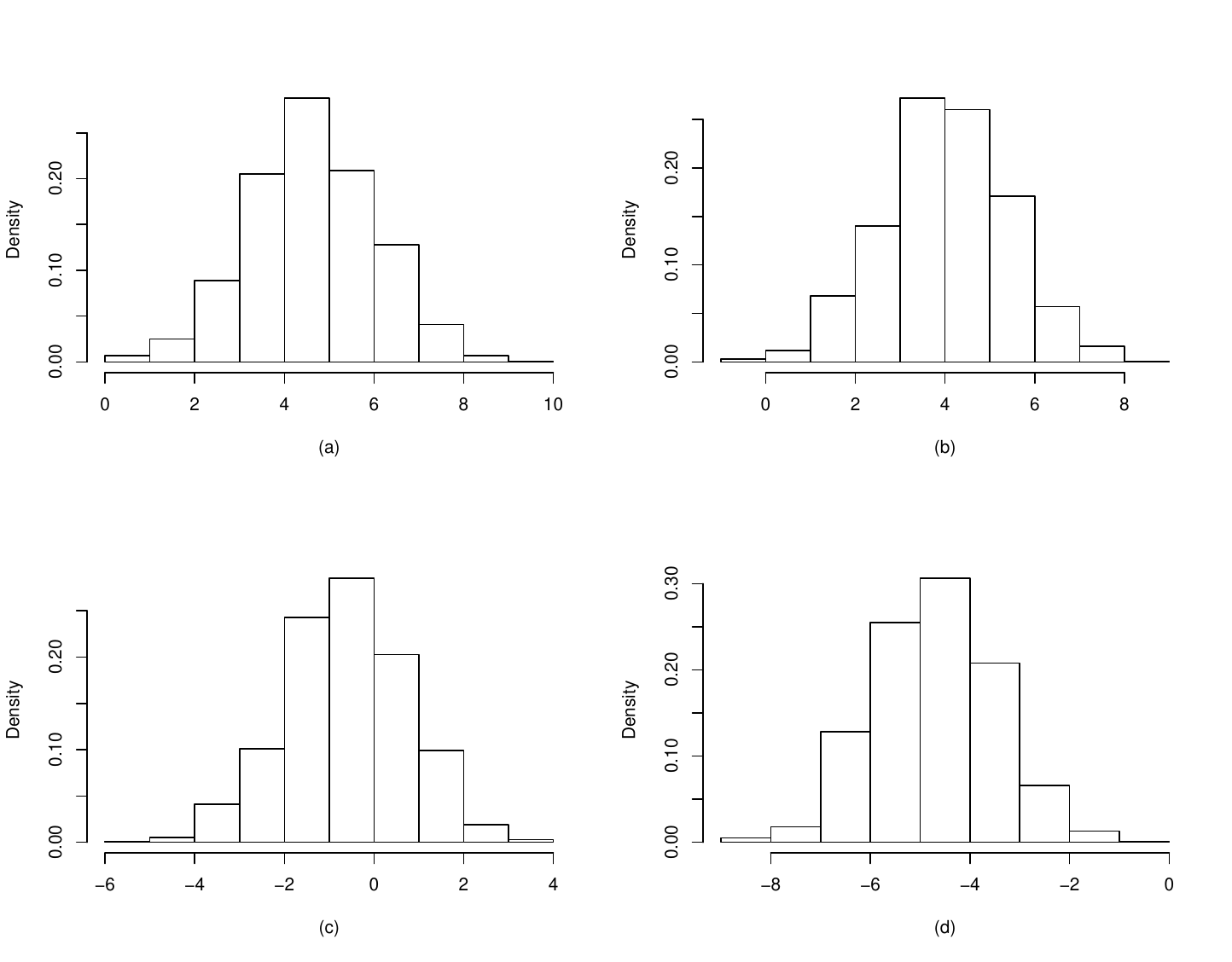}}
\caption{Empirical histograms of (a) $\sqrt{n}\bgamma_1'(\wh\bbeta(\lambda)-\bbeta)$; (b)  $\sqrt{n}\bgamma_2'(\wh\bbeta(\lambda)-\bbeta)$; (c) $\sqrt{n}\bgamma_3'(\wh\bbeta(\lambda)-\bbeta)$; and (d) $\sqrt{n}\bgamma_4'(\wh\bbeta(\lambda)-\bbeta)$ in Example 2. We set $(p,n)=(220,200)$ and $\lambda=0.1n$. 1000 replications are used in the experiments.  }\label{fig-3}
\end{center}
\end{figure}

\begin{figure}[ht]
\begin{center}
%{\includegraphics[width=8cm,height=14cm]{Vol-loadings.pdf}}
{\includegraphics[width=14cm,height=8cm]{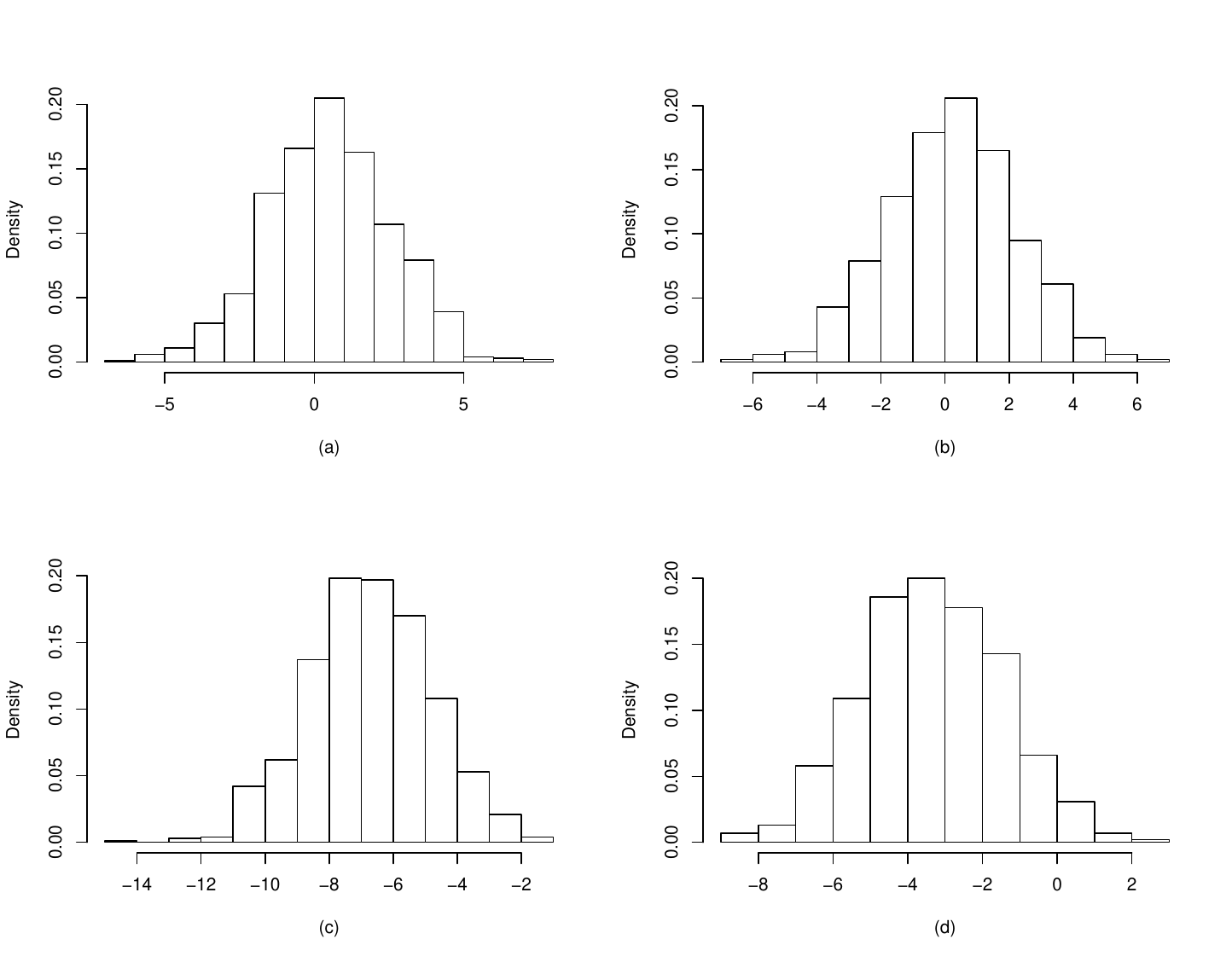}}
\caption{Empirical histograms of (a) $\sqrt{n}\bgamma_1'(\wh\bbeta_{c,k}(\lambda)-\bbeta)$; (b)  $\sqrt{n}\bgamma_2'(\wh\bbeta_{c,k}(\lambda)-\bbeta)$; (c) $\sqrt{n}\bgamma_3'(\wh\bbeta_{c,k}(\lambda)-\bbeta)$; and (d) $\sqrt{n}\bgamma_4'(\wh\bbeta_{c,k}(\lambda)-\bbeta)$ in Example 2. We set $(p,n)=(220,200)$, $k=100$, and $\lambda=0.1n$. 1000 replications are used in the experiments.  }\label{fig-4}
\end{center}
\end{figure}

Finally, we study the performance of the inference method in Theorem~\ref{thm6}(ii). We only consider the case when $(p,n)=(220,200)$, $\lambda^*=\lambda=0.1n$, and $n^*=40$, and we can produce similar results for other settings. Define
\[\bgamma_1=(0.8,-1,0.5,{\bf 0}_{p-3}')',\quad\bgamma_2=(0,-1,0.5,0.8,{\bf 0}_{p-4}')',\]
\[\bgamma_3=({\bf 0}_3',-0.9,0.4,-0.8,{\bf 0}_{p-6}')',\,\,\text{and}\,\,\bgamma_4=({\bf 0}_4',0.5,0.7,-0.8,{\bf 0}_{p-7}')'.\]
We plot the empirical histograms of  $\sqrt{n}\bgamma_1'(\wh\bbeta(\lambda)-\bbeta)$,   $\sqrt{n}\bgamma_2'(\wh\bbeta(\lambda)-\bbeta)$, $\sqrt{n}\bgamma_3'(\wh\bbeta(\lambda)-\bbeta)$, and  $\sqrt{n}\bgamma_4'(\wh\bbeta(\lambda)-\bbeta)$ in Figure~\ref{fig-3}, from which it is observed that most of the estimators are not centred around zero. Subsequently, we apply the bias-correction procedure to the original ridge estimators, and the empirical histograms of the bias-corrected estimators are presented in Figure~\ref{fig-4} as compared to Figure~\ref{fig-3}. From Figure~\ref{fig-4}, it can be seen that while some estimators become more centred around zero (e.g., (a) and (b)) due to the bias-correction procedure, the validity of inference without ridge-screening cannot be assured, as (c) and (d) still exhibit significant biases.

\begin{figure}[ht]
\begin{center}
%{\includegraphics[width=8cm,height=14cm]{Vol-loadings.pdf}}
{\includegraphics[width=14cm,height=8cm]{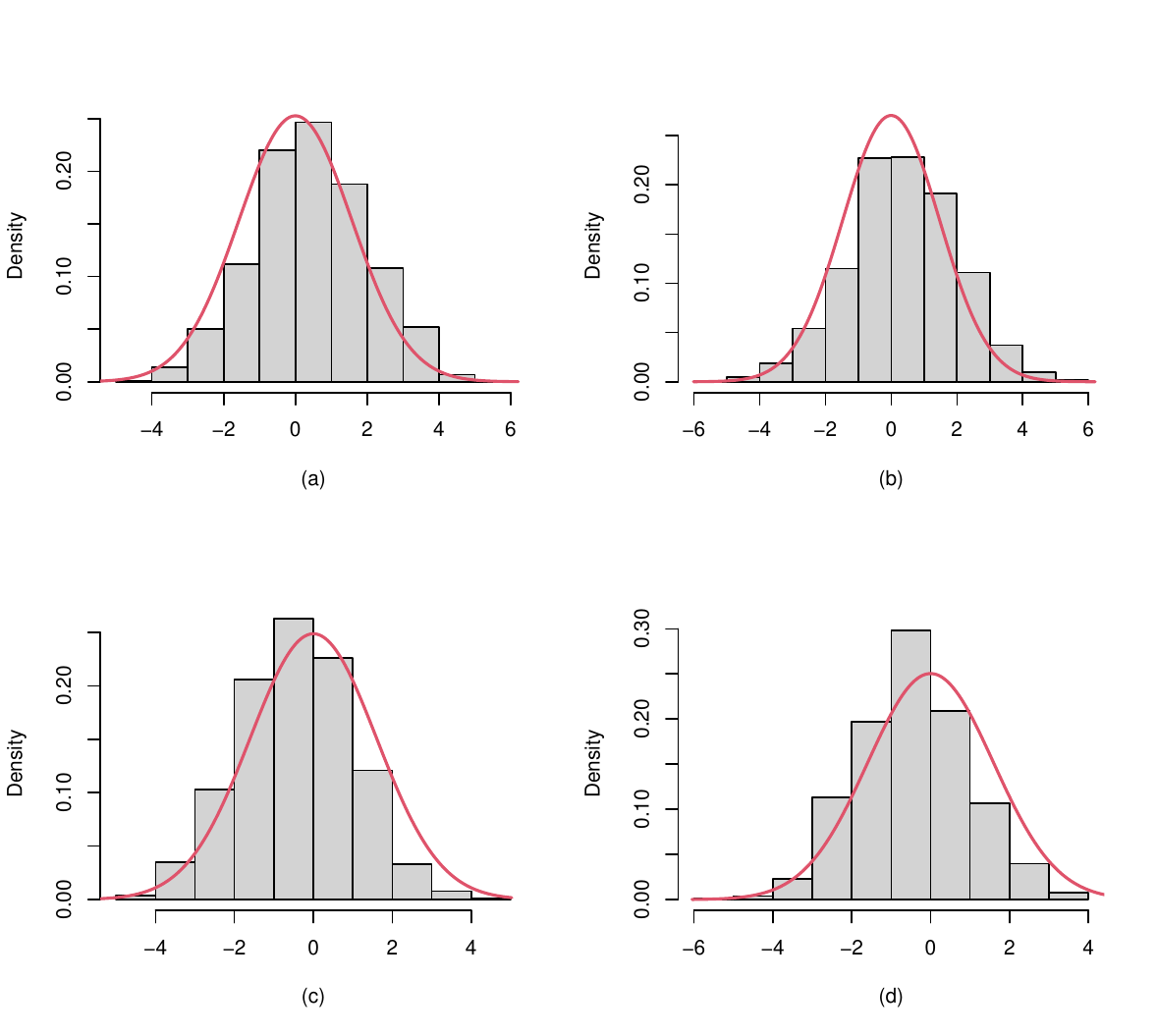}}
\caption{Empirical histograms of (a) $\sqrt{n}\bgamma_1'(\wh\bbeta_{\mathcal{M}_k,l}^*(\lambda)-\bbeta)$; (b)  $\sqrt{n}\bgamma_2'(\wh\bbeta_{\mathcal{M}_k,l}^*(\lambda)-\bbeta)$; (c) $\sqrt{n}\bgamma_3'(\wh\bbeta_{\mathcal{M}_k,l}^*(\lambda)-\bbeta)$; and (d) $\sqrt{n}\bgamma_4'(\wh\bbeta_{\mathcal{M}_k,l}^*(\lambda)-\bbeta)$ in Example 2, where $\wh\bbeta_{\mathcal{M}_k,l}^*(\lambda)$ is a $p$-dimensional vector with $\wh\bbeta_{\mathcal{M}_k,l}(\lambda)$ as its sub-vector and the other elements being zero. We set $(p,n)=(220,200)$, $k=100$, $n^*=40$, $l=100$, and $\lambda=\lambda^*=0.1n$. The density curve is plotted based on the limiting distribution in Theorem~\ref{thm6}(ii). 1000 replications are used in the experiments. }\label{fig-5}
\end{center}
\end{figure}

Furthermore, we apply the ridge-screening approach to the de-biased estimators depicted in Figure~\ref{fig-4}, and subsequently obtain restricted ridge estimators along with their de-biased counterparts. Figure~\ref{fig-5} showcases the empirical histograms of: (a) $\sqrt{n}\bgamma_1'(\wh\bbeta_{\mathcal{M}_k,l}^*(\lambda)-\bbeta)$; (b)  $\sqrt{n}\bgamma_2'(\wh\bbeta_{\mathcal{M}_k,l}^*(\lambda)-\bbeta)$; (c) $\sqrt{n}\bgamma_3'(\wh\bbeta_{\mathcal{M}_k,l}^*(\lambda)-\bbeta)$; and (d) $\sqrt{n}\bgamma_4'(\wh\bbeta_{\mathcal{M}_k,l}^*(\lambda)-\bbeta)$, where $\wh\bbeta_{\mathcal{M}_k,l}^*(\lambda)$ is a $p$-dimensional vector with $\wh\bbeta_{\mathcal{M}_k,l}(\lambda)$ as its sub-vector and the remaining elements are zero. Similar to the curves in Figure~\ref{fig-2}, the density curve in Figure~\ref{fig-5} is plotted based on the limiting distribution specified in Theorem~\ref{thm6}(ii). From Figure~\ref{fig-5}, it is evident that the de-biased restricted ridge estimators predominantly cluster around zero. The density curves for the corresponding estimators are quite satisfactory, suggesting the validity of the proposed inference approach.

%%%%%%%%%%%%%%%%%%%%%%%%%%%%%
%%%%%%%%%%%%%%%%%%%%%

%%%%%%%%%%%%%%%%%%%%%%%%%%%%%
%%%%%%%%%%%%%%%%%%%%%

%%%%%%%%%%%%%%%%%%%%%%%%%%%%%
%%%%%%%%%%%%%%%%%%%%%

\section{An Empirical Application} \label{sec40}

In this section, we apply the proposed method to forecast the U.S. macroeconomic series and study the out-of-sample prediction intervals using ridge regression. We consider the widely used macroeconomic variables studied by \cite{Stock2002}, \cite{mccracken2016fred}, \cite{giannone2021economic}, and \cite{gao2022modeling,gao2023supervised}, among many others. The data are obtained from the FRED-MD database which is maintained by St. Louis Federal Reserve Bank. See \url{https://research.stlouisfed.org/econ/mccracken/fred-databases/}. %\footnote{\url{https://research.stlouisfed.org/econ/mccracken/fred-databases/}}. 
There are 127 variables in the online data set, and we remove 4 of them because of missing values therein.  Consequently, we consider the 123 macro variables spanning from July 1962 to December
2019 as all the series have no missing values during this period. The resulting 
data set is the same as that in \cite{gao2023supervised}. The detailed variables and transformation codes to ensure the stationarity of each
macro variable are provided in Table IA.I of \cite{gao2023supervised}. The sample size is $n = 690$.

Similarly to \cite{Stock2002b}, \cite{bai2006}, and \cite{cheng2015forecasting}, we adopt the following factor-augmented regression model: 
\begin{equation}
\begin{array}{l}
     \bX_t= \bLambda\bff_t+\be_t, \\
     y_{t+h}=\alpha_1y_{t}+...+\alpha_{q}y_{t-q+1}+\gamma_1f_{1,t}+...+\gamma_rf_{r,t}+\ve_{t+h}, t=q,...,n-h,
\end{array}
\end{equation}
where $\bX_t$ consists of 122 macroeconomic variables, $y_{t}$ is the remaining target one, and $\bff_t=(f_{1,t},...,f_{r,t})'$ is an $r$-dimensional latent factor process which can be estimated by applying Principal Component Analysis (PCA) to $\bX_t$. We set $q=10$ and $r=60$, and apply the ridge regression to estimate the parameters $\alpha_i$ and $\gamma_j$, for $1\leq i\leq q$ and $1\leq j\leq r$, with estimated factors. Hence, the number of covariates in the ridge regression is $p=q+r=70$. We focus on the prediction of  the consumer price index: all (CPI-All), which is an important index related to the inflation. See also \cite{Stock2002} and \cite{gao2022modeling} for similar studies. 

First, we set $\lambda\in\{0.05n,0.1n,0.2n,...,1.5n\}$, which contains 16 candidates for the penalty parameter.  We split the sample into two subsamples, where the first one consists of the first $80\%$ of the data for modelling and the second subsample of the remaining $20\%$ of the data for out-of-sample prediction. We  adopt the rolling-window scheme as that in \cite{gao2023supervised}, that is, we train the factors and the forecasting coefficients using the first subsample to predict the next target data point. Then we repeat the above procedure after moving the next available observation of predictors and target CPI-All variable from the second subsample to the first one to obtain the next prediction. This rolling-window scheme is terminated when there is no more observation to compute the forecasting error. We choose the optimal penalty in the ridge estimator such that the out-of-sample mean-squared forecast errors (MSFEs) achieve a minimum value. Thus, this setting is in line with the advocated models combining regularised estimation and data-driven selection of regularisation parameters  in \cite{abadie2019choosing}. In this experiment, we find that the optimal penalty $\wh\lambda=0.8n$ for $h=1$ and $\wh\lambda=0.7n$ for $h=2$.

\begin{figure}[ht]
\begin{center}
%{\includegraphics[width=8cm,height=14cm]{Vol-loadings.pdf}}
{\includegraphics[width=14cm,height=7cm]{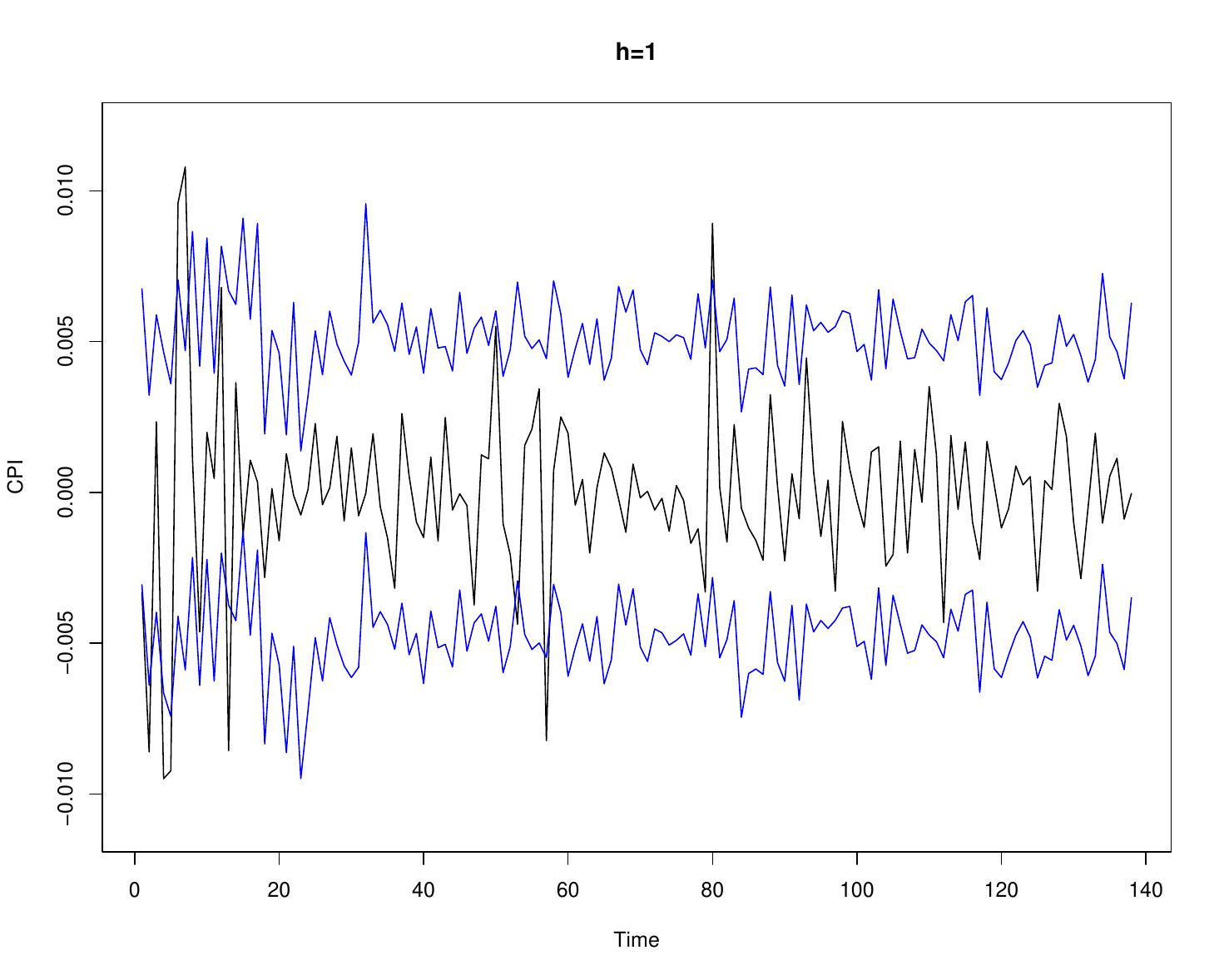}}
\caption{Out-of-sample pointwise prediction intervals for the monthly CPI-All from July 1, 2008 to Dec. 12, 2019, constructed using (\ref{l2})-(\ref{u2}) at a $95\%$ confidence level with $\wh\lambda=0.8n$ and $k=10$ for the 1-step ahead forecast. The blue lines represent the confidence intervals, while the dark lines plot the true observations.}\label{fig-8}
\end{center}
\end{figure}

\begin{figure}[ht]
\begin{center}
%{\includegraphics[width=8cm,height=14cm]{Vol-loadings.pdf}}
{\includegraphics[width=14cm,height=7cm]{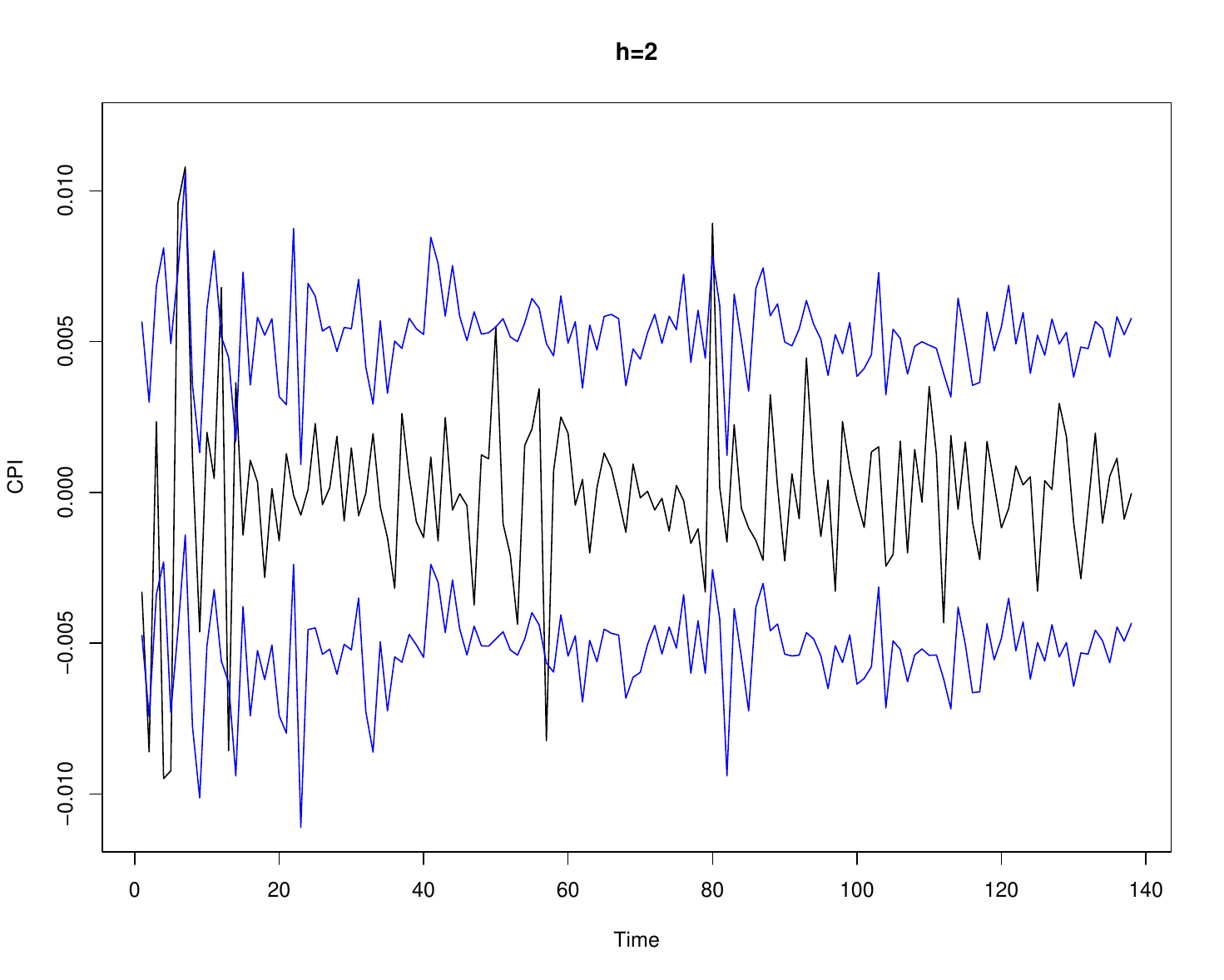}}
\caption{Out-of-sample pointwise prediction intervals for the monthly CPI-All from July 1, 2008 to Dec. 12, 2019, constructed using (\ref{l2})-(\ref{u2}) at a $95\%$ confidence level with $\wh\lambda=0.7n$ and $k=10$ for the 2-step ahead forecast. The blue lines represent the confidence intervals, while the dark lines plot the true observations. }\label{fig-9}
\end{center}
\end{figure}

Next, for each configuration of $(\wh\lambda,h)=(0.8n,1)$ and $(0.7n,2)$, we execute Algorithm~\ref{pless:a1}. We observe that the bias-correction procedure terminates within 3 to 5 iterations when setting $\eta=O(10^{-4})$  in equation (\ref{con:cri}). For simplicity, we opt for $k=10$ in the bias-correction, a choice deemed sufficiently large. Following the guidelines provided in Section~\ref{sec27}, we compute pointwise prediction intervals for the data points in the testing subsample. 
The standard errors for each prediction are determined using the method described in Remark~\ref{rm5}(iv). We present the $95\%$ pointwise prediction intervals for both 1-step ahead and 2-step ahead forecasts in Figure~\ref{fig-8} and Figure~\ref{fig-9}, respectively. It is evident that the majority of the true observations fall within these prediction intervals. Additionally, we evaluate the coverage rates of the prediction intervals across the 138 points in the testing subsample, and find a consistent coverage rate of $92.03\%$ for both the 1-step and 2-step ahead forecasts, indicating that the $95\%$ prediction intervals perform effectively in this empirical example.

 %We plot the 1-step ahead and 2-step ahead pointwise prediction intervals at $95\%$ level in Figure~\ref{fig-8} and Figure~\ref{fig-9}, respectively. We can see clearly that most of the true observations are covered by the prediction intervals. Furthermore, we also examine the coverage rates of the prediction intervals at the 138 points in the testing set, and found that the coverage rate is $92.03\%$ for both the 1-step and 2-step ahead forecasts, implying that the $95\%$ prediction intervals work well in this empirical example.

%%%%%%%%%%%%%%%%%%

%%%%%%%%%%%%%%%%%%%%%%%%%%%%%
%%%%%%%%%%%%%%%%%%%%%

%%%%%%%%%%%%%%%%%%%%%%%%%%%%%
\begin{figure}[ht]
\begin{center}
%{\includegraphics[width=8cm,height=14cm]{Vol-loadings.pdf}}
{\includegraphics[width=14cm,height=7cm]{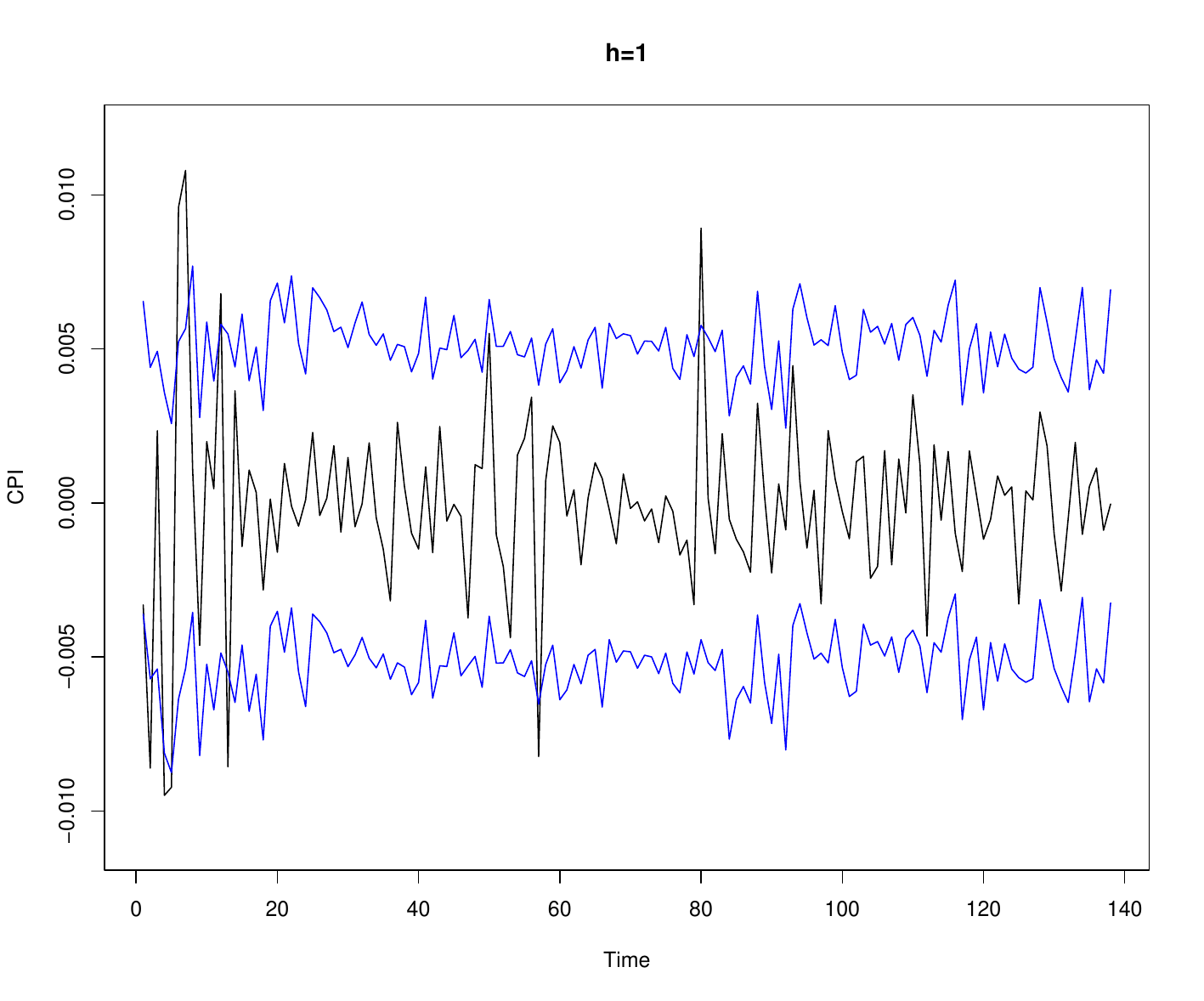}}
\caption{Out-of-sample pointwise prediction intervals for the monthly CPI-All from July 1, 2008 to Dec. 12, 2019, constructed using (\ref{l3})-(\ref{u3}) at $95\%$ level with $\wh\lambda^*=\wh\lambda=0.5n$, $k=10$, $n^*=31$, and $l=10$ for the $1$-step ahead forecast. The blue lines represent the confidence intervals, while the dark lines plot the true observations.}\label{fig-10}
\end{center}
\end{figure}

%%%%%%%%%%%%%%%%%%%%%
Finally, we investigate the effectiveness of the ridge-screening (RS) approach in out-of-sample predictions. We take the 1-step ahead prediction as an example. For each $\lambda^*\in\{0.05n,0.1n,0.2n,...,1.5n\}$, which is the same as our previous candidate set, we adopt $k=10$  and vary $n^*$ from $10$ to $70$  in (\ref{rs:sub}), that is, we select the largest $10$ to $70$ variables based on their magnitudes from the set $\{|\wh\bbeta_{c,10,1}(\lambda^*)|,...,|\wh\bbeta_{c,10,70}(\lambda^*)|\}$ and compute their corresponding out-of-sample forecast errors. By minimizing the out-of-sample MSFEs, we determined that the optimal parameters are $(\wh\lambda^*,n^*)=(0.5n,31)$. This implies that the submodel, consisting of covariates corresponding to the largest 31 elements in $|\wh\bbeta_{c,10}(0.5n)|$, yields the smallest out-of-sample MSFEs. Consequently, we chose $\wh\lambda=\wh\lambda^*=0.5n$ for the subsequent bias-correction procedure in the restricted ridge estimators, conducting it with $l=10$ iterations. Figure~\ref{fig-10} displays the $95\%$ out-of-sample prediction intervals based on the selected submodel and the corresponding restricted ridge estimators for $h=1$. Interestingly, we observe that some observations not covered by the prediction intervals in Figure~\ref{fig-8} are encompassed by the new intervals in Figure~\ref{fig-10}. The coverage rate of these $95\%$ prediction intervals stands at $93.5\%$, surpassing the $92.03\%$ achieved by the previous approach without ridge-screening. This suggests that the ridge-screening method enhances the performance of the prediction intervals.

 %Figure-\ref{fig-10} provides the $95\%$ out-of-sample prediction intervals constructed by the submodel and the restricted ridge estimators when $h=1$. We find that some observations which is not covered by the prediction intervals in Figure~\ref{fig-8} can be covered by the new ones in Figure~\ref{fig-10}. In fact, the coverage rate of the prediction intervals is $93.5\%$, which is higher than the $92.03\%$ in the previous approach without the ridge-screening, implying that the ridge-screening method can improve the performance of the prediction intervals.

%%%%%%%%%%%%%%%%%%%%%%%%%%%%%
			\section{Conclusion} \label{sec4}
  % A fundamental question in ridge regression is: Can we correct the bias term using the data? How do we make valid statistical inferences? This paper addresses these issues by proposing a simple and easy-to-implement iterative procedure and a ridge-screening method. When the dimension $p$ is less than the sample size $n$, we show that our procedure can fully correct the bias term under vey mild assumptions. When $p>n$ and hence the projection matrix is singular, we propose a ridge-screening approach to drop those coefficients that are small compared to others. We only consider the restricted model which has sufficient many parameters that contain most of the information of the data with respect to the response vector. Our bias-correction procedure can be further applied to the restricted model and the bias can be corrected. The ridge-screening method can be treated as a new variable selection approach, which is of independent interest. We derive limiting distributions of the de-biased estimators, which can be used to make statistical inferences. Simulated experiments confirm the validity of our proposes methodology.

   A fundamental question in ridge regression revolves around the possibility of correcting the bias term using available data, alongside ensuring the validity of statistical inferences without affecting the predictive performance of the original ridge estimator. This paper tackled these long-standing challenges by introducing a straightforward and readily implementable iterative procedure, along with a ridge-screening method. In cases where the dimension $p$ is smaller than the sample size $n$, our procedure effectively corrects the bias term under some mild assumptions. However, when $p>n$ and the projection matrix becomes singular, we propose a ridge-screening approach to eliminate coefficients that are relatively insignificant compared to others. We then specifically focus on the restricted model, which retains a sufficient number of parameters to encapsulate most of the information regarding the response vector. Our bias-correction procedure can be further applied to the restricted model, enabling the correction of the bias. Moreover, the ridge-screening method offers a novel approach to variable selection, which is of independent interest beyond bias correction. We derive the limiting distributions of the de-biased estimators, to facilitate making statistical inferences. Simulated  and real data examples are used to corroborate the effectiveness and validity of our proposed methodology. The proposed inference solution for ridge regression serves as an illustrative example addressing the inference challenges of regularised machine learning methods outlined in Section 2.8 of \cite{athey2019machine}  without adversely affecting the predictive performance.  %Furthermore, our method offers a transformative solution to bias challenges in ridge regression inferences, with broad applications across various disciplines.

		% Furthermore, we would like to study the prediction of future outcomes based on the state-varying factor model. 
		
	%\end{onehalfspacing}

	% \printbibliography 
%	\bibliographystyle{apalike} 
%	\bibliography{reference}
%	
	
		\singlespacing
\bibliographystyle{econometrica-3}
\let\oldbibliography\thebibliography
\renewcommand{\thebibliography}[1]{%
  \oldbibliography{#1}%
  \setlength{\itemsep}{3pt}%
}
%\bibliographystyle{jof}
%\biboptions{authoryear}
%\bibliographystyle{elsarticle-harv}
%\biboptions{authoryear}
%\renewcommand*{\bibfont}{\normalsize}
	{\footnotesize
		\bibliography{reference}
	}

 %\end{document}
%%%%%%%%%%%%%%from here 
 %\onehalfspacing
 \setstretch{1.8}%1.4
%	

%%%%%%%%%%%%%%%%%%%%%%%%%%%%%%
%%%%%%%%%%%%%%%%%%%%%%%%%%%%%%%%%%%
%%%%%%%%%%%%%%%%%%%%%%%%%%Appendix
\newpage
\setcounter{page}{1}
%%%environment
\appendix
\restoregeometry

 \setcounter{equation}{0}
	\renewcommand{\theequation}{A.\arabic{equation}}
	\renewcommand{\thefigure}{A.\arabic{figure}} \setcounter{figure}{0}
	\renewcommand{\thetable}{A.\arabic{table}} \setcounter{table}{0}

	\begin{center}
 {\Large Supplementary Material for \\``Optimal Bias-Correction and Valid Inference in High-Dimensional Ridge Regression: A Closed-Form Solution”}
 \vskip 0.3 cm

 {\Large Zhaoxing Gao$^1$ and Ruey. S. Tsay$^2$\\$^1$University of Electronic Science and Technology of China\\ $^2$University of Chicago}
	\end{center}
\vskip 0.5cm
 Section~\ref{sec5} in this Supplementary Material contains all the proofs and technical details related to the theoretical results in the main text, and Section~\ref{secB} provides one additional figure and table illustrating the simulation results discussed in Example 2 of Section~\ref{sec3}.

  %\section*{Online Appendix: Proofs of the Theorems}\label{sec5}
 % \section{Online Appendix: Proofs of the Theorems}\label{sec5}
 \section{Proofs of the Theorems}\label{sec5}
%\begin{small}
%\onehalfspacing%spacing
%\end{small}
 In this section, we will use $C$ or $c$ to denote a generic constant the value of which may change at different places.
\vskip 0.5cm

 {\bf Proof of Theorem \ref{thm1}.} To see the bias of the ridge estimator, it follows from (\ref{v:hlm}) that
\begin{equation}\label{decom}
    \wh\bbeta(\lambda)=\bbeta-\lambda(\bX'\bX+\lambda\bI_p)^{-1}\bbeta+(\bX'\bX+\lambda\bI_p)^{-1}\bX'\bve.
\end{equation}
Conditioning on $\bX$, by the assumption that $E\bve={\bf 0}$, it follows immediately that
\[\bbeta-E(\wh\bbeta(\lambda))=\lambda(\bX'\bX+\lambda\bI_p)^{-1}\bbeta.\]
This completes the proof. $\Box$
\vskip 0.6 cm
{\bf Proof of Theorem~\ref{thm2}}. By (\ref{decom}) and an elementary argument, $\wh\bbeta_{c,k}(\lambda)$ can be written as
\begin{align}\label{btck}
   \wh\bbeta_{c,k}(\lambda)=&\wh\bbeta(\lambda)+\sum_{j=1}^{k}\lambda^j(\bX'\bX+\lambda\bI_p)^{-j}\wh\bbeta(\lambda)\notag\\
   =&\bbeta-\lambda^{k+1}(\bX'\bX+\lambda\bI_p)^{-(k+1)}\bbeta+\sum_{j=0}^k\lambda^j(\bX'\bX+\lambda\bI_p)^{-(j+1)}\bX'\bve.
\end{align}
Then, it follows that
\[\bbeta-E\wh\bbeta_{c,k}=\lambda^{k+1}(\bX'\bX+\lambda\bI_p)^{-(k+1)}\bbeta.\]
When $p<n$ and $\bX'\bX$ is invertible, by Assumption~\ref{asm1}, the singular-value decomposition of $\bX'\bX$ is
\[\bX'\bX=\bU_1\bD_1^2\bU_1,\,\,\bD_1=\diag(d_1,...,d_p),\,\, d_j>0.\]
Then, it is not hard to see that
\[(\bX'\bX+\lambda\bI_p)^{-(k+1)}=\bU_1(\bD_1^2+\lambda\bI_p)^{-(k+1)}\bU_1',\]
and
\[\bb_{\lambda,k}=\lambda^{k+1}\bU_1(\bD_1^2+\lambda\bI_p)^{-(k+1)}\bU_1'\bbeta.\]
Note that, for $1\leq j\leq p$, each diagonal element in $\lambda^{k+1}(\bD_1^2+\lambda\bI_p)^{-(k+1)}$ is
\[\left(\frac{\lambda}{d_j^2+\lambda}\right)^{k+1}\rightarrow 0,\,\, \text{as $k\rightarrow\infty$},\]
at a rate of exponential decay. Therefore, for any given configuration of $(p,n)$ with $p<n$, if the number of iteration $k$ satisfies $\max_{1\leq j\leq p}C_{n,p}(\frac{\lambda}{d_j^2+\lambda})^{k+1}\rightarrow 0$ with $\|\bbeta\|_2<C_{p,n}$, we can show that
\[\|\bb_{\lambda}\|_2\rightarrow 0,\,\,\text{as $k\rightarrow\infty$}.\]
As a matter of fact, the aforementioned result applies to any given and fixed $(p, n)$. Furthermore, when considering the scenario where both $p$ and $n$ tend towards infinity in an asymptotic framework, it is not hard to see that the above convergence also holds under the condition that $d_j^2\asymp \lambda \asymp n$ in relation to the sample size $n$ within an asymptotic setting. 
This completes the proof. $\Box$
\vskip 0.6cm

{\bf Proof of Theorem~\ref{thm3}}. Note that
\[\bb_{\lambda,k}=\lambda^{k+1}(\bX'\bX+\lambda\bI_p)^{-(k+1)}\bbeta.\]
When $p>n$, by Assumption~\ref{asm2}, the singular-value decomposition of $(\bX'\bX+\lambda\bI_p)$ is
\[\bX'\bX+\lambda\bI_p=\bU\bD\bU',\,\,\bU=[\bU_1,\bU_2],\,\,\bD=\diag(\bD_1^2+\lambda\bI_{p^*},\lambda\bI_{p-p^*}).\]
Then, 
\begin{align}\label{blk2}
 \bb_{\lambda,k}=&\lambda^{k+1}\left\{\bU_1(\bD_1+\lambda\bI_{p^*})^{-(k+1)}\bU_1'+\lambda^{-(k+1)}\bU_2\bU_2'\right\}\bbeta\notag\\
 =&\bU_1\lambda^{k+1}(\bD_1+\lambda\bI_{p^*})^{-(k+1)}\bU_1'\bbeta+\bU_2\bU_2'\bbeta.
\end{align}
By a similar argument as that in the proof of Theorem~\ref{thm2}, the first term in (\ref{blk2}) vanishes as $k\rightarrow \infty$. Hence, 
\[ \bb_{\lambda,k}\rightarrow \bU_2\bU_2'\bbeta,\,\,\text{as $k\rightarrow\infty$}.\]
If we further allow $n,p\rightarrow\infty$, under a similar setting in the proof of Theorem~\ref{thm2} above, we can show that
\[ \bb_{\lambda,k}-\bU_2\bU_2'\bbeta\rightarrow {\bf 0},\,\,\text{as $n,p,k\rightarrow\infty$},\]
which is discussed in Remark~\ref{rm3}(iii).
This completes the proof. $\Box$
\vskip 0.6cm

{\bf Proof of Theorem~\ref{thm4}}. We prove it by contradiction. Suppose $\bX'\bX$ is singular and there exists a transformation matrix $\bS$ such that
\begin{equation}\label{sy}
 E(\bS\bY)=\bU_2\bU_2'\bbeta.
\end{equation}
In other words, the remaining bias term in Theorem~\ref{thm3} can be corrected by $\bS\bY$.
(\ref{sy}) implies that
\begin{equation}\label{syy}
E(\bS\bX)\bbeta=\bU_2\bU_2'\bbeta.    
\end{equation}
Since a $p$-dimensional space can be spanned by the columns of $[\bU_1,\bU_2]$, there exist vectors $\balpha_1\in R^{p^*}$ and $\balpha_2\in R^{p-p^*}$ such that
\[\bbeta=\bU_1\balpha_1+\bU_2\balpha_2.\]
We only talk about the case when $\balpha_1\neq{\bf 0}$ and $\balpha_2\neq 0 $ since either $\balpha_1={\bf 0}$ or $\balpha_2\neq 0 $ will lead the bias term to be zero, implying that we do not need to correct it.
We plug it into (\ref{syy}) and obtain
\begin{equation}\label{eq:sy}
 E(\bS\bX)\bU_1\balpha_1=\bU_2\balpha_2.
\end{equation}
Note that $\bU_1\balpha_1$ is a vector in the hyperplane spanned by the columns of $\bU_1$, and $\bU_2\balpha_2$ is a vector in the hyperplane spanned by the columns of $\bU_2$. Since $\bU_1$ is orthogonal to $\bU_2$, we cannot find any linear transformation matrix $\bS$ such that the equation of (\ref{eq:sy}) holds. This contradicts our assumption. This completes the proof. $\Box$
\vskip 0.6 cm

{\bf Proof of Theorem~\ref{thm5}}. (i) We first consider the case when $p<n$. By (\ref{btck}) in the proof of Theorem~\ref{thm2} above,
\[\wh\bbeta_{c,k}(\lambda^*)=\bbeta+\sum_{j=0}^k{\lambda^*}^j(\bX'\bX+\lambda^*\bI_p)^{-(j+1)}\bX'\bve+o(1)=\bbeta+\bgamma+o(1),\]
where $o(1)$ can be made arbitrarily small at an exponential rate, and $\bbeta$ has $s^*$ nonzero element, which can be smaller than $p$. We use the SVD of $\bX'\bX$ when $p<n$ as that in the proof of Theorem~\ref{thm2} and obtain
\[\sum_{j=0}^k{\lambda^*}^{j+1}(\bX'\bX+\lambda^*\bI_p)^{-(j+1)}=\bU_1\bD_1^*\bU_1',\]
where $\bD_1^*=\diag(d_1^*,...,d_p^*)$ with
\[d_i^*=\sum_{j=0}^k\left(\frac{\lambda^*}{\lambda^*+d_i^2}\right)^{j+1}=\frac{\lambda^*}{d_i^2}+o(1)\asymp O(1),\,\,1\leq i\leq p,\]
where $o(1)$ is also an arbitrarily small term at an exponential rate, and $\lambda^*$ and $d_i^2$ are all of order $n$ by Assumption \ref{asm3}. Then, 
\[\bgamma=\sum_{j=0}^k{\lambda^*}^j(\bX'\bX+\lambda^*\bI_p)^{-(j+1)}\bX'\bve=\bU_1\bD_1^*\bU_1'\bV\frac{\bD_1}{\sqrt{\lambda^*}}\bU_1'\frac{\bve}{\sqrt{\lambda^*}}.\]
Note that the $\ell_2$-norm of each row of $\bU_1\bD_1^*\bU_1'\bV\frac{\bD_1}{\sqrt{\lambda^*}}\bU_1'$ is
\[\|\be_i'\bU_1\bD_1^*\bU_1'\bV\frac{\bD_1}{\sqrt{\lambda^*}}\bU_1'\|_2^2\leq C<\infty,\]
where we used the fact that each element in the diagonal matrices $\bD_1^*$ and $\frac{\bD_1}{\sqrt{\lambda^*}}$ is bounded from above and below. By Assumption \ref{asm5}, we have
\[\max_{1\leq i\leq p}|\gamma_i|=O_p(\sqrt{\frac{1}{\lambda^*}}\sqrt{\log(p)}=O_p(\sqrt{\frac{{\log(p)}}{n}}).\]
For $i\in\mathcal{M}_0$, by Assumption \ref{asm4} and $\log(p)/n^{1-2\tau}\rightarrow 0$,
\[\min_{i\in\mathcal{M}_0}|\wh\beta_{c,k,i}(\lambda^*)|\geq \min_{i\in\mathcal{M}_0}|\beta_i|-O_p(\sqrt{\frac{{\log(p)}}{n}})\geq C n^{-\tau},\]
and
\[\max_{i\in\mathcal{M}_0^c}|\wh\beta_{c,k,i}(\lambda^*)|\leq C\sqrt{\frac{{\log(p)}}{n}}.\]
It is obvious that 
\[\min_{i\in\mathcal{M}_0}|\wh\beta_{c,k,i}(\lambda^*)|>\max_{i\in\mathcal{M}_0^c}|\wh\beta_{c,k,i}(\lambda^*)|,\]
with probability tending to one. This implies that the magnitudes of the parameters   $\wh\beta_{c,k,i}(\lambda^*)$'s for $i \in \mathcal{M}_0$ are of larger orders than the remaining ones. Therefore, we prove that
\[P(\mathcal{M}_0\subset\mathcal{M}_k(\lambda^*))\rightarrow 1,\,\, \text{as}\,\, k\rightarrow\infty,\]
if we choose $n^*\geq s^*$ in (\ref{rs:sub}) of the main article. For instance, we may simply choose $n^*=p$ for $p<n$, which creates a dense model suitable for bias-correction.\\

Next, we consider the case when $p>n$. Note that Theorem~\ref{thm3} implies that
\begin{equation}\label{betk:rp}
    \wh\bbeta_{c,k}(\lambda^*)=\bbeta-\bU_2\bU_2'\bbeta+\sum_{j=0}^k{\lambda^*}^j(\bX'\bX+\lambda^*\bI_p)^{-(j+1)}\bX'\bve+o(1)=\bbeta-\bdelta+\bgamma+o(1),
\end{equation}
where the term $o(1)$ can be arbitrarily small at a rate of exponential decay as $k\rightarrow\infty$.
The idea of proving the result is as follows. We first investigate the lower bound of $(\min_{i\in\mathcal{M}_0}|\wh\beta_{c,k,i}(\lambda^*)|)^2$ and the upper bound of $\|\wh\bbeta_{c,k,i}(\lambda^*)\|_2^2$. If the number of coordinates selected in $\mathcal{M}_k(\lambda^*)$ is greater than 
the ratio between the upper bound and the lower bound, the submodel $\mathcal{M}_k(\lambda^*)$ must contain the true $\mathcal{M}_0$ as a subset.

For $i\in\mathcal{M}_0$, by Assumption \ref{asm4}, the magnitude of the $i$-th coordinate of $\bU_2\bU_2'\bbeta$ is less than that of $\bbeta$. If $\log(p)/n^{1-2\tau}\rightarrow 0$, we have 
\[\min_{i\in \mathcal{M}_0}|\wh\beta_{c,k,i}(\lambda^*)|\geq C_2n^{-\tau}-C_1n^{-\tau}-O_p(\sqrt{\log(p)/n})\geq O_p(n^{-\tau}),\]
where $C_2>C_1>0$. In addition,
\[\|\wh\bbeta_{c,k}(\lambda^*)\|_2^2\leq C\|\bU_1\bU_1'\bbeta\|_2^2+C\|\sum_{j=0}^k{\lambda^*}^j(\bX'\bX+\lambda^*\bI_p)^{-(j+1)}\bX'\bve\|_2^2\leq Cs^*+Cn^{-2}p\log(p).\]
Then,
\begin{equation}\label{ratio}
    \frac{\|\wh\bbeta_{c,k}(\lambda^*)\|_2^2}{(\min_{i\in \mathcal{M}_0}|\wh\beta_{c,k,i}(\lambda^*)|)^2}\leq Cn^{-2\tau}s^*+Cn^{2\tau-2}p\log(p).
\end{equation}
The upper bound in (\ref{ratio}) is important for us to show the result because it implies that the number of elements in $|\wh\bbeta_{c,k}(\lambda^*)|$ that are greater than $\min_{i\in \mathcal{M}_0}|\wh\beta_{c,k,i}(\lambda^*)|$ is at most $Cn^{-2\tau}s^*+Cn^{2\tau-2}p\log(p)$. Note that all elements with indexes $i\in \mathcal{M}_0$ are greater than or equal to  $\min_{i\in \mathcal{M}_0}|\wh\beta_{c,k,i}(\lambda^*)|$. Therefore, if the choice of  $n^*$ in (\ref{rs:sub}) satisfies 
\[\frac{n^*}{Cn^{-2\tau}s^*+Cn^{2\tau-2}p\log(p)}\rightarrow\infty,\]
or equivalently,  the number of largest elements selected in $|\wh\bbeta_{c,k}(\lambda^*)|$, $n^*$, is more than the total number of parameters in $|\wh\bbeta_{c,k}(\lambda^*)|$ that is greater than $\min_{i\in \mathcal{M}_0}|\wh\beta_{c,k,i}(\lambda^*)|$,
we must have that $\mathcal{M}_k(\lambda^*)$ consists of all the indexes in $\mathcal{M}_0$. This proves (i).\\

(ii) Since $\mathcal{M}_0\subset\mathcal{M}_k(\lambda^*)$ with probability tending to one, on the event of  $\{\mathcal{M}_0\subset\mathcal{M}_k(\lambda^*)\}$, the proof is the same as that for Theorem~\ref{thm2}. This completes the proof. $\Box$

\vskip 0.6cm
{\bf Proof of Theorem~\ref{thm6}}. By (\ref{btck}), we can obtain that
\[ \wh\bbeta_{c,k}(\lambda)-\bbeta=-\lambda^{k+1}(\bX'\bX+\lambda\bI_p)^{-(k+1)}\bbeta+\sum_{j=0}^k\lambda^j(\bX'\bX+\lambda\bI_p)^{-(j+1)}\bX'\bve,\]
and
\[ \wh\bbeta_{\mathcal{M}_k,l}(\lambda)-\bbeta_{\mathcal{M}_k}=-\lambda^{l+1}(\bX_{\mathcal{M}_k}'\bX_{\mathcal{M}_k}+\lambda\bI_{n^*})^{-(l+1)}\bbeta_{\mathcal{M}_k}+\sum_{j=0}^l\lambda^j(\bX_{\mathcal{M}_k}'\bX_{\mathcal{M}_k}+\lambda\bI_{n^*})^{-(j+1)}\bX_{\mathcal{M}_k}'\bve.\]
Under the assumption that $\bve\sim N({\bf 0},\bSigma_\ve)$, the exact distributions in Theorem~\ref{thm6}(i)-(ii) are straightforward. 

Moreover, by a similar argument as that in the proof of Theorem~\ref{thm2} above, it is not hard to see that
\[\sqrt{n}\bmu_{1,k}(\lambda)=o(1)\quad\text{and}\quad \sqrt{n}\bmu_{2,k,l}(\lambda)=o(1),\]
where $o(1)$ can be arbitrarily small at an exponential rate if $k$ and $l$ are sufficiently large. Consequently, for any given and fixed $(p,n)$, the means $\sqrt{n}\bmu_{1,k}(\lambda)$ and $\sqrt{n}\bmu_{2,k,l}(\lambda)$ are asymptotically vanishing as $k,l\rightarrow\infty$, and the asymptotic normal distributions in Theorem~\ref{thm6} hold. 

As discussed in Remark~\ref{rm5}(i), if the normality assumption that $\bve\sim N({\bf 0},\bSigma_\ve)$ is  relaxed to the case that $\{\ve_i,i=1,...,T\}$ is a martingale difference sequence with finite variances, under the assumption that $\lambda\asymp\lambda^*\asymp n$ and the nonzero singular values of $\bX$ are of order $\sqrt{n}$,  the limiting distributions in Theorem~\ref{thm6} can also be established as $n\rightarrow\infty$, following the argument in \cite{hall2014martingale}. We omit the details. This completes the proof of Theorem~\ref{thm6}. $\Box$

	\vskip 0.6 cm
{\bf Proof of Theorem~\ref{thm7}}. (i)	Assume $\bSigma_\ve=\sigma^2\bI_n$,	by Assumption~\ref{asm1} and (\ref{mse:bt}), the MSE of $\wh\bbeta_{c,k}(\lambda)$ can be expressed as
\begin{align}\label{mse:beck:i}
    \text{MSE}(\wh\bbeta_{c,k}(\lambda))=&\bbeta'\lambda^{k+1}(\bX'\bX+\lambda\bI_p)^{-(k+1)}\lambda^{k+1}(\bX'\bX+\lambda\bI_p)^{-(k+1)}\bbeta\notag\\
    &+E[\bve'\bX\sum_{j=0}^k\lambda^j(\bX'\bX+\lambda\bI_p)^{-(k+1)}\sum_{j=0}^k\lambda^j(\bX'\bX+\lambda\bI_p)^{-(k+1)}\bX'\bve]\notag\\
    =&\bbeta'\bU_1\bLambda_{k}(\lambda)\bU_1'\bbeta+E[\bve'\bV_1\bD_{k}(\lambda)\bV_1'\bve]\notag\\
    =&\bbeta'\bU_1\bLambda_{k}(\lambda)\bU_1'\bbeta+\sigma^2\tr[\bD_{k}(\lambda)],
\end{align}
		where $\bLambda_{k}(\lambda)=\diag(\gamma_{k,1}(\lambda),...,\gamma_{k,p}(\lambda))$ with
\[\gamma_{k,i}(\lambda)=[\frac{\lambda}{\lambda+d_i^2}]^{2(k+1)},\]
and $\bD_{k}(\lambda)=\diag(d_{k,1}(\lambda),...,d_{k,p}(\lambda))$ with
\[d_{k,i}(\lambda)=\frac{1}{d_i^2}\left[1-(\frac{\lambda}{\lambda+d_i^2})^{k+1}\right]^2.\]
Let $\bU_1'\bbeta=(\delta_1,...,\delta_p)'$, then, 	
\begin{equation}\label{mse:btck:final}
    \text{MSE}(\wh\bbeta_{c,k}(\lambda))=\sum_{i=1}^p\delta_i^2[\frac{\lambda}{\lambda+d_i^2}]^{2(k+1)}+\sigma^2\sum_{i=1}^p\frac{1}{d_i^2}\left[1-(\frac{\lambda}{\lambda+d_i^2})^{k+1}\right]^2.
\end{equation}
Note that $\text{MSE}(\wh\bbeta(\lambda))=\text{MSE}(\wh\bbeta_{c,0}(\lambda))$, and the MSE will become stable if $k\rightarrow\infty$, then, for any finite $k\geq 1$, define 
\[f_i(k)=\delta_i^2[\frac{\lambda}{\lambda+d_i^2}]^{2(k+1)}+\sigma^2\frac{1}{d_i^2}\left[1-(\frac{\lambda}{\lambda+d_i^2})^{k+1}\right]^2,\]
and $\text{MSE}(\wh\bbeta_{c,k}(\lambda))=\sum_{i=1}^pf_i(k)$. It suffices to investigate the derivatives of $f_{i}(k)$ for $1\leq i\leq p$ and finite $k>0$. Note that
\begin{equation}\label{fip}
    f_i'(k)=2\delta_i^2(k+1)\log(\frac{\lambda}{\lambda+d_i^2})-\frac{2\sigma^2}{d_i^2}(k+1)[1-(\frac{\lambda}{\lambda+d_i^2})^{k+1}]\log(\frac{\lambda}{\lambda+d_i^2}).
\end{equation}
  Then $f_i'(k)<0$ if and only if
\begin{equation}\label{cond:iff}
    \frac{\delta_i^2d_i^2}{\sigma^2}>1-(\frac{\lambda}{\lambda+d_i})^{k+1}.
\end{equation}
It follows from (\ref{cond:iff}) that
\begin{equation}\label{k:upper}
    k\leq\frac{\log(1-\frac{\delta_i^2d_i^2}{\sigma^2})}{\log(\frac{\lambda}{\lambda+d_i^2})}-1, 1\leq i\leq p.
\end{equation}
On the other hand, we also expect that $k\geq k^*\geq 1$ such that we can achieve a minimum  before the MSE increases, then we also require that
\[\frac{\delta_i^2d_i^2}{\sigma^2}+(\frac{\lambda}{\lambda+d_i^2})^{k^*+1}\geq 1.\]
Since the MSE consists of $p$ terms in the summation, then it achieves the global minimum at $k$ which is between the minimum and the maximum integers of all the inflection points in the upper bounds of (\ref{k:upper}) for all $1\leq i\leq p$. This completes the proof of Theorem~\ref{thm7}(i).\\

(ii) The result of Theorem~\ref{thm7}(ii) is straightforward  from the inequality in (\ref{cond:iff}). We omit the details here. This completes the proof. $\Box$

\vskip 0.5cm

%%%%%%%%%%
\section{Additional Table and Figure for Simulations}\label{secB}
\begin{figure}[ht]
\begin{center}
%{\includegraphics[width=8cm,height=14cm]{Vol-loadings.pdf}}
{\includegraphics[width=14cm,height=8cm]{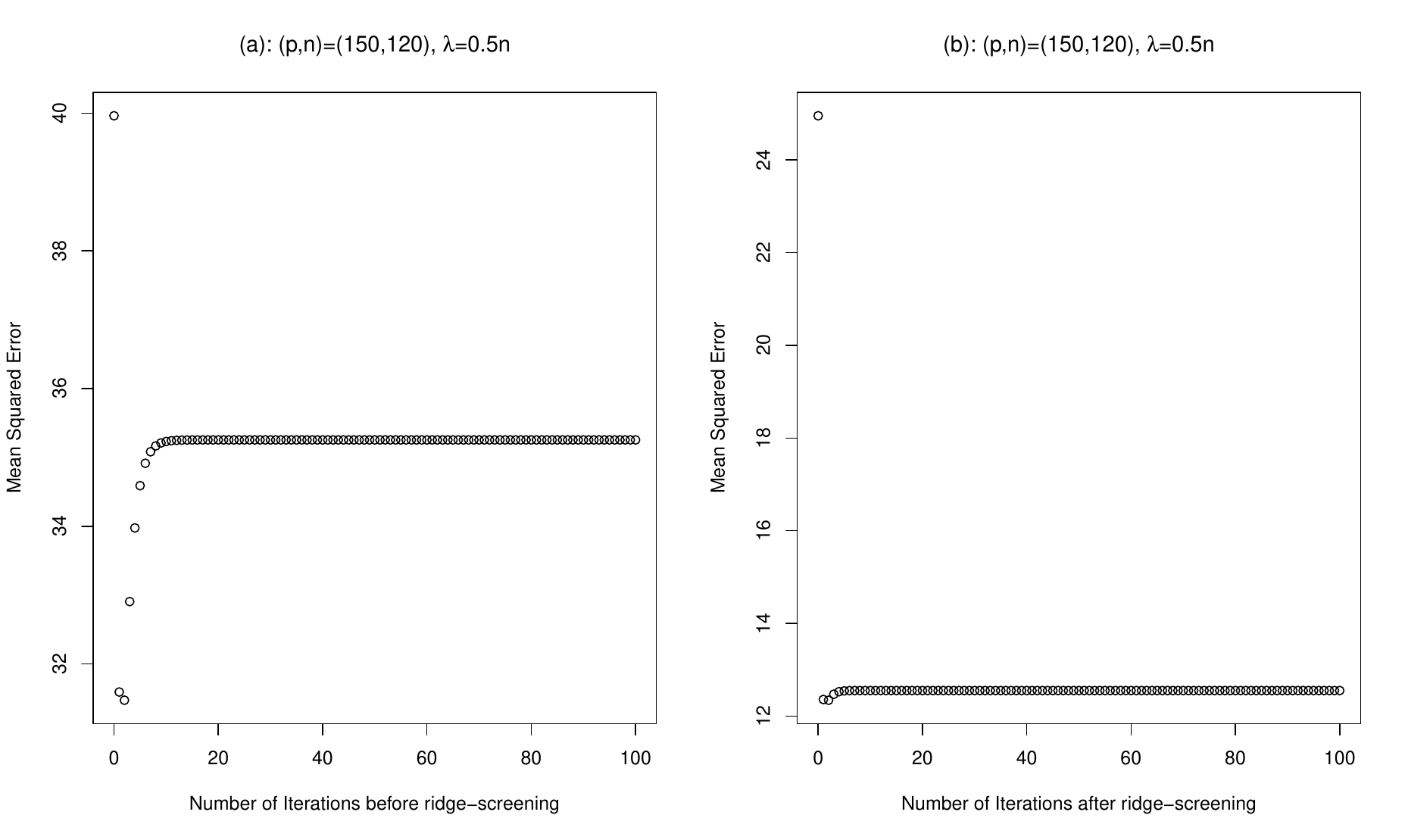}}
\caption{  The bias-variance trade-off shown by the empirical MSEs of the de-biased ridge estimators for different number of iterations in Example 2, where we consider $\lambda=0.5n$ for $(p,n)=(150,120)$. Part (a) plots the MSEs of the de-baised ridge estimators  before applying the ridge-screening method, and Part (b) provides the MSEs of the de-biased  ridge estimators after the ridge-screening.  1000 replications are used in the experiments.  }\label{fig-20}
\end{center}
\end{figure}

In this section, 	we present one additional table and figure used in Sec.~\ref{sec3} of the main article. In Figure~\ref{fig-20}, we plot the pattern of the bias-variance trade-off shown by the empirical MSEs of the de-biased ridge estimators for different number of iterations under the settings in Example 2 of Sec.~\ref{sec3}. Table~\ref{Table-a30} below reports the Empirical Probability (EP) of correct recoveries of the true models for $\lambda^*=0.1n$, $0.3n$, and $0.8n$, under the settings in Example 2 of Sec.~\ref{sec3} of the main article.

%%%%%%%%%%%%%%%%%%%%%%%%%%%%%
%%%%%%%%%%%%%%%%%%%%%

 %%%
\begin{table}[htp]\footnotesize
\caption{Empirical probability (EP) of correct recoveries of the true models for $\lambda^*=0.1n$, $0.3n$, and $0.8n$ in Example 2, where we conduct 100 iterations in the bias-correction procedure and set $n^*=40$ in the ridge-screening. 1000 replications are used in the experiments.} 
          \label{Table-a30}
{\begin{center}
\begin{tabular}{ccccc}
\toprule
\multicolumn{5}{c}{$\lambda^*=0.1n,k=100$}\\
\hline
&\multicolumn{4}{c}{$(p,n)$}\\
\cline{2-5}
EP&(150,200)&(150,140)&(220,180)&(220,200)\\
\hline
$EP(\mathcal{M}_0\subset \mathcal{M}_{k}(\lambda^*))$&100\%&100\%&100\%&100\%\\
\midrule
\multicolumn{5}{c}{$\lambda^*=0.3n,k=100$}\\
\hline
&\multicolumn{4}{c}{$(p,n)$}\\
\cline{2-5}
EP&(150,200)&(150,140)&(220,180)&(220,200)\\
\hline
$EP(\mathcal{M}_0\subset \mathcal{M}_{k}(\lambda^*))$&100\%&100\%&100\%&100\%\\
\midrule
\multicolumn{5}{c}{$\lambda^*=0.8n,k=100$}\\
\hline
&\multicolumn{4}{c}{$(p,n)$}\\
\cline{2-5}
EP&(150,200)&(150,140)&(220,180)&(220,200)\\
\hline
$EP(\mathcal{M}_0\subset \mathcal{M}_{k}(\lambda^*))$&100\%&100\%&100\%&100\%\\
\bottomrule
\end{tabular}
  \end{center}}
\end{table}
%%%%%%%%%%%%%%%%%%%%%%%%%%%%%%%%%%%%%%%%%%%%%%%%%%

\end{document}